\documentclass[10pt,twocolumn,oneside,final]{IEEEtran}
\pdfoutput=1
\usepackage{cite}

\usepackage[numbers,sort&compress]{natbib}

\usepackage{graphicx}
\usepackage{amsmath}
\usepackage{times}
\usepackage{latexsym}
\usepackage{bm}
\usepackage{amssymb}
\usepackage[center,footnotesize]{caption2}
\usepackage{array}
\usepackage{fancyhdr}
\usepackage{cite,graphicx,amsmath,amssymb}
\usepackage{subfigure}
\usepackage{citesort}
\usepackage{psfrag}
\usepackage{multirow}

\usepackage{algorithm}               
\usepackage{algorithmic}             
\usepackage{multirow}                
\usepackage{amsmath}
\usepackage{xcolor}

\ifCLASSINFOpdf

\else

\fi

\makeatother

\begin{document}
\title{
Automatic Registration of Images with Inconsistent Content Through Line-Support Region Segmentation and Geometrical Outlier Removal}

\author{\IEEEauthorblockN{Ming Zhao, Yongpeng Wu, \emph{Senior Member IEEE}, Shengda Pan, Fan Zhou, Bowen An, Andr\'{e} Kaup, \emph{Fellow, IEEE}}

\thanks{This work was supported in part by National Natural Science Foundation of China under Grant 61302132, 41701523, 61504078, the Shanghai Educational Development Foundation under Grant 13CG51, the Scientific Research Foundationx of Guangxi Education Department under Grant YB2014207.}

\thanks{M. Zhao, S. Pan, F. Zhou, and B. An are with the Department of Information Engineering, Shanghai Maritime University, Shanghai, 201306, China (e-mail: mingzhao@shmtu.edu.cn).}

\thanks{Y. Wu is  with the Department of Electronic Engineering, Shanghai Jiao Tong University, Shanghai, 200240, China (e-mail: yongpeng.wu@sjtu.edu.cn). }

\thanks{A. Kaup is with the Chair of Multimedia Communications and Signal Processing, Friedrich-Alexander University Erlangen-N$\ddot{u}$rnberg,
Cauerstr. 7, 91058 Erlangen, Germany (e-mail: andre.kaup@fau.de). }

}
\maketitle

\begin{abstract}
The implementation of automatic image registration is still difficult in various applications.
In this paper, an automatic image registration approach through line-support region segmentation and geometrical outlier removal (ALRS-GOR) is proposed.
This new approach is designed to address the problems associated with the registration of images with affine deformations and inconsistent content, such as  remote sensing images with different spectral content or noise interference, or map images with inconsistent annotations.
To begin with, line-support regions, namely a straight region whose points share roughly the same image gradient angle, are extracted to address the issues of inconsistent content existing in images.
To alleviate the incompleteness of line segments, an iterative strategy with multi-resolution is employed to preserve global structures that are masked at full resolution by image details or noise.
Then, Geometrical Outlier Removal (GOR) is developed to provide reliable feature point matching, which is based on affine-invariant geometrical classifications for corresponding matches initialized by SIFT. The candidate outliers are selected by comparing the disparity of accumulated classifications among all matches, instead of conventional methods which only rely on local geometrical relations.
Various image sets have been considered in this paper for the evaluation of the proposed approach, including aerial images with simulated affine deformations, remote sensing optical and synthetic aperture radar images taken at different situations (multispectral, multisensor, and multitemporal), and map images with inconsistent annotations.
Experimental results demonstrate the superior performance of the proposed method over the existing approaches for the whole data set.

\end{abstract}
\providecommand{\keywords}[1]{\textbf{\textit{Index terms---}} #1}
\begin{keywords}
Linear features, scale invariant feature transformation, feature point matching, image segmentation, automatic image registration.
\end{keywords}

\section{Introduction}
Image registration is a vital yet challenging task, which aims at aligning two or more images with overlapping scenes captured at different times, by different sensors, or from different viewpoints. It has been widely applied in many fields, such as computer vision, remote sensing, medical image analysis, pattern matching, but far from being commonly automatized \cite{J_BZitova_2003_IVC}, \cite{J_AW_2007_ITGRS}.

Automatic image registration is still challenging due to the presence of particular difficulties as follows:

\begin{itemize}
\item {Images to be registered are usually acquired by different sensors or from different viewpoint, which causes geometrical deformations, such as translation, rotation, scaling, and sheared. The scenes exited in the reference images do not always stay in the corresponding sensed images.  }
\item {Spectral content difference and illumination changes usually exist in multispectral/multisensor images/multitemporal images. The inconsistent spectral content increases the difficulty of corresponding feature matching in automatic registrations.  }
\item { The particular interferences cause the scene content to be inconsistent between images to be registered. For example, the speckle noises inevitably presented in SAR images make the feature extraction and identification difficult. For a better visualization, the interest icons and texts of street names existing in map images don't always keep the same transformations with the whole map images \cite{J_GY_2014_ITCSVT}, \cite{C_GY_2013_PII}.}
\end{itemize}

Numerous previous works have been proposed for the high desired automatic image registration. These methods can be generalized into two major categories: intensity-based and feature-based \cite{J_AW_2007_ITGRS}, \cite{J_ZW_2012_ITIP}. The intensity-based methods compare the similarity between pixel intensities to determine the image alignments. The widely used similarity measures include normalized cross-correlation coefficient \cite{C_WS_2012_PII}, mutual information \cite{J_HC_2004_IJRS}, \cite{J_SS_2010_ITGRS}, and maximum likelihood \cite{J_WL_2004_ITIP}. However, the computational complexities of these intensity-based methods are expensive. Moreover, the performance of these methods declines significantly when applied to images with significant geometrical deformations, images acquired by the sensors with different modalities, or images taken in different illumination conditions. Feature-based methods attempt to extract salient features from the images to be registered, and establish corresponding matches between these features. Salient points, lines, curves, edges, line intersections, and regions around each feature are the most commonly used image features \cite{J_HS_2015_ITGRS}.
The feature-based methods are capable of handling significant geometry inconsistency betweens scenes.
Moreover, they have low implementation complexity for limited numbers of pixels associated with extracted features.
Although feature-based methods are effective to most of homologous image registration, they have limited performance when directly applied to register the images with illumination changes, difference of spectral contents, or inconsistent objects.

As a widely used local feature descriptor, Scale Invariant Feature Transform (SIFT) has been proved to be a powerful feature point matching approach for images with geometrical deformations, noise, and illumination change in a certain extent \cite{J_DGL_2004_IJCV}.
Various adapted versions of SIFT have been proposed to improve the performance of
SIFT, such as Principal component analysis (PCA)-SIFT \cite{J_FD_2015_ITGRS}, Bilateral filter SIFT (BF-SIFT) \cite{J_SHW_2012_IGRSL}, and Adaptive binning (AB-SIFT) \cite{J_AS_2015_ITGRS}.
Nevertheless, SIFT-like methods cannot easily produce meaningful matching results when directly applied to significantly different spectral contents.
Moreover, SIFT feature points are easily concentrated on the scene with salient texture details, such as the interest icons and texts of street names in map images \cite{J_GY_2014_ITCSVT}.
However, these features are usually inconsistent between the images with different views.
To preserve salient and consistent features, the segmentation stage is a good alternative to exclude the effects of illumination changes and inconsistent content for feature point extractions and matching \cite{J_HG_2011_ITGRS}.

Image segmentation partitions an image into regions according to given criteria, and transforms the image to a binary image to distinguish objects and background. Various image segmentation methods have been proposed, such as feature space clustering, region-based approaches, edge detection approaches, histogram thresholding \cite{J_HDC_2001_PR,J_OJT_2002_ITIP,J_SKP_2000_IJPS,J_NRP_1993_PR,J_DMYS_2012_ITIP}.
However, they have been scarcely adopted into image registration except the followings. Goshtasby \emph{et al.} \cite{J_AG_1986_ITGRS} proposed a region refinement to obtain similar corresponding close-boundary regions by iterative thresholding segmentations. The correspondences are determined between the centers of gravity of close-boundary regions according to the clustering technique \cite{J_GS_1982_ITPAMI}. Knowing that the performance of clustering-based matching highly depends on the corresponding samples of regions, the method in \cite{J_AG_1986_ITGRS} is only appropriate for simple image contents. Troglio \emph{et al.} \cite{J_GT_2012_IGRSL} proposed a region-based approach to extract ellipsoidal features for planetary image registration purposes. The watershed segmentation algorithm is adopted to identify the structures of rocks and craters according to the intensity gradients. The optimal transformation matrix for registration is obtained by a genetic algorithm.
However, this segmentation method is only adequate for simple ellipsoidal objects, such as rocks and craters entirely contained in the images. Gon\c{c}alves \emph{et al.} \cite{J_HG_2011_ITIP} developed an automatic image registration method called HAIRIS through histogram-based image segmentation. This method utilizes a relaxation parameter on the histogram mode delineation for segmentation.
The extracted objects at the segmentation stage are characterized by four attributes,  which allow for their adequate morphological description. Then, the transformation parameters are determined by restricting possible values on a statistical basis. Although leading to a subpixel accuracy, HAIRIS only applies for the registration of image pairs with geometrical differences in rotations and translations.
Morago \emph{et al.} \cite{J_BM_2015_ITIP}presented a contextual framework using an ensemble feature.
The surrounding regions of keypoints are described in terms of salient structural features and the rich texture information by line segment extraction and histograms of gradients (HOG) respectively.
Maximally stable extremal regions (MSER) are adopted to determine the neighborhood sizes.
The iterative and global refinement stages using corner, edge, and gradient information across the entire image planes are implemented after combining several local keypoint and regional template matching techniques.
If only a small percentage of the identified ensemble features are actually inliers, it is very unlikely to find a correct image alignment.

Regarding feature correspondence techniques to support the initial SIFT matches, various approaches have tried to explore the geometrical relations between feature points for solving the feature point matching problem.
A popular group of methods based on geometric transformation models is to formulate this problem in terms of correspondence matrix between initial corresponding feature points with parametric or nonparametric geometrical constraint.
Examples of this strategy include the classical RANSAC algorithm that typically rely on parametric models, and Vector Field Consensus (VFC) \cite{J_JM_2014_ITIP}, robust point matching via L2E \cite{J_JM_2015_ITIP}, Locally Linear Transforming (LLT) that rely on nonparametric models \cite{J_JM_2015_ITGRS}.
Ma \emph{et al.} \cite{J_JM_2015_ITGRS} developed a local geometrical constraint to preserve local structures among neighboring SIFT points.
The basic idea in \cite{J_JM_2015_ITGRS} is to formulate the feature point matching problem as a maximum-likelihood estimation of a Bayesian model with hidden/latent variables to indicate whether matches in the putative sets are inliers or outliers.
It is worth mentioning that the transformation between non-rigid images is modeled in a reproducing kernel Hilbert space (RKHS), and a sparse approximation is applied to the transformation that reduces the method computation complexity to linearithmic.
Another group of graph-based methods for feature point matching tries to explore the similarity of graph structures between feature points to reject outliers.
Aguilar \emph{et al.} \cite{J_WAr_2009_IVC} proposed a point matching method named Graph Transformation Matching (GTM) based on finding consensus K Nearest Neighbor (KNN) graphs. This method iteratively eliminates dubious matches by selecting the maximal disparities of edges connecting with KNN points.
Izadi \emph{et al.} \cite{J_MI_2012_ITGRS} proposed Weighted Graph Transformation Matching (WGTM) algorithm, which not only adopts KNN as the geometrical relation, but also utilizes the angular distances between edges that connect a feature point to its KNN as the weight.
Besides of KNN, graph structures such as bilateral KNN \cite{J_MZ_2013_IGRSL}, Delaunay triangulation \cite{J_MZ_2015_IGRSL}, and triangle area \cite{J_ZLS_2014_ITGRS} are also explored. Candidate outliers are distinguished from initial matches by comparing their corresponding graph structures.
The proposed graph based methods mentioned above are invariant with respect to translation, scales, and rotations, but variant to shear deformations.
As a result, they have limited performance for images with shear deformations. Moreover, only local geometrical relations for each of candidates are considered in these methods. Therefore, they may easily fail in obtaining reliable matches when outliers have the same local structures. Also, the inliers with outlier existing in their local structures are likely to be mistakenly removed.

In this paper, we propose an automatic image registration through line-support region segmentation and geometrical outlier removal (ALRS-GOR). The main contributions of this paper are as follows:

\begin{itemize}
\item[1)] The line-support region, namely a straight region whose points share roughly the same image gradient angle, is first explored to segment images to be registered, which can alleviate the challenges of inconsistent contents in image registration.
\item[2)] Geometrical Outlier Removal (GOR) is developed to eliminate outliers and preserve inliers based on the affine-invariant geometrical classifications for candidate matches. The directed edges connected by any two feature points are utilized to classify all of initial feature points according to their locations. The candidate outliers are selected by comparing the disparity of accumulated classifications for each matched pair.
\item[3)] To deal with the incompleteness of detected line segments, an iterative strategy with multi-resolution is employed to preserve global structures that masked at full resolution by image details or noise.
\end{itemize}

ALRS-GOR allows for the registration of image pairs with affine deformations and inconsistent content, such as multispectral remote sensing images and map images with inconsistent annotations.
The experimental results demonstrate the superior performance of the proposed method over the existing design for the whole dataset.

The remainder of this paper is organized as follows:
Section II interprets the motivation of each process of the proposed approach. Section III describes the proposed approach in detail.
Section IV presents the performance evaluation of the proposed algorithm and illustrates experimental results with representative applications in image registration.
Finally, Section V presents the concluding remarks.

\section{MOTIVATION}\label{sec:model}
In this section, we interpret the motivation of the proposed registration approach. First, we state the reason of utilizing line-support regions to segment image before feature point extraction.  The benefit of the iterative strategy to re-segment and re-match features with multi-resolution is explained.
Then, we describe the advantage of the proposed geometrical outlier removal for feature point matching.

\subsection{Motivation of Utilizing Iterative Strategy of Line-support Regions as Segmentation}
Image segmentation utilized as a previous step for image registration allows to simplify the image representations, and significantly reduce the inconsistent appearance of the same scene in the images to be registered. In contrast to other features, linear segments offer important information about geometrical contents in images. Also, elaborated shapes in scenes can be easily analyzed and detected through the basic line segments.

To the best of our knowledge, line segment detector (LSD) has been mainly applied to extract linear features \cite{J_RG_2010_ITPAMI}. It aims to detect line segments from images on the consensus that most shapes accept an economic description by straight lines. The line segments are extracted from the line-support regions, whose points roughly share the same image gradient angles. Despite of well describing linear features for images, line segments are not sufficient to support SIFT extraction and matching. This is because it is difficult to extract SIFT feature points from line segments with limited texture details.
In this paper, we adopt the line-support region as the segmentation stage for image registration for the following reasons:

\begin{itemize}
\item[1)] Line-support regions are extracted relying on the similarity of gradients rather than pixel intensities.
    Regardless of different pixel intensities, the corresponding regions with inconsistent spectral content have similar gradients.
    Therefore, the corresponding line-support regions from multispectral/multisensor images or multitemporal images with illumination changes are capable of being identified and matched.
\item[2)]
The image details without linear features are discarded during line segment detections. As proved in \cite{J_RG_2010_ITPAMI} and \cite{J_GY_2014_ITCSVT}, the subsequent feature matching would not be affected by the unexpected details, such as speckle noise in SAR images or inconsistent annotations in map images.
\item[3)] The implementation of line-support region extraction is efficient and fully automatic without human assistance.
\end{itemize}

\begin{figure}[htb]
\centering
 \setlength{\abovecaptionskip}{0pt}
 \setlength{\belowcaptionskip}{0pt}
 \setlength{\intextsep}{8pt plus 3pt minus 2pt}
  \subfigure[]{
    \label{fig:mini:subfig:a}
    \begin{minipage}[c]{0.15\textwidth}
      \centering
      \includegraphics[width=1.0in]{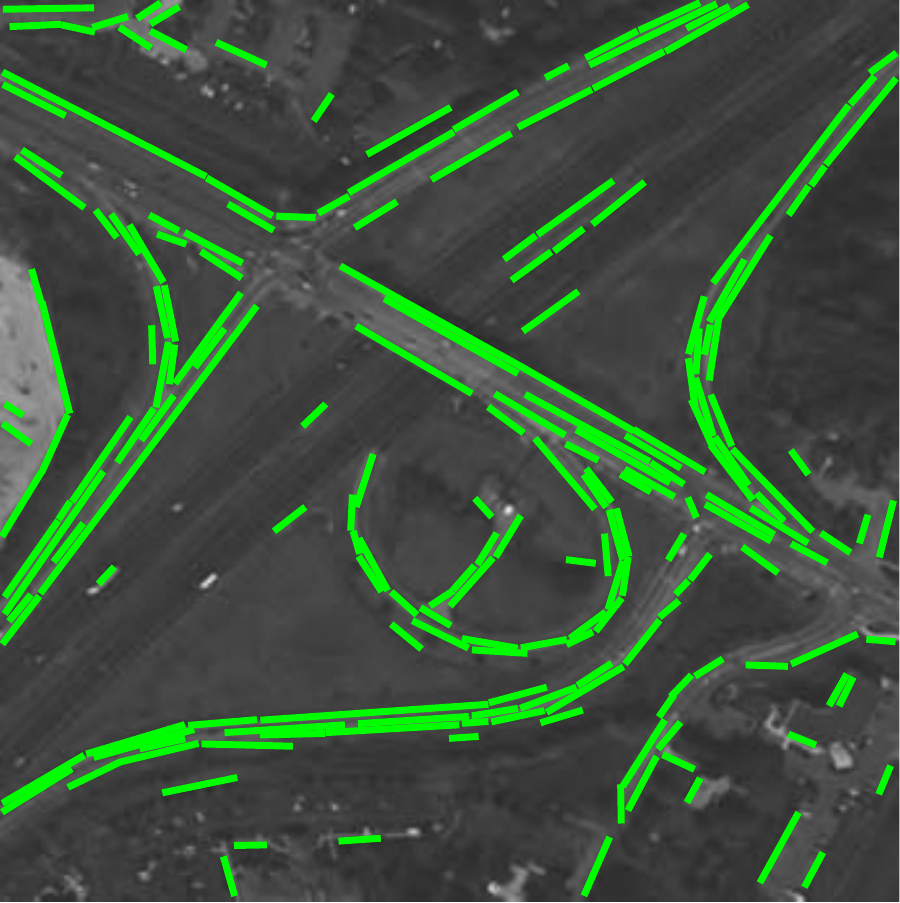}
    \end{minipage}}
  \subfigure[]{
    \label{fig:mini:subfig:b}
    \begin{minipage}[c]{0.15\textwidth}
      \centering
      \includegraphics[width=1.0in]{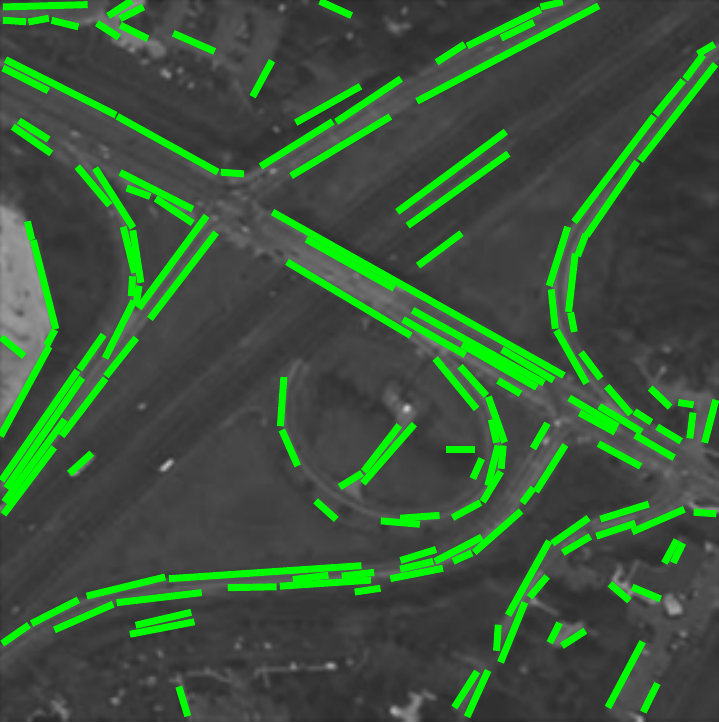}
    \end{minipage}}
  \subfigure[]{
    \label{fig:mini:subfig:a}
    \begin{minipage}[c]{0.15\textwidth}
      \centering
      \includegraphics[width=1.0in]{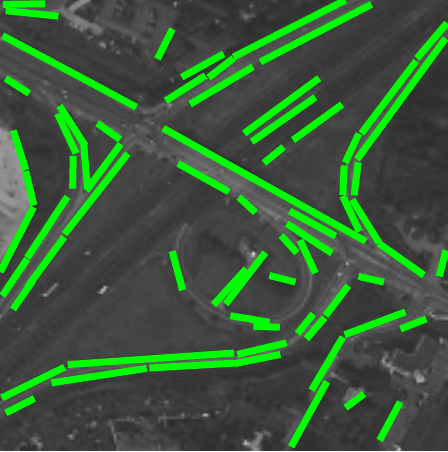}
    \end{minipage}}\\
  \subfigure[]{
    \label{fig:mini:subfig:b}
    \begin{minipage}[c]{0.15\textwidth}
      \centering
      \includegraphics[width=1.0in]{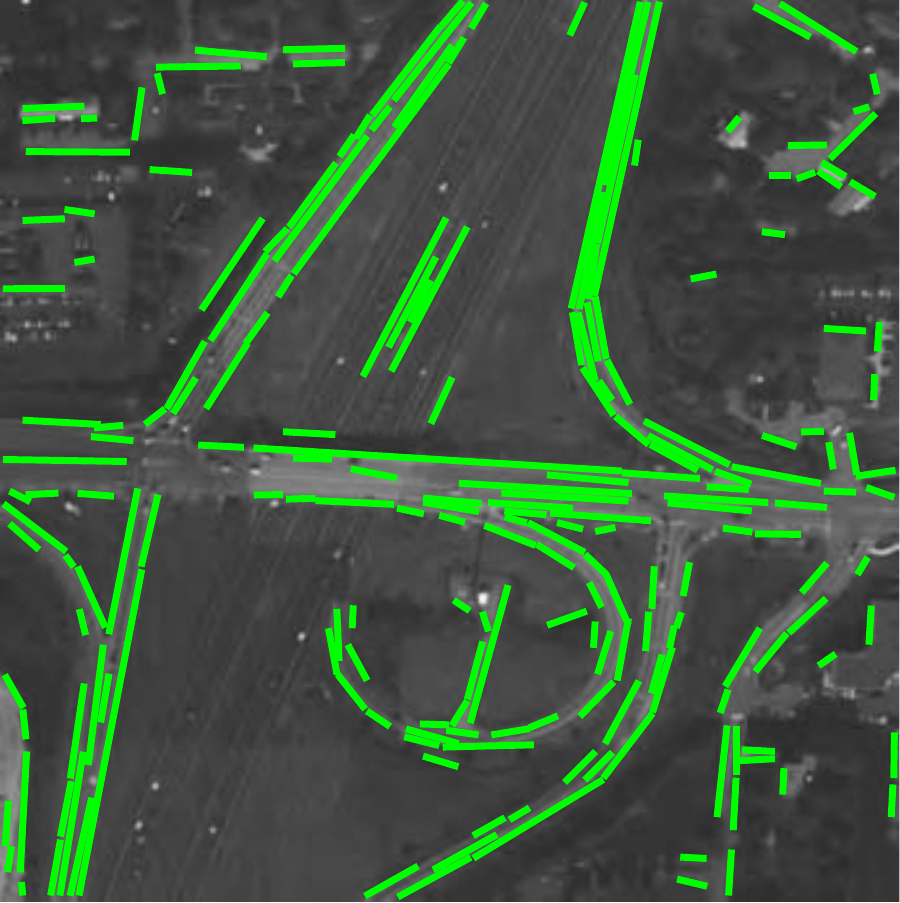}
    \end{minipage}}
  \subfigure[]{
    \label{fig:mini:subfig:a}
    \begin{minipage}[c]{0.15\textwidth}
      \centering
      \includegraphics[width=1.0in]{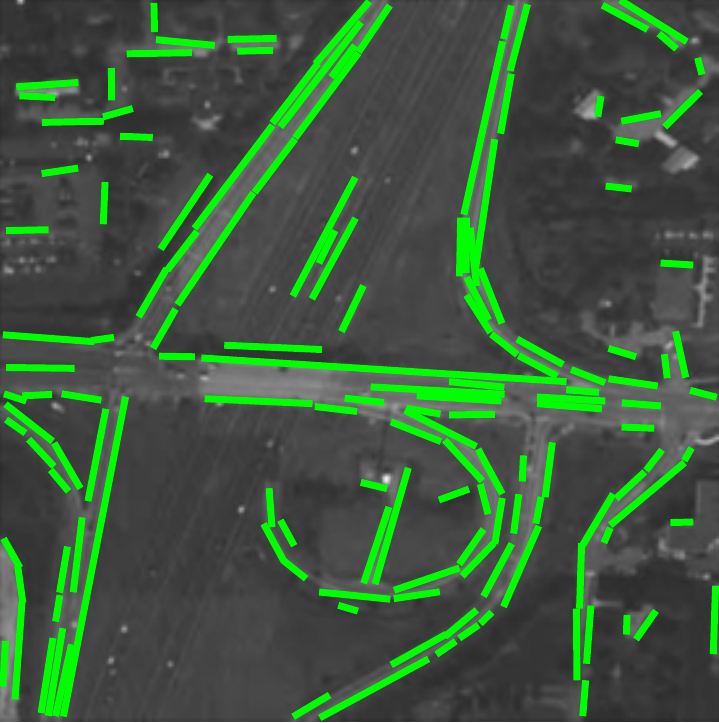}
    \end{minipage}}
  \subfigure[]{
    \label{fig:mini:subfig:b}
    \begin{minipage}[c]{0.15\textwidth}
      \centering
      \includegraphics[width=1.0in]{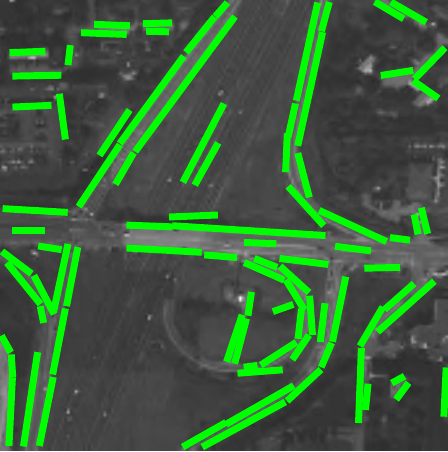}
    \end{minipage}}\\
  \subfigure[]{
    \label{fig:mini:subfig:a}
    \begin{minipage}[c]{0.15\textwidth}
      \centering
      \includegraphics[width=1.0in]{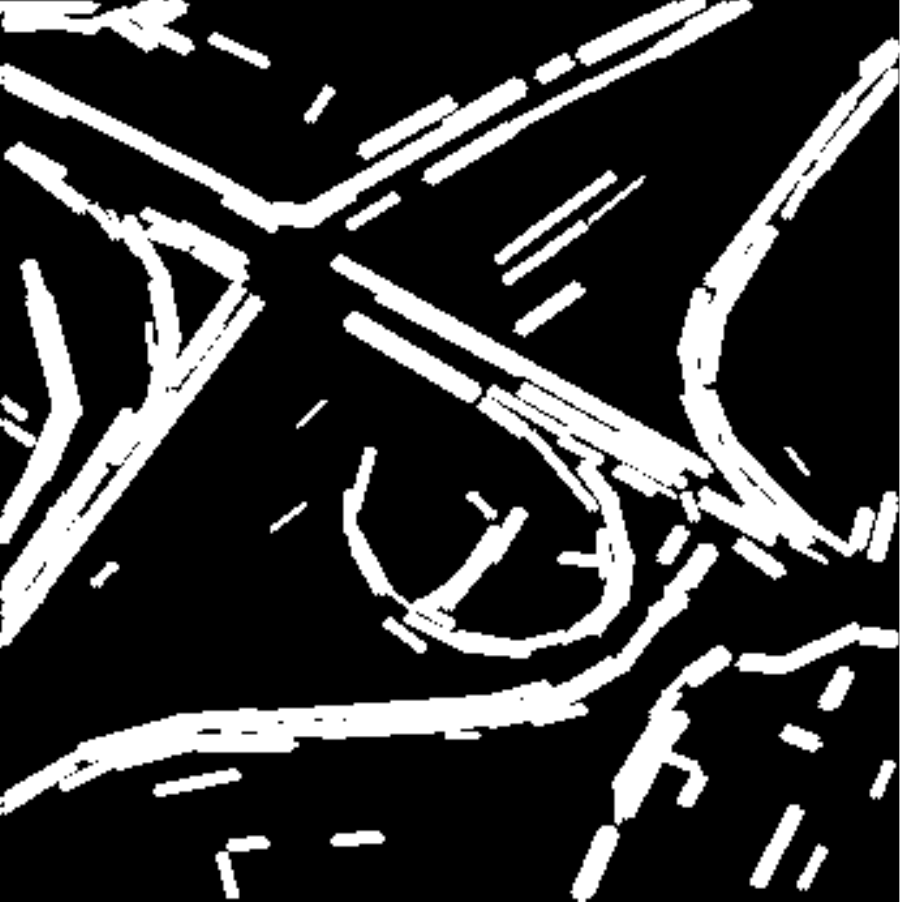}
    \end{minipage}}
  \subfigure[]{
    \label{fig:mini:subfig:b}
    \begin{minipage}[c]{0.15\textwidth}
      \centering
      \includegraphics[width=1.0in]{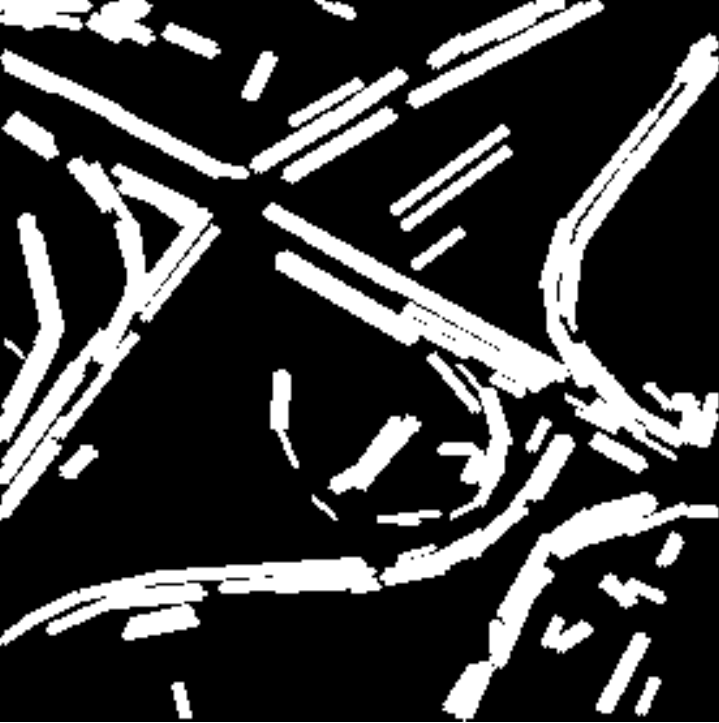}
    \end{minipage}}
        \subfigure[]{
    \label{fig:mini:subfig:a}
    \begin{minipage}[c]{0.15\textwidth}
      \centering
      \includegraphics[width=1.0in]{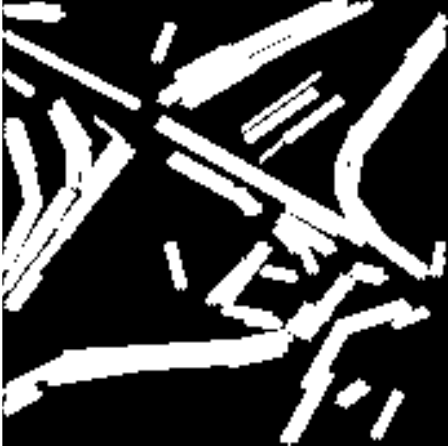}
    \end{minipage}}\\
  \subfigure[]{
    \label{fig:mini:subfig:b}
    \begin{minipage}[c]{0.15\textwidth}
      \centering
      \includegraphics[width=1.0in]{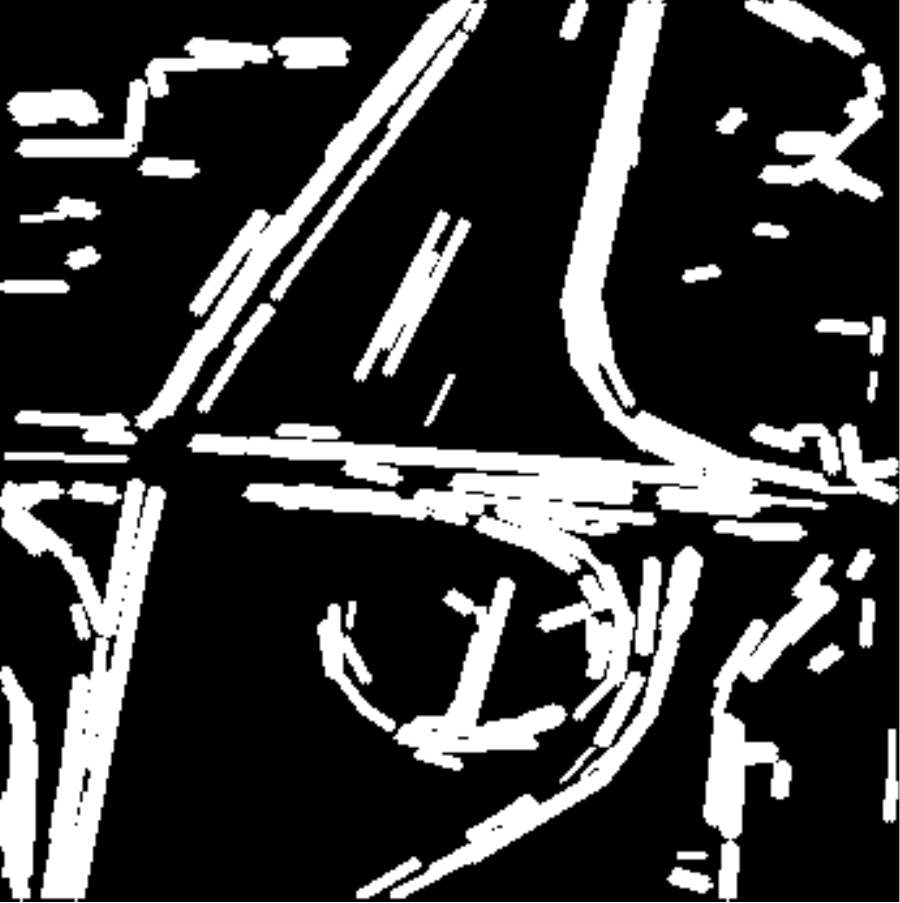}
    \end{minipage}}
  \subfigure[]{
    \label{fig:mini:subfig:a}
    \begin{minipage}[c]{0.15\textwidth}
      \centering
      \includegraphics[width=1.0in]{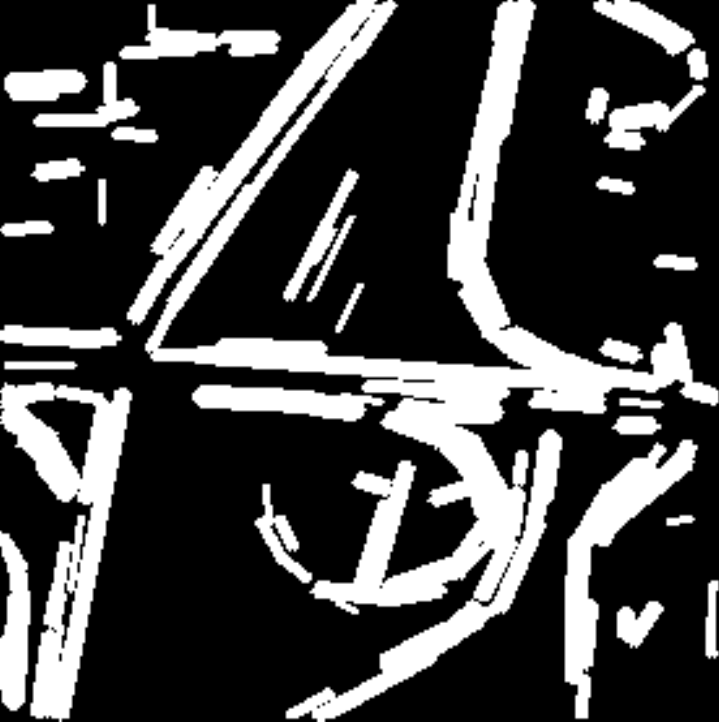}
    \end{minipage}}
  \subfigure[]{
    \label{fig:mini:subfig:b}
    \begin{minipage}[c]{0.15\textwidth}
      \centering
      \includegraphics[width=1.0in]{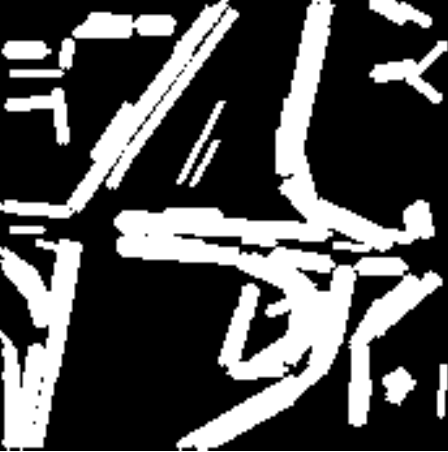}
  \end{minipage}}\\
  \captionstyle{normal}
  \caption{Demonstration of iterative line segment detection and line-support regions. (a) and (d) Line segments from the reference and sensed image of 700$\times$700 pixels. (b) and (e) Line segments from the reference and sensed image of 350$\times$350 pixels. (c) and (f) Line segments from the reference and sensed image of 175$\times$175 pixels.  (g) and (j) Line-support regions of the reference and sensed image of 700$\times$700 pixels. (h) and (k) Line-support regions of the reference and sensed image of 350$\times$350 pixels. (i) and (l) Line-support regions of the reference and sensed image of 175$\times$175 pixels.}
  \label{fig-1}
\end{figure}

However, the line segments detected by LSD are incomplete, in the sense that line features with details or noise are easily detected as overlapped segments or few broken segments.
It has been proved in \cite{J_RG_2010_ITPAMI} that analyzing at a coarser resolution by LSD helps to detect global structures, which are masked at the full resolution by image details or noise.
It can provide better initial conditions for feature extraction and matching, and also filter out the slightly different details in the corresponding images.
This is because line-support regions are extracted relying on the similarity of gradients, which are easily affected by noise.
The noise in the full images can be decreased at coarser scales.
Also, the coarser images lose some details with the decimation.
The multi-resolution strategy has been explored in current state-of-the-art registration methods to decompose an image into fine and coarse details based on scale.
For instance, wavelet and shearlet transforms are multi-resolution, so that images can be decomposed into subimages with features of progressively finer scales.
Different levels of wavelets and shearlets in multi-resolution pyramids are  used both for invariant feature extraction and for representing images at multiple spatial resolutions to accelerate registration  and increases the robustness of the algorithms \cite{J_IZ_2005_ITIP,J_JL_2002_ITGRS,J_JM_2016_ITGRS,C_JM_2015_PII}.
In this paper, we propose the iterative strategy to ensure the accuracy of registration by down-sampling the image to a coarser resolution, and then re-segmenting and re-matching until an expected accuracy is achieved. The implementation of the iterative strategy is described in detail in Section III-C.
Fig. \ref{fig-1} displays an example of extracting line segments and line-support regions from an aerial image pair of a circular road at different resolutions.
At the first iteration, the original image pair of 700$\times$700 pixels are downsampled into 350$\times$350 pixels. The x-axis and y-axis are each reduced to 50$\%$ of the original size.
As demonstrated in Fig. \ref{fig-1} (i)-(j) with 175$\times$175 pixels, the original images are downsampled twice by scaling into 25$\%\times$25$\%$ of the original size.
It can be observed that the incompleteness of line segment extraction can be alleviated by dowsampling the original images.
Most of isolated fragments disappear when the images are downsampled into a coarser resolution.
Besides that, many overlapped or broken segments extracted in the original images are merged into complete segments in the dowsampled images.

\subsection{Motivation of Utilizing SIFT with Geometrical Outlier Removal as Feature Point Matching}
The geometrical distortions always exist in the images acquired by different sensors, or from different viewing angles. Compared by Mikolajzyk \cite{J_KM_2005_ITPAMI}, SIFT is one of powerful intensity-based descriptors for local interest regions, which is invariant to image scaling and rotations, and also partially invariant to intensity changes and shear deformations.

However, even after the identification of matching candidates by SIFT matching, many false SIFT matches arise between the similar local features of different scenes.
It leads to a further incorrect geometrical correction \cite{J_MGG_2014_ITGRS}.
Therefore, a reliable outlier removal is needed to ensure the accuracy of feature point matching.
In this paper, a geometrical outlier removal is proposed based on the consensus that the feature points clustering in the same side of a line remain in the same side after the geometrical transformation.
The directed edges $\left\langle {p_i  \mapsto p_j } \right\rangle $ are explored to connect any two feature points $p_i$ and $p_j$.
The rest of feature points are classified according to their relative locations with these directed edges, i.e., lying on the left side of $\left\langle {p_i  \mapsto p_j } \right\rangle $, on the right side of $\left\langle {p_i  \mapsto p_j } \right\rangle $, or exactly on the directed edge.
Then, the unreliable candidate matches are excluded iteratively relying on the disparity of corresponding classifications.

\begin{figure*}[htb]
\centering
 \setlength{\abovecaptionskip}{0pt}
 \setlength{\belowcaptionskip}{0pt}
 \setlength{\intextsep}{8pt plus 3pt minus 2pt}
  \subfigure[]{
    \label{fig:mini:subfig:a}
    \begin{minipage}[c]{0.32\textwidth}
      \centering
      \includegraphics[width=2.2in]{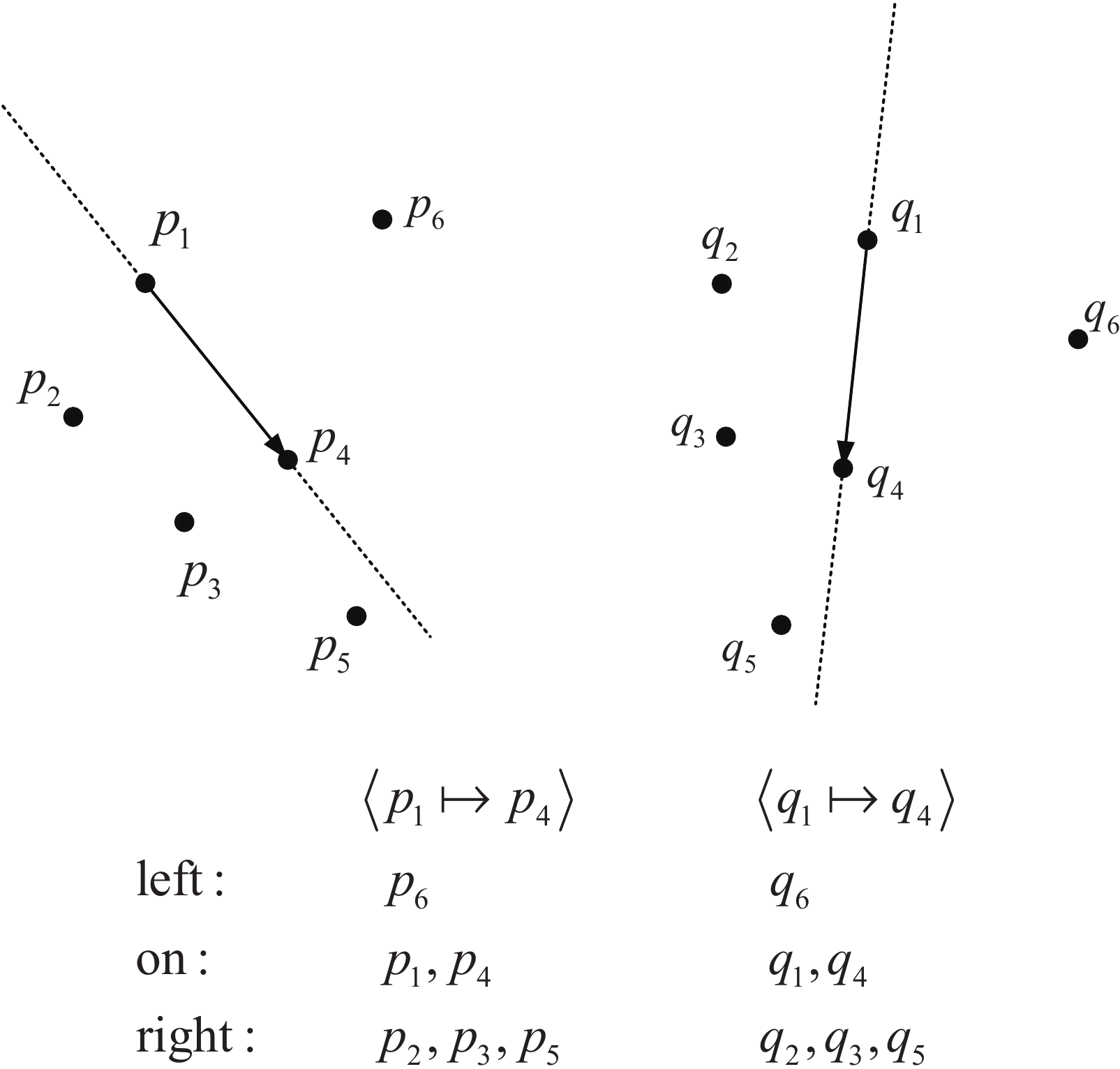}
    \end{minipage}}
  \subfigure[]{
    \label{fig:mini:subfig:b}
    \begin{minipage}[c]{0.32\textwidth}
      \centering
      \includegraphics[width=2.2in]{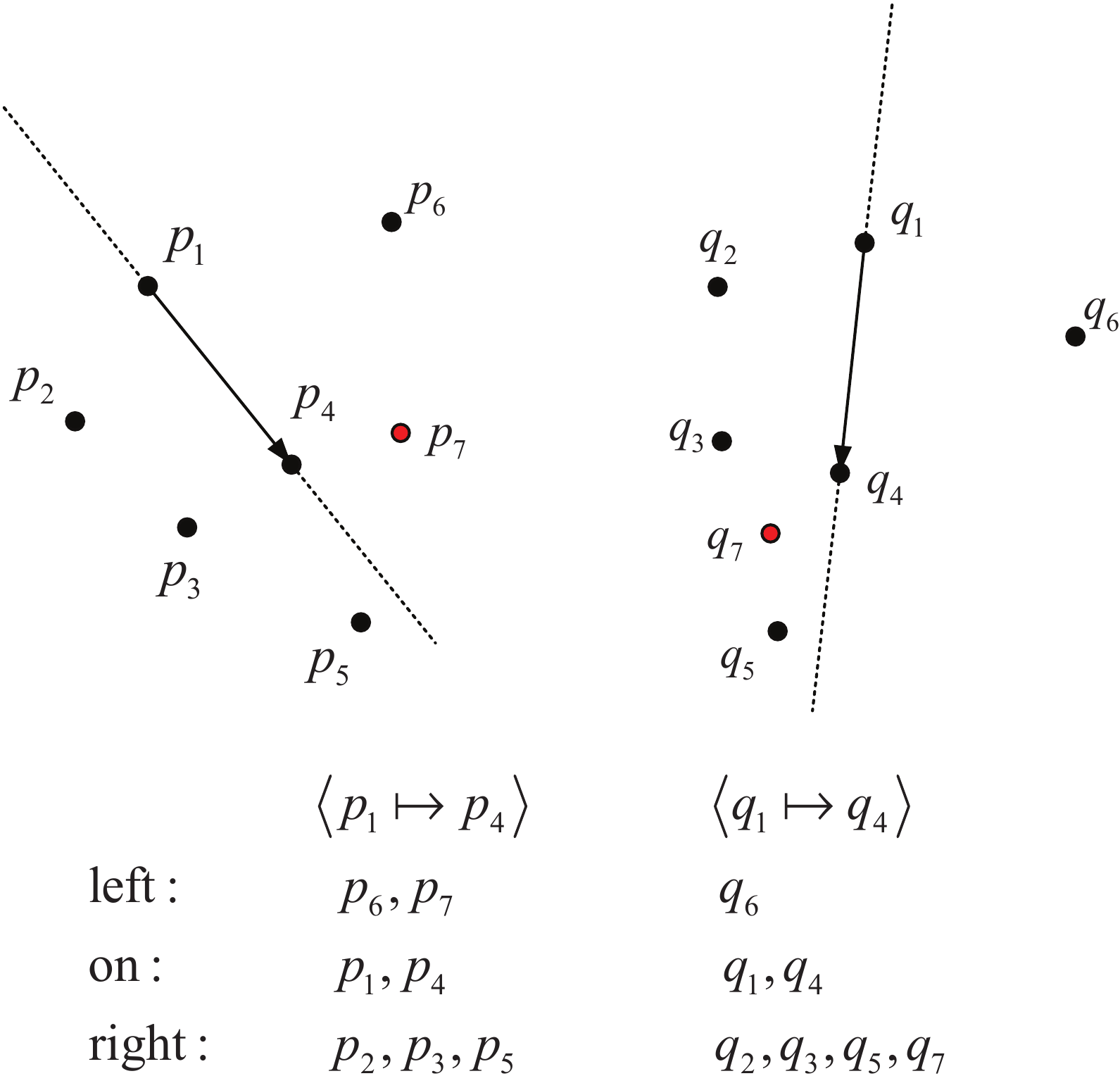}
    \end{minipage}}
  \subfigure[]{
    \label{fig:mini:subfig:a}
    \begin{minipage}[c]{0.32\textwidth}
      \centering
      \includegraphics[width=2.2in]{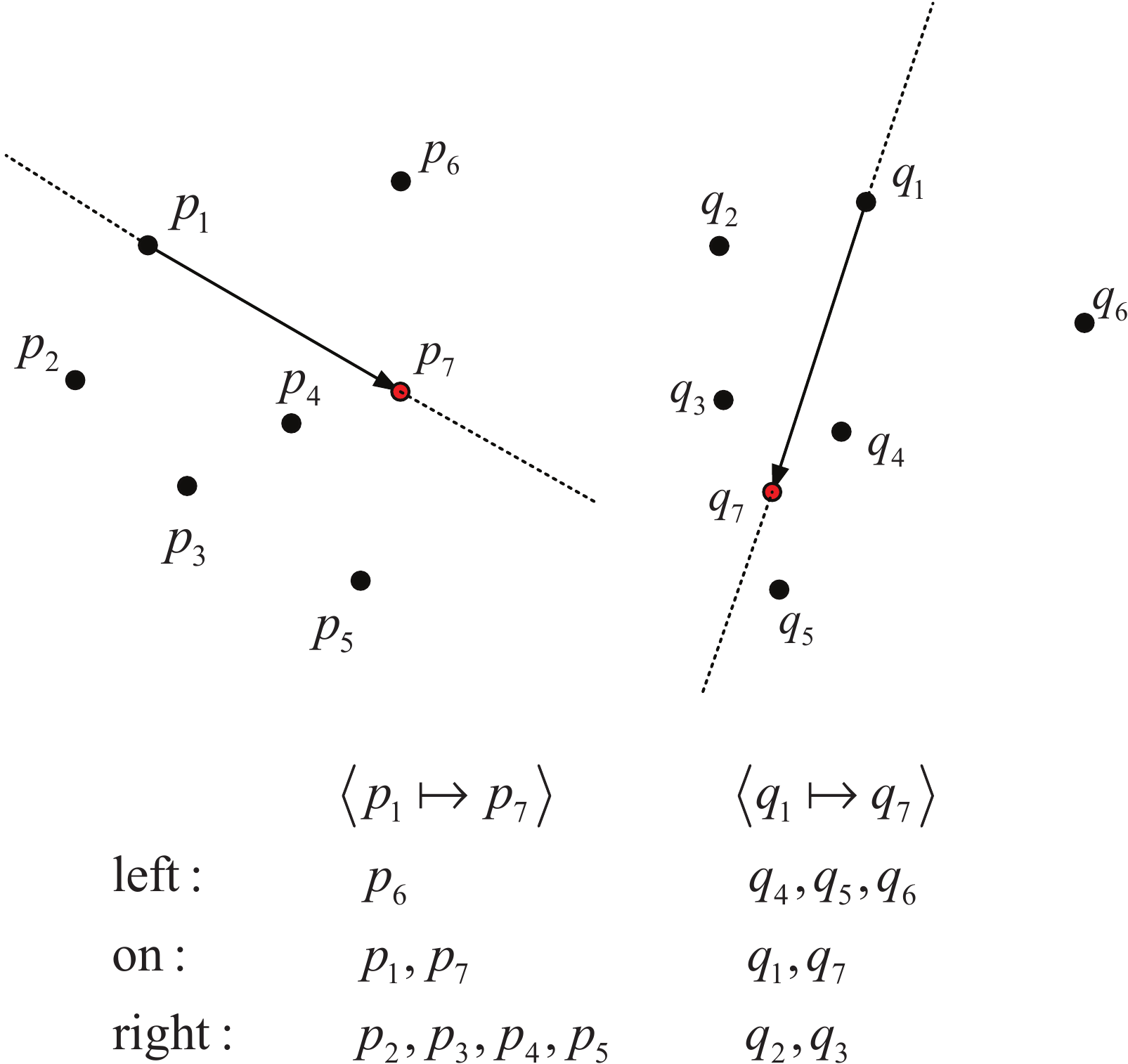}
    \end{minipage}}\\
  \captionstyle{normal}
  \caption{Demonstrations for geometrical graphs with rotations. (a) The classifications by  the directed edges $\left\langle {p_1  \mapsto p_4 } \right\rangle $  and $\left\langle {q_1  \mapsto q_4 } \right\rangle $  without outliers. (b) The classifications by  the directed edges $\left\langle {p_1  \mapsto p_4 } \right\rangle $ and  $\left\langle {q_1  \mapsto q_4 } \right\rangle $  with the outlier $\left( {p_7 ,q_7 } \right)$. (c) The classifications by the directed edges $\left\langle {p_1  \mapsto p_7 } \right\rangle $  and $\left\langle {q_1  \mapsto q_7 } \right\rangle $ with the outlier $\left( {p_7 ,q_7 } \right)$.}
  \label{fig-2}
\end{figure*}

Fig. \ref{fig-2} provides a demonstration of geometrical outlier removal with rotated point sets. If the feature points are all correctly matched, the classifications by any directed edges should be identical.
$\{p_1,p_2,p_3,p_4,p_5,p_6\}$ and $\{q_1,q_2,q_3,q_4,q_5,q_6\}$ are the corresponding matched feature point sets, where $\{p_i\}$ and $\{q_i\}$ are correctly matched.
Any corresponding directed edges $\left\langle {p_i  \mapsto p_j } \right\rangle $ and $\left\langle {q_i  \mapsto q_j } \right\rangle $, $\forall i,j \in \left\{ {1,2,3,4,5,6} \right\}$ divide the rest of corresponding feature points from the respective point sets into the same classifications.
An example of the classifications for $\left\langle {p_1  \mapsto p_4 } \right\rangle $ and $\left\langle {p_1  \mapsto p_4 } \right\rangle $ is shown in Fig. \ref{fig-2} (a).
Both  $\{p_2,p_3,p_5\}$ and $\{q_2,q_3,q_5\}$ are located on the right side of $\left\langle {p_1  \mapsto p_4 } \right\rangle $ and $\left\langle {q_1  \mapsto q_4 } \right\rangle $, while $\{p_6\}$ and $\{q_6\}$ are on the left side.
However, the classifications for the corresponding feature points will be different with any additions of outliers.
Fig. \ref{fig-2} (b) demonstrates the geometrical graphs  with the addition of outlier $(p_7,q_7)$, which displays in red.
Only the outlier $(p_7,q_7)$  is located on the two different sides of $\left\langle {p_1  \mapsto p_4 } \right\rangle $  and $\left\langle {q_1  \mapsto q_4 } \right\rangle $, while the rest corresponding points are on the same side.
Compared to the directed edge established by inliers, the disparity of feature point classifications increases when the directed edges are established by outliers.
An example of the classifications for $\left\langle {p_1  \mapsto p_7 } \right\rangle $ and $\left\langle {q_1  \mapsto q_7 } \right\rangle $ is shown in Fig. \ref{fig-2} (c).
 $\{p_2,p_3,p_4,p_5\}$ and $\{q_2,q_3\}$ are located on the right side of $\left\langle {p_1  \mapsto p_7 } \right\rangle $ and $\left\langle {q_1  \mapsto q_7 } \right\rangle $ respectively, while $\{p_6\}$ and $\{q_4,q_5,q_6\}$ are on the left side.
Regarding the fact that the classifications of the directed edges established by outliers are  more different than those directed edges established by inliers, we propose an iterative outlier removal
method GOR in this paper. The disparities of classifications are accumulated for all directed edges associated with each corresponding feature points. The matches with the maximum disparity are selected as candidate outliers and removed at each iteration. The implementation of feature point matching with GOR is described in details in Section III-B.

\section{METHODOLOGY}\label{sec:model}
\subsection{Line-Support Region Segmentation}
 Given an input image, a line-support region is defined as a straight region whose points roughly share the same image gradient angle. The line-support regions are extracted by grouping connected pixels that share the same gradient angle up to a certain tolerance. Each region $R$  starts with one randomly selected pixel $p_i$. The region angle $\theta _R$ orthogonal to the gradient angle is defined as the level-line angle  \cite{J_RG_2010_ITPAMI}:

\begin{equation} \label{eqn:achievable_7}
{\theta _R  = \arctan \left( {\frac{{\sum\nolimits_{p_i  \in R} {\sin \left( {\theta _{p_i } } \right)} }}{{\sum\nolimits_{p_i  \in R} {\cos \left( {\theta _{p_i } } \right)} }}} \right)}.
\end{equation}%

Then, the adjacent pixels $\overline p _i$  whose level-line orientations are equal to the region angle up to a certain precision $\tau $ are added to the region $R$:

\begin{multline} \label{eqn:achievable_7}
{R\!\leftarrow \! R  \cup \left\{ {\overline p _i \left| {\left| {\arctan \left( {\frac{{\sin \left( {\theta _{\overline p _i } } \right)}}{{\cos \left( {\theta _{\overline p _i } } \right)}}} \right)\! } \right. } \right. } \right. } \\
  -  {\left.{\left.{\left. { \arctan \left( {\frac{{\sum\nolimits_{p_i  \in R} {\sin \left( {\theta _{p_i } } \right)} }}{{\sum\nolimits_{p_i  \in R} {\cos \left( {\theta _{p_i } } \right)} }}} \right)} \right| \! < \! \tau } \right.} \right\}}.
\end{multline}%

The process is repeated until no new pixels can be added.
The value of $\tau $ is the angle tolerance used in the search for line-support regions.
A small value of $\tau $ leads to an over-partition of line segments, while a large one results in large regions.
As suggested by Burns et al. \cite{J_RG_2010_ITPAMI}, the value of $\tau $ in this paper is set to be 22.5$^ \circ$, which corresponds to eight different angle bins.
There is no theory behind this parameter value, but supported by the results on thousands of images in \cite{J_RG_2010_ITPAMI}.

In this paper, we utilize the extracted line-support regions to represent the segmentations of the scene.
A line segment associated with the corresponding line-support region is approximated by a rectangle region $L_i = \left( {o_i ,\theta _i ,l_i ,w_i } \right)$, which is determined by its center point $o_i $, region angle $\theta _i$, length $l_i $, and width $w_i $.
For the binary images to be described, the pixels belonging to the rectangular approximations of line-support regions are assigned as 1, and others are assigned as 0.

\subsection{SIFT Feature Point Matching with Geometrical Outlier Removal}
\begin{itemize}
\item[1)] \emph{Initial Correspondence Obtention}:
The initial SIFT corresponding matches are established from the segmented  line-support regions of the reference and sensed images.
SIFT matching includes five major steps: scale-space extrema detection, keypoint localization, orientation assignments, keypoint descriptor, and keypoint matching.
As mentioned previously, SIFT matching is effectively implemented through the nearest neighbor approach. The nearest neighbor is defined as the keypoint with minimum Euclidean distance between the 128-element SIFT vectors.
The effective measure for matching is the ratio $d_{ratio} $, which denotes the ratio between the distance of the nearest neighbor and that of the second nearest neighbor. More details about SIFT can be found in \cite{J_DGL_2004_IJCV}. Knowing that the image is segmented by line-support regions, it is highly probable that the correct matches only occur between the keypoints with the same binary values. Therefore, the inital matching for line-support regions can be simplified by refining the matching between the keypoints with the same pixel values respectively from the two corresponding feature points sets.

\item[2)] \emph{Geometrical Outlier Removal}: Incorrect matches may still exist after SIFT matching as mentioned before. Therefore, we propose a reliable geometrical outlier removal to exclude false matches in segmented images. Given two initial corresponding sets of   matched keypoints  $\left\{ {p_i } \right\}$ and $\left\{ {q_i } \right\}$ belonging to the reference and sensed segmented images respectively. For each targeted correspondence $\left( {p_i ,q_i } \right)$, $N - 1$ directed edges  $\left\langle {p_i  \mapsto p_j } \right\rangle$ and $\left\langle {q_i  \mapsto q_j } \right\rangle$ starting  from $\left( {p_i ,q_i } \right)$ to any other corresponding keypoints $\left( {p_j ,q_j } \right)$ exist.
    There are generally three cases for classifying all of $N$  keypoints in the feature point sets relying on the relations between the keypoints and the directed edges.
    Take $\left\langle {p_i  \mapsto p_j } \right\rangle$ for example. All of keypoints can be classified into three groups, i.e., lying on the left side of $\left\langle {p_i  \mapsto p_j } \right\rangle$, on the right side of $\left\langle {p_i  \mapsto p_j } \right\rangle$, and in the directed edge.
    A quick judge to distinguish the location relations between $\left\langle {p_i  \mapsto p_j } \right\rangle$ and $p_k$ is provided as follows:

            \begin{equation} \label{eqn:achievable_9}
            {\det \left( {p_i  \mapsto p_j ,p_k } \right) = \left| {\begin{array}{*{20}c} {x_i } & {x_j } & {x_k }  \\
               {y_i } & {y_j } & {y_k }  \\
                  1 & 1 & 1  \\
            \end{array}} \right|}
            \end{equation}%
    where  $\left( {x,y} \right)$ denotes the keypoint location and $\left|  \cdot  \right|$  solves the determinant of the matrix. The three different relations mentioned above can be represented by different symbol signs of  $\det \left( {p_i  \mapsto p_j ,p_k } \right)$, including plus, minus, and zero. We use ${\bf{diff}}_{i \mapsto j} \left( k \right)$  to measure the disparity of classifications for each keypoint $\left( {p_k ,q_k } \right)$ by the corresponding directed edges $\left\langle {p_i  \mapsto p_j } \right\rangle $ and $\left\langle {q_i  \mapsto q_j } \right\rangle$:

    \begin{equation} \label{eqn:achievable_9}
  {\bf{diff}}_{i \mapsto j} \left( k \right)=
  \left\{ {\begin{array}{*{20}l}
   {0,\ {\mathop{\rm sign}\nolimits} \left( {\det \left( {p_i \mapsto  p_j ,p_k } \right)} \right)} \\
   { \quad \ = {\mathop{\rm sign}\nolimits} \left( {\det \left( {q_i \! \mapsto \!q_j ,q_k } \right)} \right)}  \\
   {1,\ otherwise}  \\   \end{array}} \right..
  \end{equation}%

If all the keypoints in $\left\{ {p_i } \right\}$ and $\left\{ {q_i } \right\}$
 are matched correctly,  ${\bf{diff}}_{i \mapsto j} \left( k \right)$ should be zero for the identical location relationships. Thus, the candidate outliers $i^{out}$ that achieve the maximum accumulated difference of classifications are selected:

  \begin{equation} \label{eqn:achievable_9}
  {i^{out}  = \mathop {\arg \max }\limits_{i = 1,2,...,N} \sum\limits_j^N {\sum\limits_k^N {{\bf{diff}}_{i \mapsto j} \left( k \right)} }}.
  \end{equation}%

When the candidate outliers are identified, all directed edges related to the candidate outliers should be removed accordingly.
A new iteration begins with the decrement of residual correspondences to classify the residual keypoints through the location relations.
The iteration stops when ${\bf{diff}}_{i \mapsto j} \left( k \right) = 0$, $\forall i,j,k \in \left\{ {1,2,...,N} \right\}$.
It indicates that there is no difference between the corresponding location relations, and no candidate outlier needs to be removed.

\end{itemize}

\subsection{Iterative Strategy of ALRS-GOR}
\begin{figure}[htb]
\centering
 \setlength{\abovecaptionskip}{0pt}
 \setlength{\belowcaptionskip}{0pt}
 \setlength{\intextsep}{8pt plus 3pt minus 2pt}
  \subfigure[]{
    \begin{minipage}[c]{0.45\textwidth}
      \centering
      \includegraphics[width=3.25in]{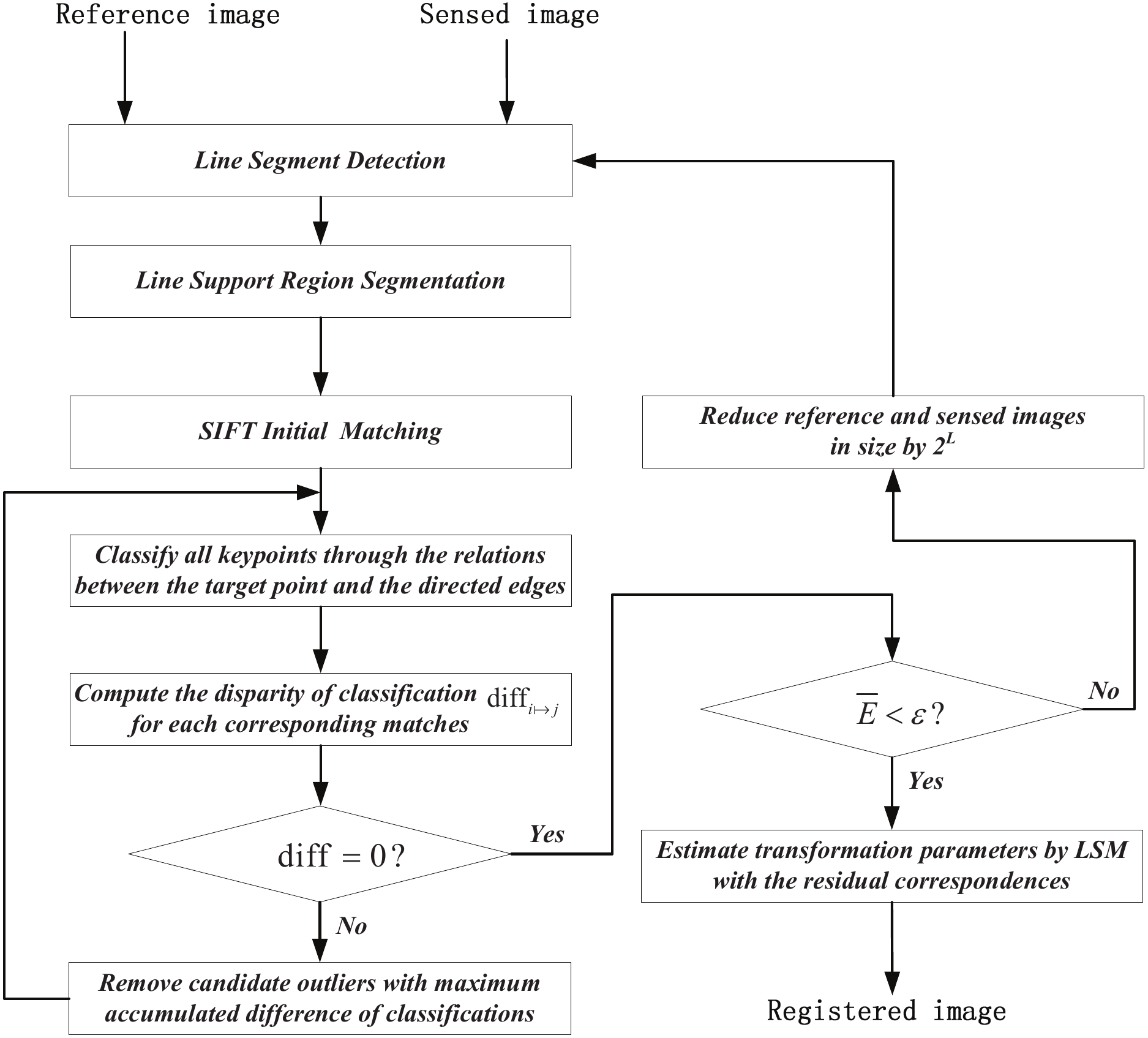}
    \end{minipage}}\\
  \captionstyle{normal}
  \caption{The framework of ALRS-GOR.}
  \label{fig-framework}
\end{figure}

As analyzed in Section II-A, an iterative strategy with multi-resolution is proposed to detect the global structures that masked at full resolution by image details or noise. Moreover, slight differences in the corresponding images with inconsistent content can be filtered out in coarser resolutions.
The framework of the the proposed iterative strategy is demonstrated in Fig. \ref{fig-framework}.
First, the reference and sensed images at full resolution are segmented by  line-support regions. Then, the corresponding matches are obtained from the segmented images by SIFT with geometrical outlier removal.
The root mean square error (RMSE) related to the current resolution $\overline E(L) $ is adopted to estimate the accuracy of corresponding matches:
  \begin{equation} \label{eqn:achievable_9}
  {\bar E(L) = 2^L \times \sqrt {\frac{1}{N}\sum\limits_{k = 1}^N {\left\| {T\left( {p_k ,\theta ^ *  } \right) - q_k } \right\|^2 } } }
  \end{equation}%
where $T\left(  \cdot  \right)$ is the transformation model, and the parameters $\theta ^ *  $  are estimated by the residual matches through the common model parameter estimation approach least squares method (LSM) \cite{J_SU_1991_ITPAMI}. $L$ is the number of iterations.

Given an original image, the coarser image at iteration $L$ is generated by averaging the image at full resolution with $2^L \times 2^L$ windows.
Re-extraction and re-matching are implemented for the images with coarser resolution.
The multiresolution framework works iteratively until  $\bar E(L)< \epsilon$.

\subsection{Complexity Analysis}
In this section, we present the issues related to the execution complexity of the proposed methodology.
Line-support region segmentation can be divided into two kinds of stages, i.e., computing the gradient angles and summing the regions.
The time computation of the gradient angles is $O(m)$,where $m$ is the number of image pixels.
Summing the region is proportional to the total number of pixels involved in all regions, which also requires $O(m)$.
We assume $n$ as the number of corresponding matches initialized by SIFT from the line-support region images.
The first iteration for geometrical outlier removal is breakdown into classifying feature points and searching for outliers.
Classifying feature points involves two stages, i.e., establishing $n(n-1)/2$ directed edges and distinguishing the location relations for the rest of $(n-2)$ points.
This is the most time consuming step in the proposed algorithm.
In implementing this step, we provide an improvement by updating the directed edges and location relations in each iteration, instead of re-establishing and re-distinguishing them.
Then, the time complexity of classifying feature points can be reduced to $O(n^2logn)$.
Searching for the candidate outliers with maximum accumulated disparity of classifications requires $O(n)$.
The number of iterations depends on initial  matches and the percentage of the outliers.

\section{EXPERIMENTS AND ANALYSIS}\label{sec:model}
In this section, we present experiments to validate the effectiveness of the proposed methodology in a laptop with 2-GHz CPU and 8-GB RAM (Intel Core i5). Here, the dataset of experiments are composed by three image sets (see Table I), i.e.,
\begin{itemize}
\item[1)] ImgSet1: simulated images with different affine deformations, including rotation, scaling, and shear transformations;
\item[2)] ImgSet2: multispectral remote sensing images, including multispectral images with similar patterns, multispectral with significant spectral difference, multispectral images with temporal difference, and multispectral with SAR noise;
\item[3)] ImgSet3: images with inconsistent annotations, including navigation maps, computer-generated graphics maps, and maps overlaid with satellite images.
\end{itemize}

First, the performance of feature point extraction and matching is validated with the simulated images in ImgSet1.
ImgSet2-1 and ImgSet2-2 with poor initial inliers are provided to prove the necessity of iterative strategy.
Then, the registration applications of the proposed approach are illustrated to images with affine transformation and inconsistent content, including the real remote sensing images in ImgSet2 with different spectral content or speckle noise, map images in ImgSet3 with inconsistent annotations.
Finally, the proposed registration approach is compared  with other representative registration methods in terms of registration accuracy and computation complexity.

\begin{table*}[t]
\centering
  \captionstyle{normal}
  \setlength{\abovecaptionskip}{0pt}
  \setlength{\belowcaptionskip}{10pt}
\caption{SPECIFICATIONS OF IMAGE DATASET FOR EXPERIMENTS}
\newsavebox{\tablebox}
\begin{lrbox}{\tablebox}
\begin{tabular}{|c|c|c|c|c|c|c|}
\hline
\multicolumn{2}{|c|}{Dataset}  & Sensor/platform &	Size(pixel)& Spatial Resolution& Date & Description\\
\hline
\multirow{2}{*}{ImgSet1} & 1 & Aerial & 512$\times$512 & 20m & 1977 & Aerial image for simulation, from USC, San Diego\\
\cline{2-7}
& 2 & Aerial & 400$\times$400 & 15m & 1977 & Aerial image for simulation, from USC, airport\\
\hline

\multirow{10}{*}{ImgSet2}
& 1 & Landsat TM band 5 & 600$\times$600 & 30m & 1986 & Multispectral images with similar patterns \\
\cline{3-6}
& & Landsat TM band 7	& 600$\times$600 & 30m & 1988 & from UCSB\\
\cline{2-7}

& 2 & SPOT band 3 & 256$\times$256 & 20m & 1995 & Multispectral images with small size sparking features \\
\cline{3-6}
& & Landsat TM band 4 & 256$\times$256 & 30m & 1994 & Brasilia, Brazil \\
\cline{2-7}

& 3 & SPOT panchromatic mode & 669$\times$539 & 10m & unknown & Multispectral images with significant spetral difference\\
\cline{3-6}
& & SPOT XS3 band & 329$\times$278 & 20m & unknown & unknown\\
\cline{2-7}

& 4 & ASTER L1B band 1 & 512$\times$512 & 15m & 1999 & Multispectral images with speckle noise\\
\cline{3-6}
& & PALSAR fine mode & 512$\times$512 & 18m & 2006 & Tokyo bay, Japan\\
\cline{2-7}

& 5 & Orthophotograph green band & 512$\times$512 & 1m & unknown & Multispectral images with temporal difference\\
\cline{3-6}
&  & IKONOS panchromatic mode & 512$\times$512 & 1m & unknown & Porto, Portugal\\
\hline

\multirow{6}{*}{ImgSet3} & 1  & Google map & 700$\times$700 & 200m & 2016 & Google navigation map for mobile phone\\
\cline{3-6}
& & Google map & 	700$\times$700 & 200m & 2016 & auto rotated with direction changes, Erlangen, Germany\\
\cline{2-7}

& 2	 & Google map & 650$\times$650 & 100m & 2016 & Google map with computer-generated graphics\\
\cline{3-6}
& & Google map	& 650$\times$650	& 50m & 2016 & zoom in, Shanghai Oriental Pearl TV Tower, China\\
\cline{2-7}

& 3  & Google map & 650$\times$650 & 1000m & 2016 & Maps overlaid with real satellite images\\
\cline{3-6}
& & Google map	& 650$\times$650	& 500m	& 2016 & rotate $90^ \circ$  and zoom in, Erlangen, Germany\\
\hline
\end{tabular}
\end{lrbox}
\scalebox{0.8}{\usebox{\tablebox}}
\end{table*}

\subsection{Feature Point Extraction and Matching Comparison}
To evaluate the effectiveness of feature extraction and point matching for affine distortions, the two aerial images of ImgSet1-1 and ImgSet1-2 are simulated with different affine transformations, namely, rotation, scale, and shear deformations.
The initial SIFT feature points are extracted and matched with $d_{radio}  = 0.8$ for the original images and segmented line-support regions (LSR), respectively.
The subsequent outlier removals are implemented from the initial sets generated by the segmented line-support regions.
The feature point matching of the proposed approach are compared with three matching methods RANSAC, WGTM (K=10), and LLT (K=15) in terms of recall and precision \cite{J_ZXLiu_2012_ITGRS}. Here, $recall$ measures the proportion of actual matches that correctly identified. $precision$ represents the correctly matches with respect to the residual matches.

  \begin{equation} \label{eqn:achievable_9}
recall = \frac{{residual\ correct\ matches}}{{initial\ correct\ matches}}
  \end{equation}%

  \begin{equation} \label{eqn:achievable_9}
precision = \frac{{residual\ correct\ matches}}{{residual\ matches}}
  \end{equation}%

\begin{figure*}[htb]
\centering
 \setlength{\abovecaptionskip}{0pt}
 \setlength{\belowcaptionskip}{0pt}
 \setlength{\intextsep}{8pt plus 3pt minus 2pt}
  \subfigure[]{
    \label{fig:mini:subfig:a}
    \begin{minipage}[c]{0.11\textwidth}
      \centering
      \includegraphics[width=0.8in]{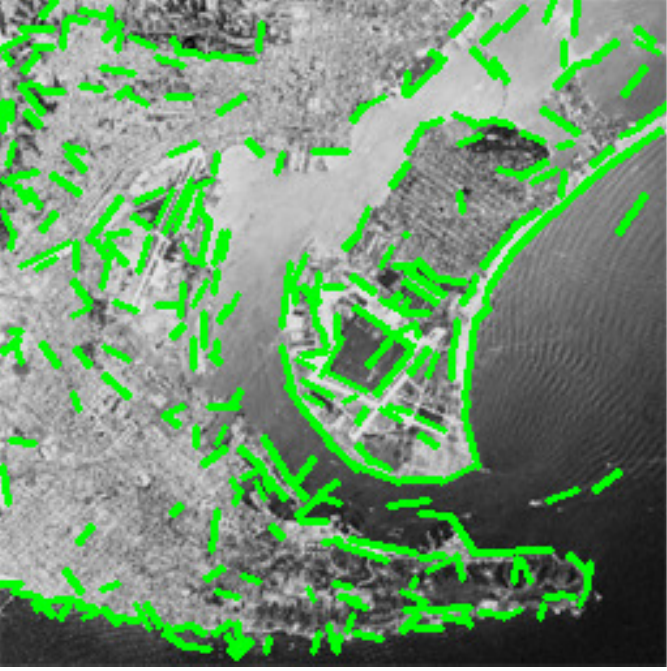}
    \end{minipage}}
  \subfigure[]{
    \label{fig:mini:subfig:b}
    \begin{minipage}[c]{0.11\textwidth}
      \centering
      \includegraphics[width=0.8in]{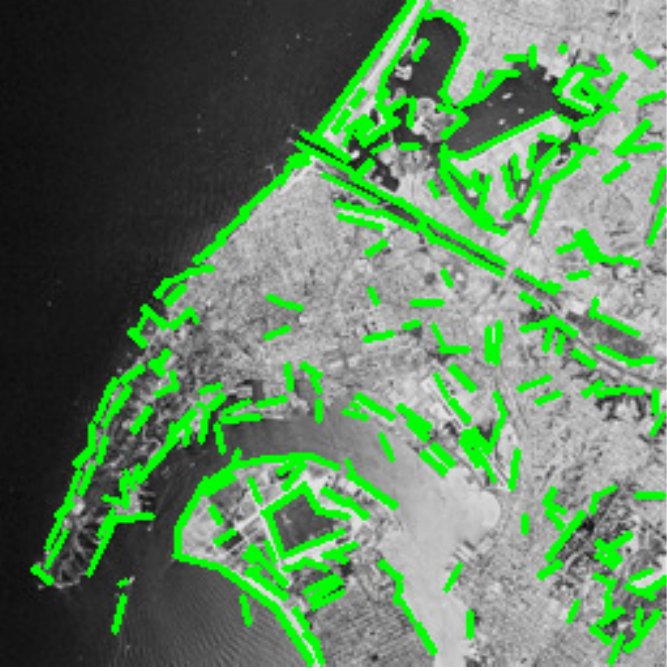}
    \end{minipage}}
  \subfigure[]{
    \label{fig:mini:subfig:a}
    \begin{minipage}[c]{0.11\textwidth}
      \centering
      \includegraphics[width=0.8in]{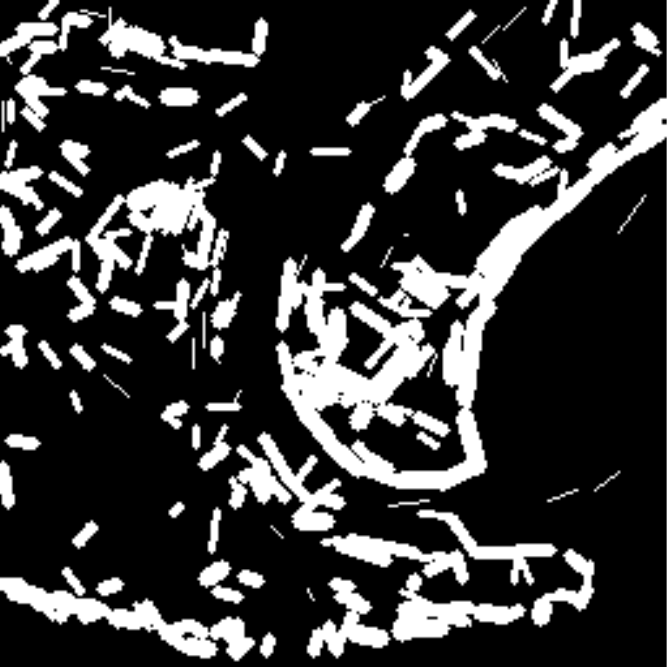}
    \end{minipage}}
  \subfigure[]{
    \label{fig:mini:subfig:b}
    \begin{minipage}[c]{0.11\textwidth}
      \centering
      \includegraphics[width=0.8in]{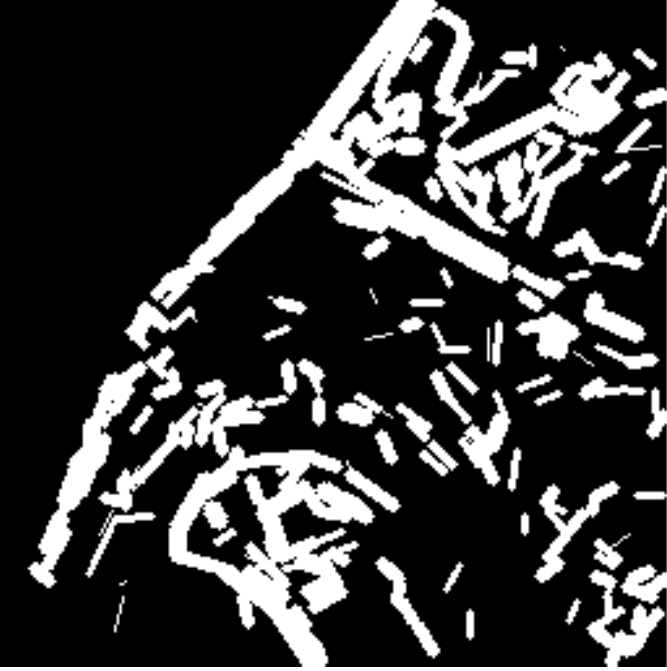}
    \end{minipage}}
  \subfigure[]{
    \label{fig:mini:subfig:a}
    \begin{minipage}[c]{0.22\textwidth}
      \centering
      \includegraphics[width=1.6in]{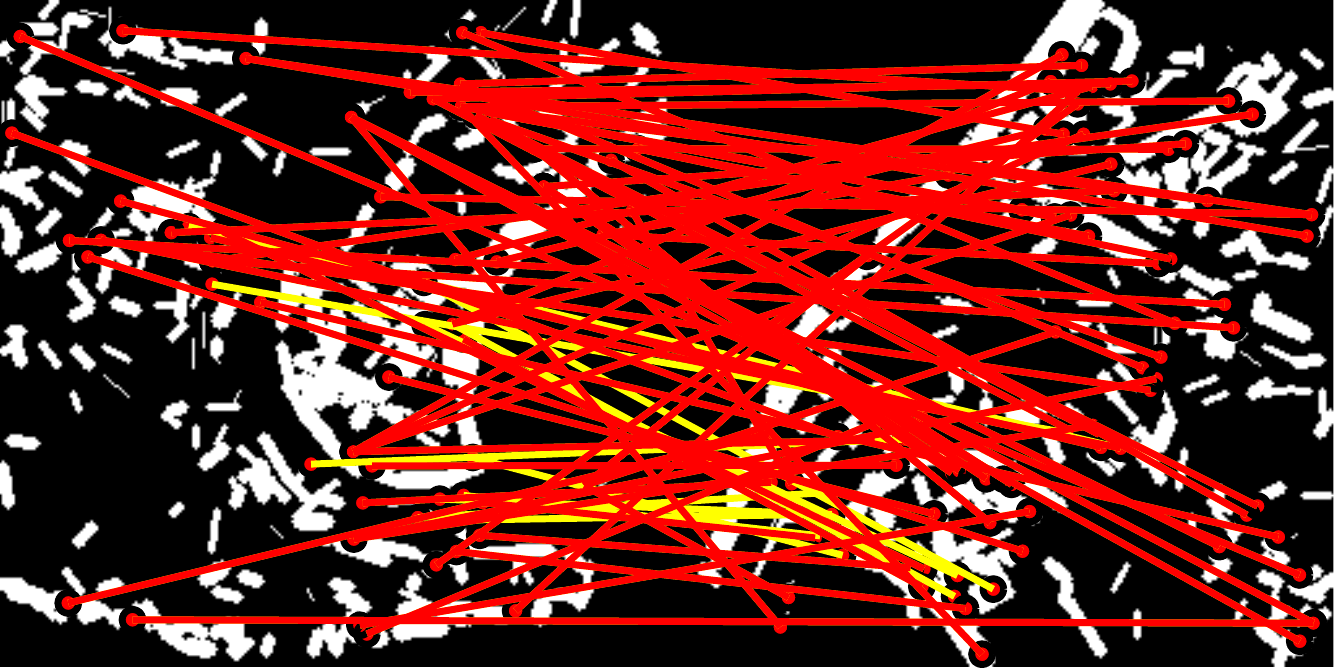}
    \end{minipage}}\\
  \subfigure[]{
    \label{fig:mini:subfig:b}
    \begin{minipage}[c]{0.22\textwidth}
      \centering
      \includegraphics[width=1.6in]{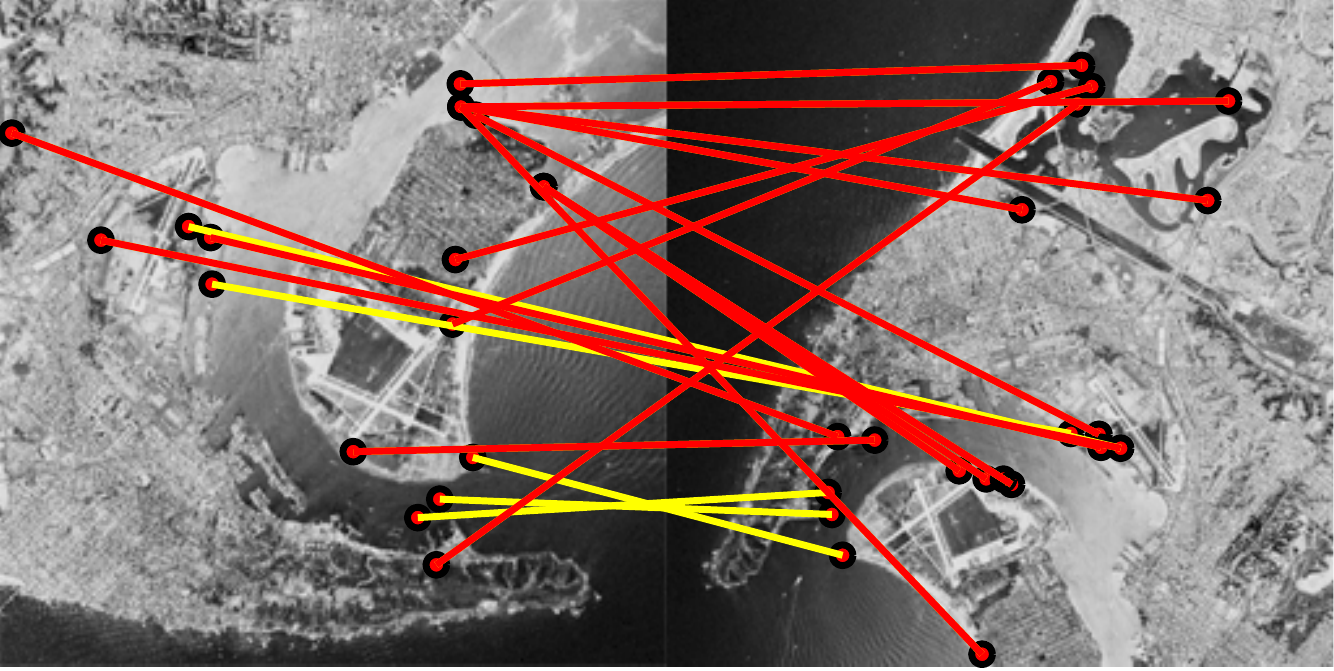}
    \end{minipage}}
  \subfigure[]{
    \label{fig:mini:subfig:a}
    \begin{minipage}[c]{0.22\textwidth}
      \centering
      \includegraphics[width=1.6in]{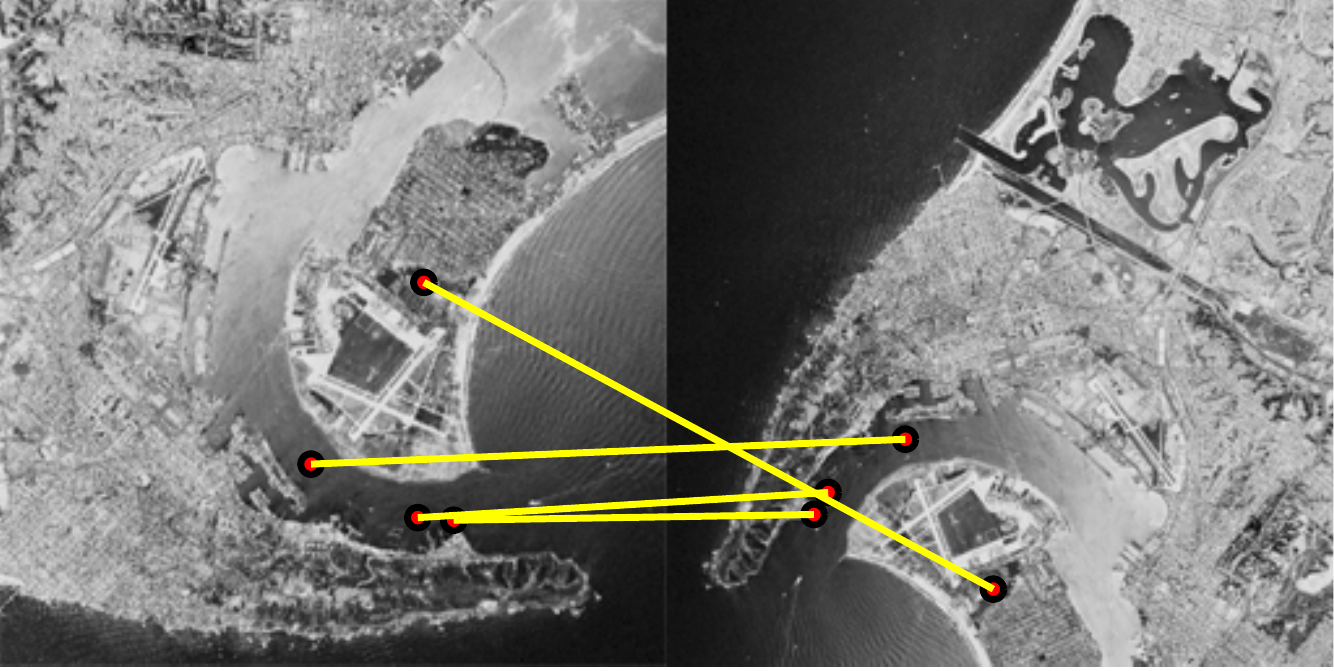}
    \end{minipage}}
  \subfigure[]{
    \label{fig:mini:subfig:b}
    \begin{minipage}[c]{0.22\textwidth}
      \centering
      \includegraphics[width=1.6in]{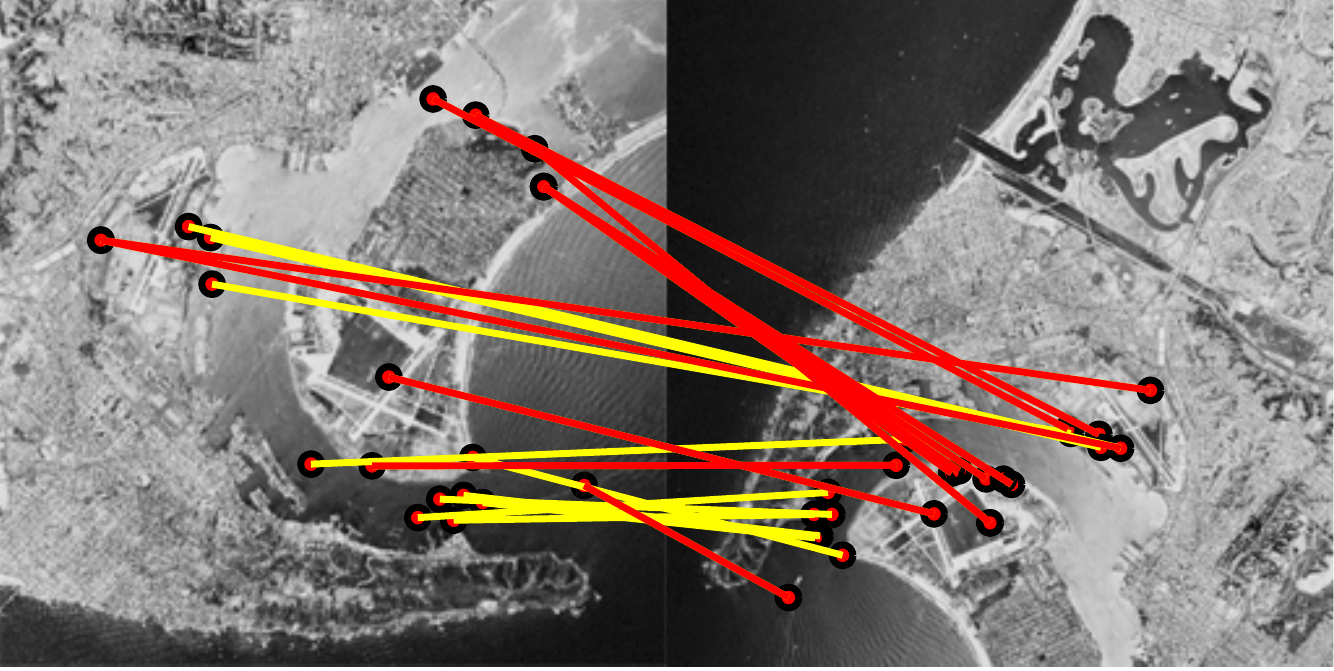}
    \end{minipage}}
  \subfigure[]{
    \label{fig:mini:subfig:b}
    \begin{minipage}[c]{0.22\textwidth}
      \centering
      \includegraphics[width=1.6in]{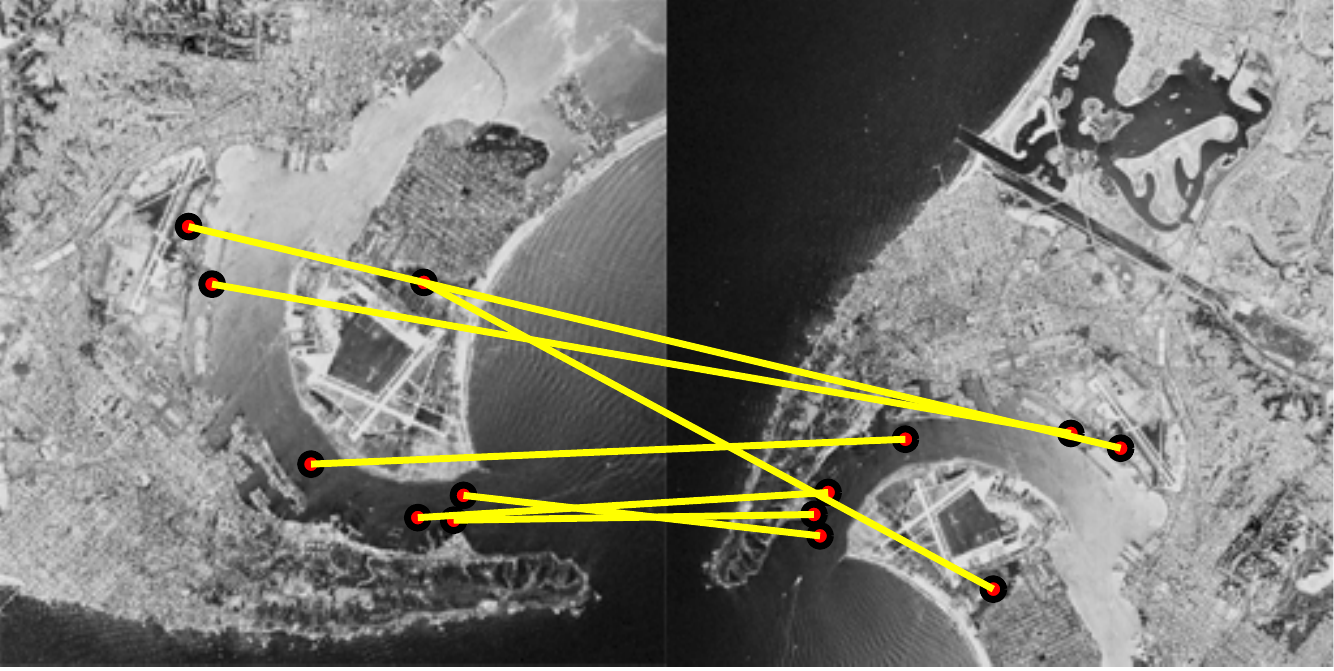}
    \end{minipage}}\\
  \captionstyle{normal}
  \caption{Feature point matching results for ImgSet1-1 with rotation and scale deformations. (a) Line segments of the reference image. (b) Line segments of the sensed image. (c) Line-support regions of the reference image. (d) Line-support regions of the sensed image. (e) Initial correspondences by SIFT. (f) Point correspondences by RANSAC. (g) Point correspondences by WGTM. (h) Point correspondences by GOR. (i) Point correspondences by LLT.}
  \label{fig-4}
\end{figure*}

\begin{figure*}[htb]
\centering
 \setlength{\abovecaptionskip}{0pt}
 \setlength{\belowcaptionskip}{0pt}
 \setlength{\intextsep}{8pt plus 3pt minus 2pt}
  \subfigure[]{
    \label{fig:mini:subfig:a}
    \begin{minipage}[c]{0.11\textwidth}
      \centering
      \includegraphics[width=0.8in]{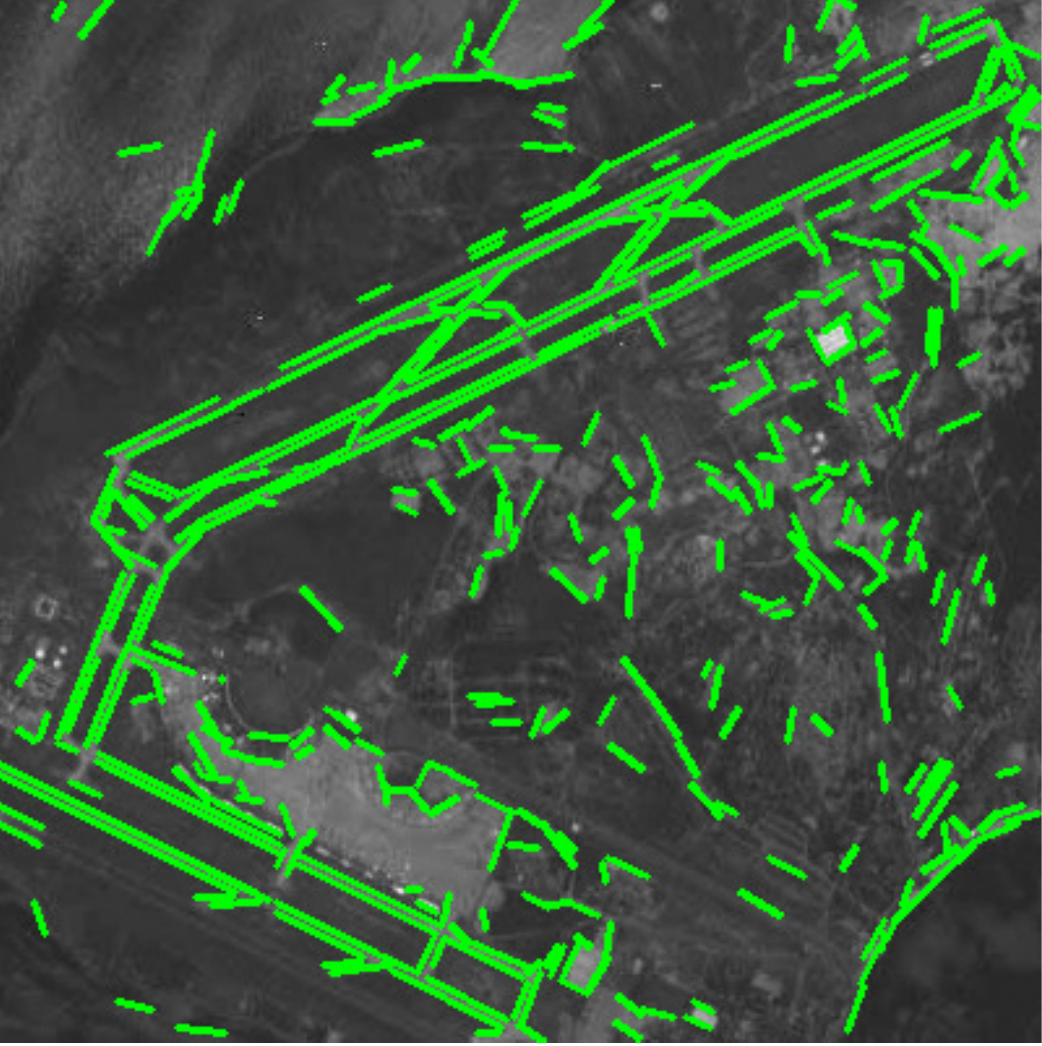}
    \end{minipage}}
  \subfigure[]{
    \label{fig:mini:subfig:b}
    \begin{minipage}[c]{0.11\textwidth}
      \centering
      \includegraphics[width=0.8in]{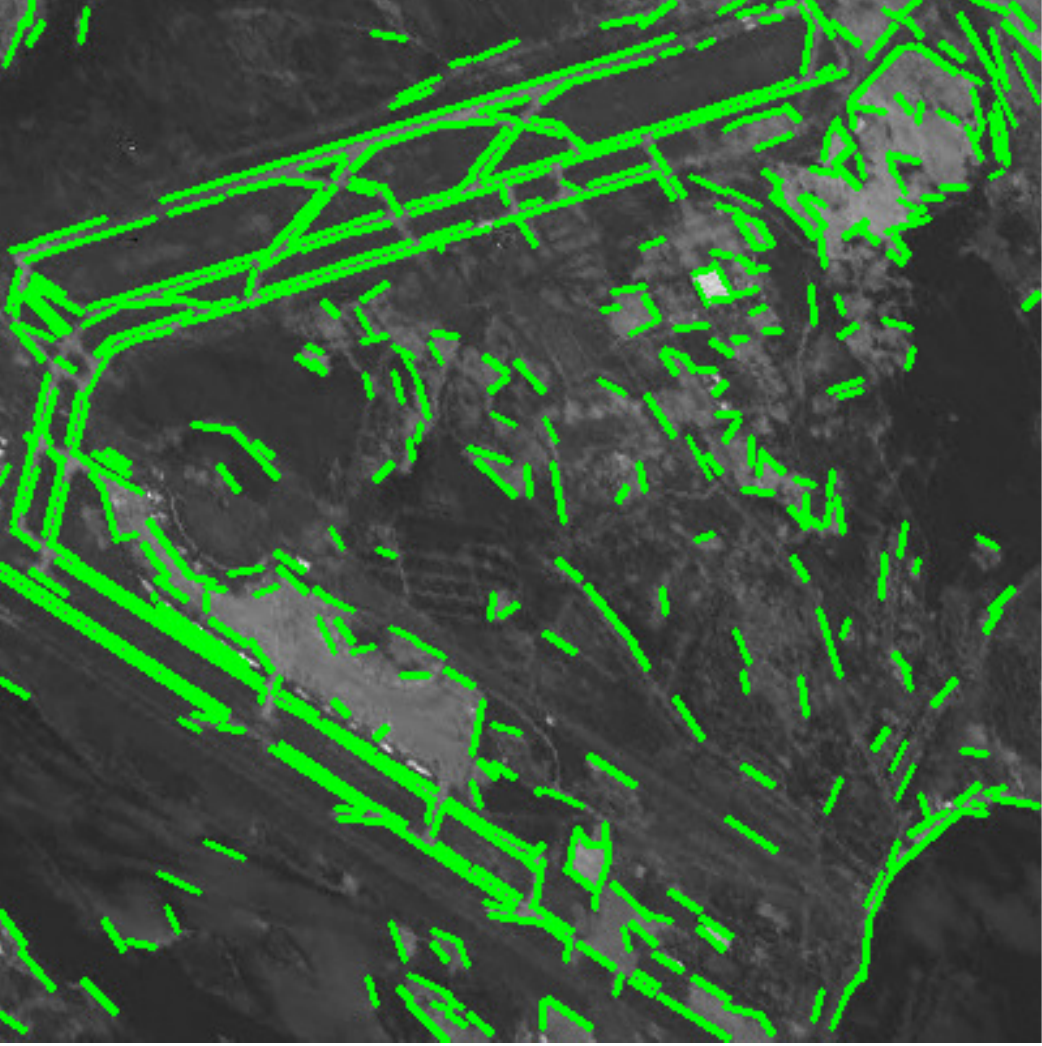}
    \end{minipage}}
  \subfigure[]{
    \label{fig:mini:subfig:a}
    \begin{minipage}[c]{0.11\textwidth}
      \centering
      \includegraphics[width=0.8in]{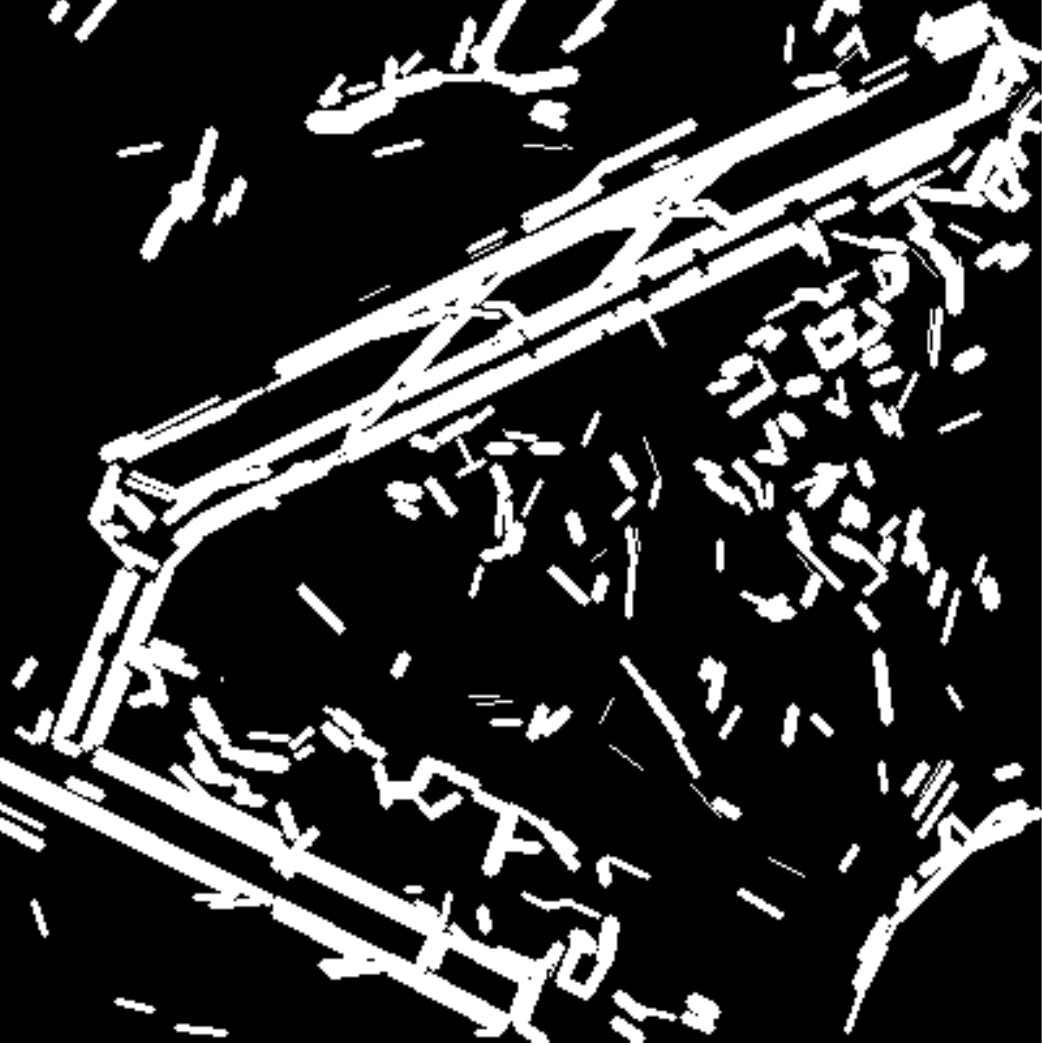}
    \end{minipage}}
  \subfigure[]{
    \label{fig:mini:subfig:b}
    \begin{minipage}[c]{0.11\textwidth}
      \centering
      \includegraphics[width=0.8in]{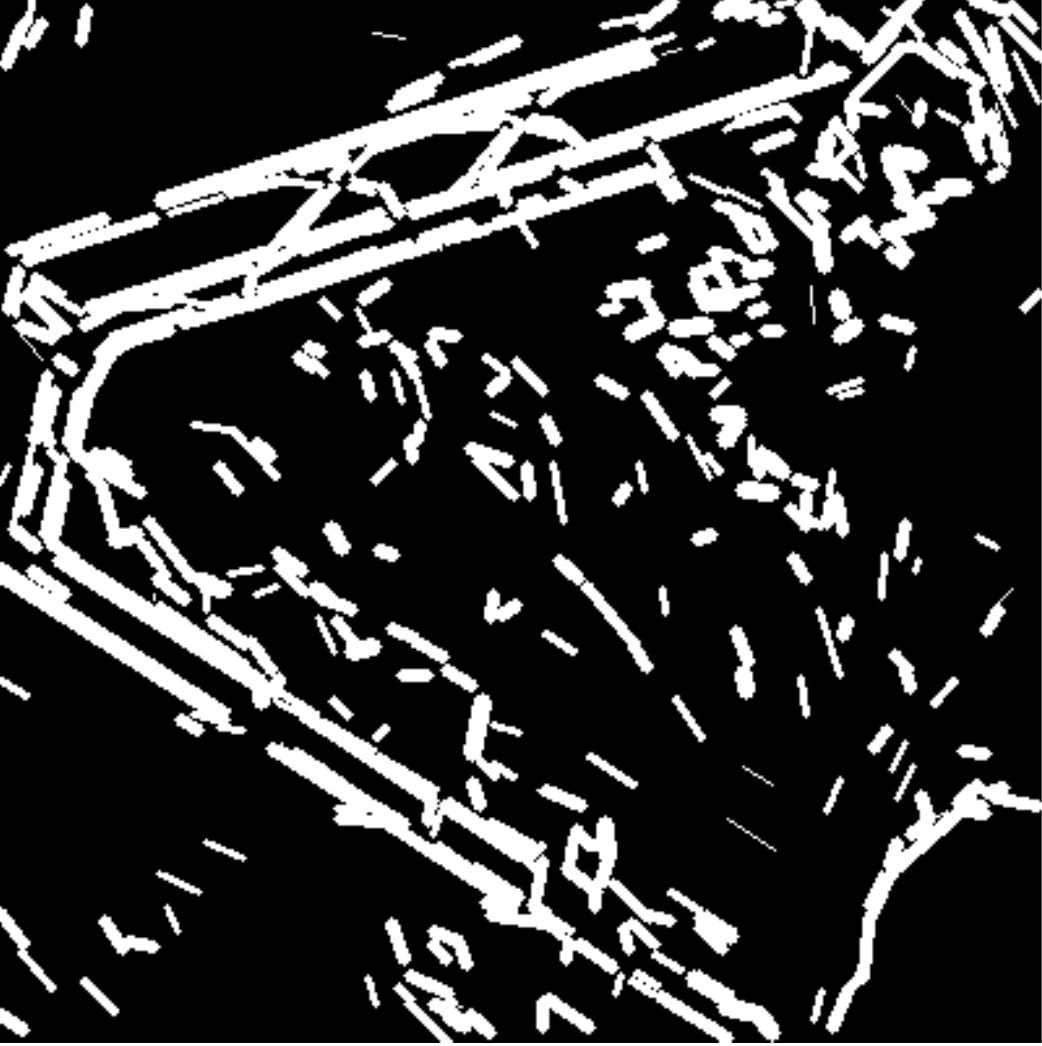}
    \end{minipage}}
  \subfigure[]{
    \label{fig:mini:subfig:a}
    \begin{minipage}[c]{0.22\textwidth}
      \centering
      \includegraphics[width=1.6in]{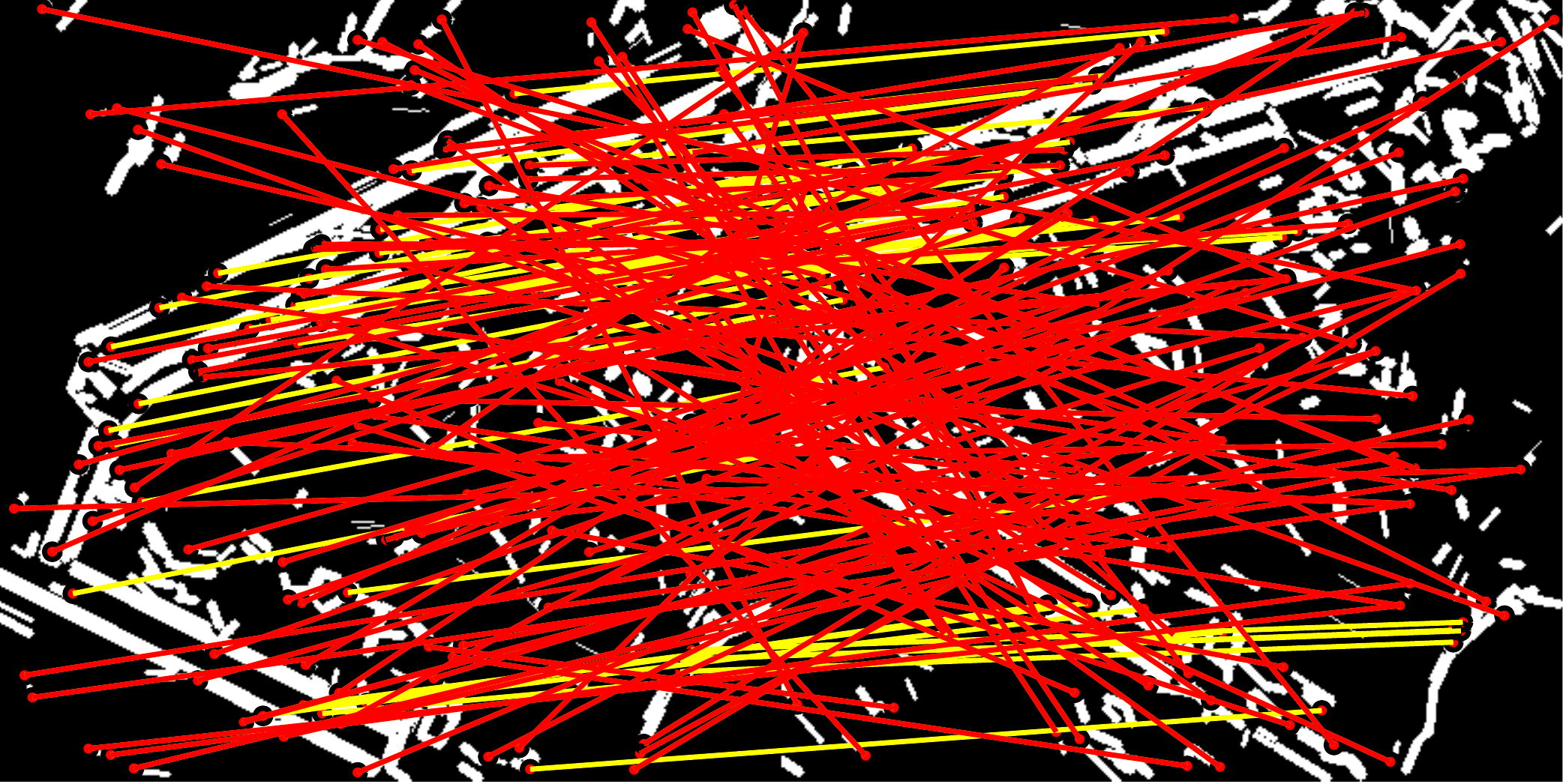}
    \end{minipage}}\\
  \subfigure[]{
    \label{fig:mini:subfig:b}
    \begin{minipage}[c]{0.22\textwidth}
      \centering
      \includegraphics[width=1.6in]{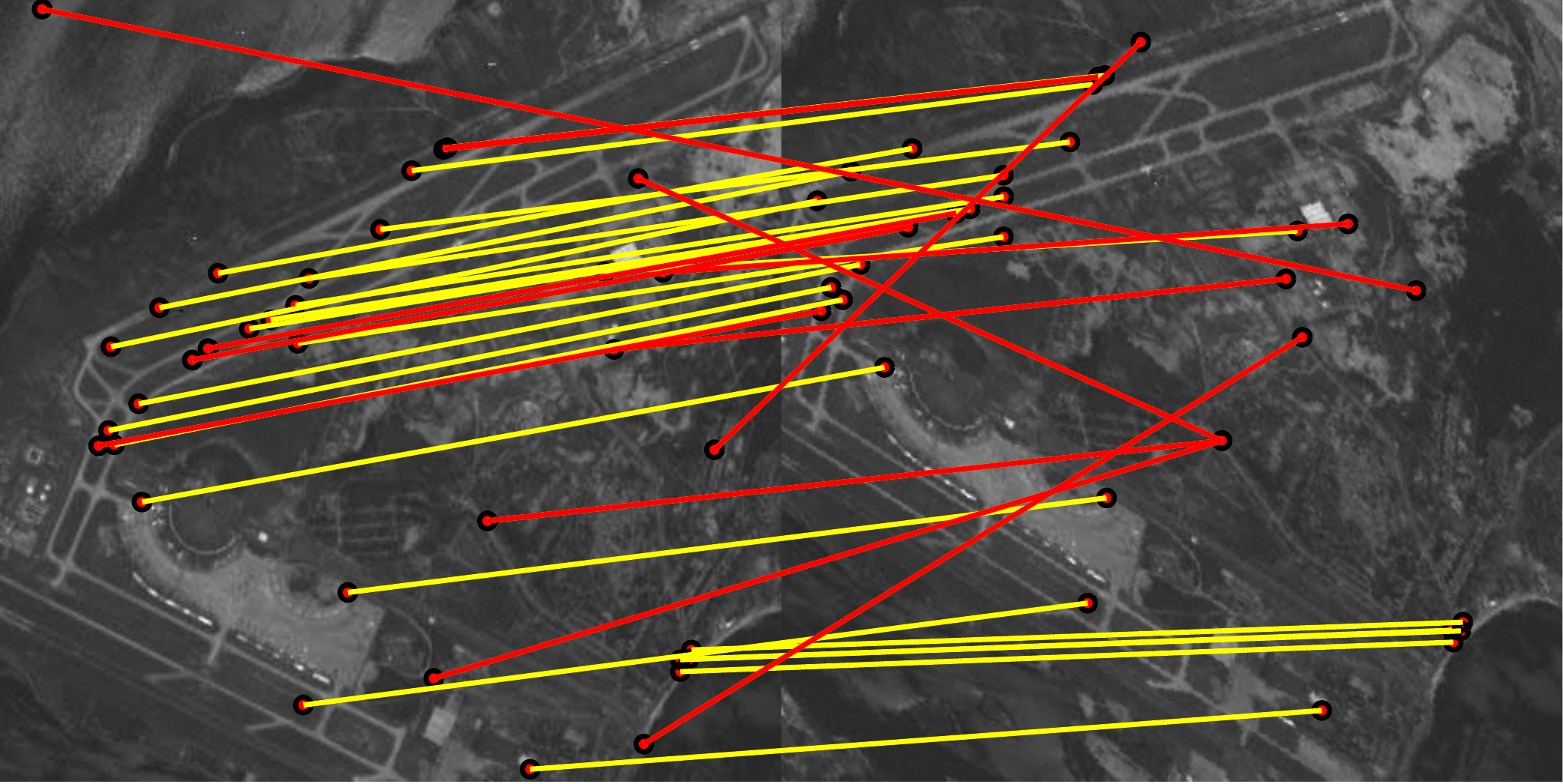}
    \end{minipage}}
  \subfigure[]{
    \label{fig:mini:subfig:a}
    \begin{minipage}[c]{0.22\textwidth}
      \centering
      \includegraphics[width=1.6in]{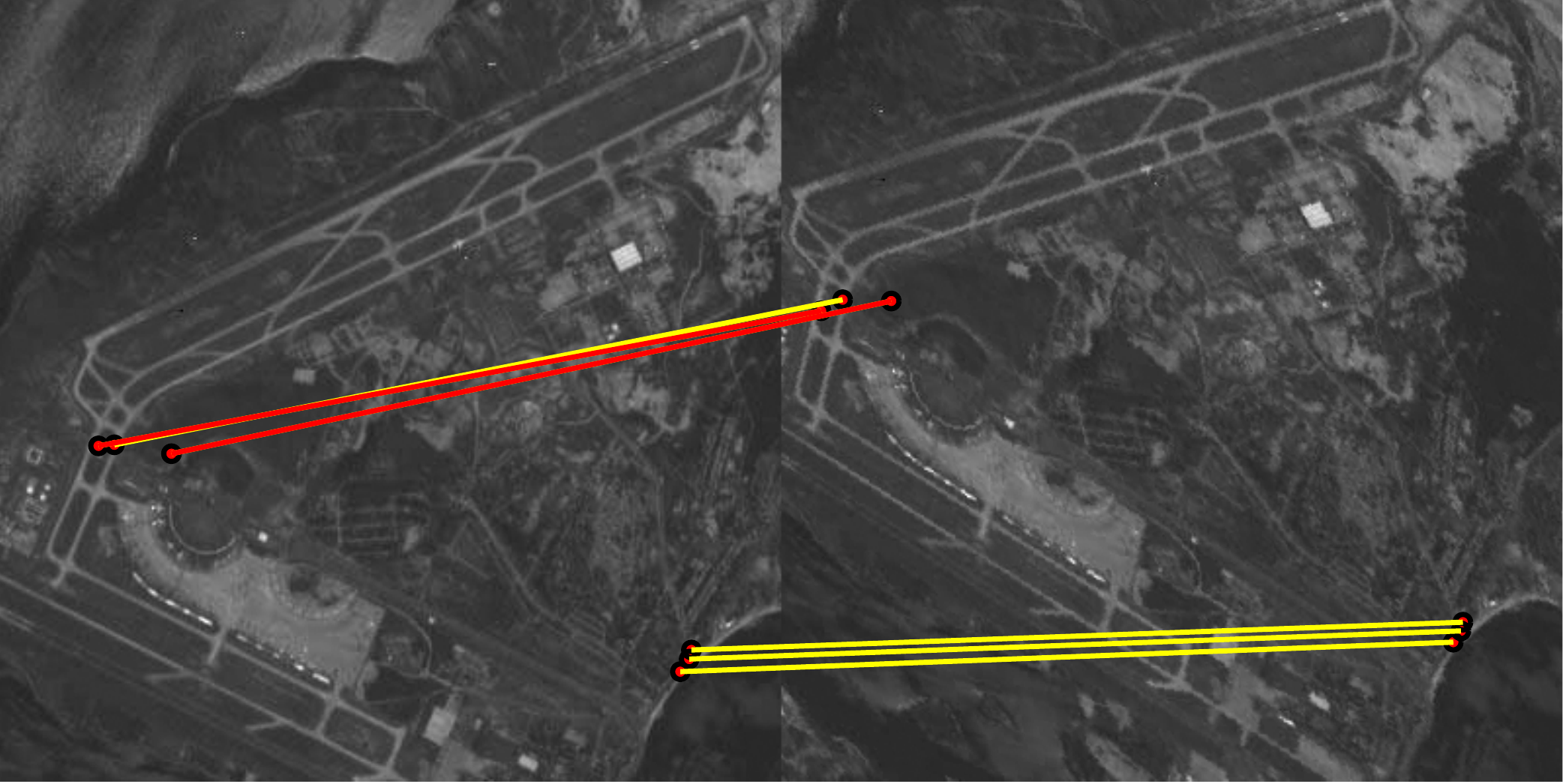}
    \end{minipage}}
  \subfigure[]{
    \label{fig:mini:subfig:a}
    \begin{minipage}[c]{0.22\textwidth}
      \centering
      \includegraphics[width=1.6in]{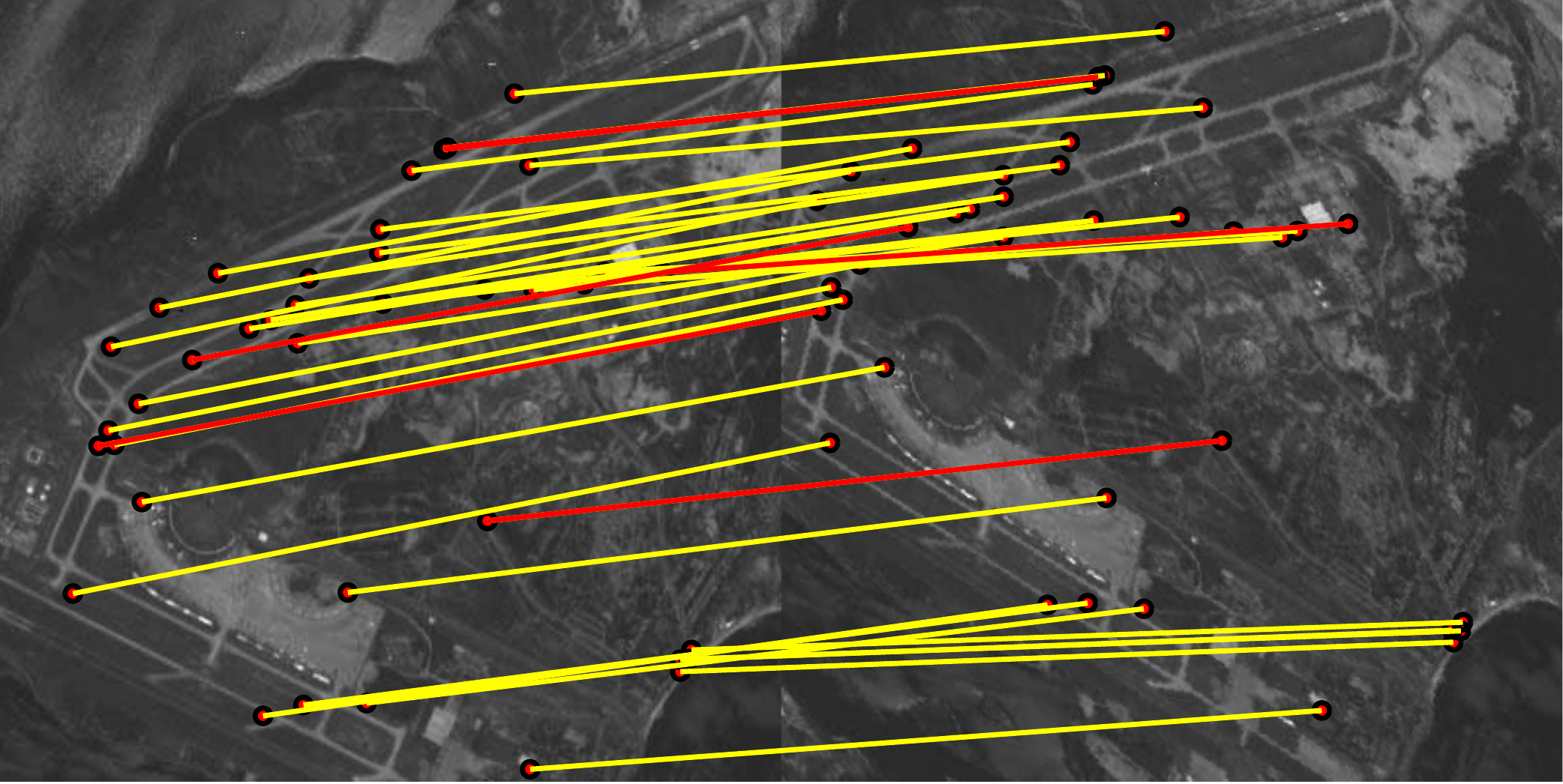}
    \end{minipage}}
  \subfigure[]{
    \label{fig:mini:subfig:b}
    \begin{minipage}[c]{0.22\textwidth}
      \centering
      \includegraphics[width=1.6in]{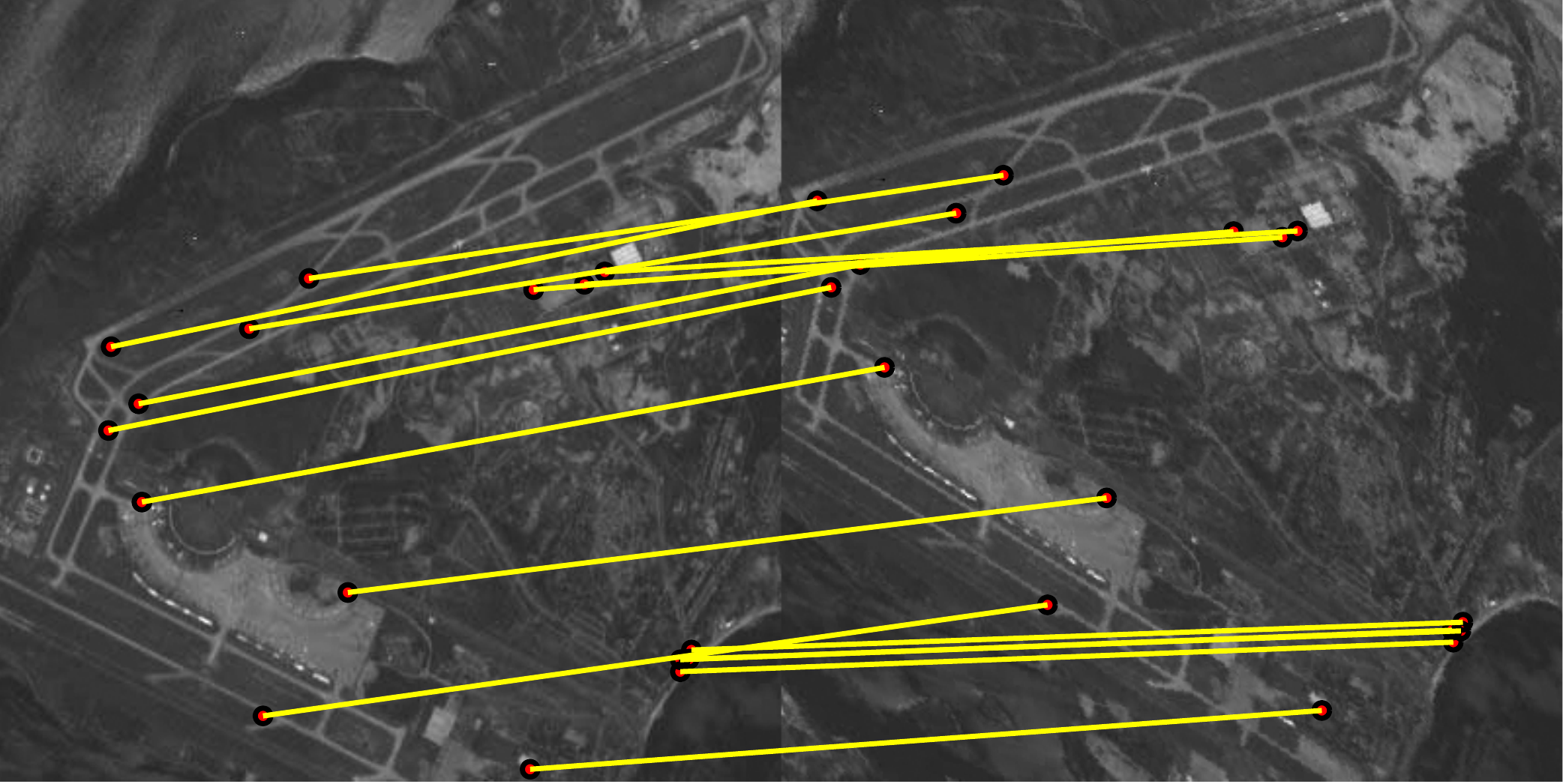}
    \end{minipage}}\\
  \captionstyle{normal}
  \caption{Feature point matching results for ImgSet1-2 with shear deformations. (a) Line segments of the reference image. (b) Line segments of the sensed image. (c) Line-support regions of the reference image. (d) Line-support regions of the sensed image. (e) Initial correspondences by SIFT. (f) Point correspondences by RANSAC. (g) Point correspondences by WGTM. (h) Point correspondences by GOR. (i) Point correspondences by LLT.}
  \label{fig-5}
\end{figure*}

For the aerial image of ImgSet1-1, a segment of 512$\times$512 pixels from the original image is selected as the reference image. A segment of same size is taken as the sensed image after a simulated rotation of $120^ \circ$ clockwise and a simulated scale factor of 0.8. As shown in Fig. \ref{fig-4} (a) and (b), the line segments extracted from the reference image and the sensed image respectively are highlighted in green. The segmented line-support regions are depicted in Fig. \ref{fig-4} (c) and (d). The initial correspondences are established between the line-support regions as shown in Fig. \ref{fig-4} (e), which consist of 11 inliers and 79 outliers. The matching performance of RANSAC, WGTM, LLT, and GOR is demonstrated in Fig. \ref{fig-4} (f)-(i). Inliers of residual correspondences are depicted by yellow lines, whereas the outliers are represented by red lines.
All outliers in the initial correspondences are removed by WGTM and GOR.
Despite of preserving as many inliers as GOR, RANSAC is unable to remove a large amount of outliers.
WGTM removes not only all of outliers but also a lot of inliers.
LLT preserves more inliers than others.
However, some stubborn outliers with the similar local structures are still preserved by LLT.

For the aerial image of ImgSet1-2, the original image sheared in both of the horizontal (h) and vertical (v) directions with the factors of h=0.1 and v=0.1 is regarded as the sensed image.
The detected line segments are shown in Fig. \ref{fig-5} (a) and (b).
34 inliers and 144 outliers matched by SIFT are used as initial correspondences.
Compared with RANSAC and WGTM, LLT and GOR have good performance in preserving inliers.
Several outliers that close to the correct locations have not been removed by LLT.
The compared results in Fig. \ref{fig-5} (f)-(i) indicate the advantage of the proposed GOR matching on sheared deformations.

Table II summarizes the results of SIFT feature point extraction and matching  for ImgSet1, ImgSet2, and ImgSet3.
SIFT feature points are extracted and matched from the original images and the line support region (LSR-SIFT), respectively.
Compared to SIFT extracting from original images with inconsistent content, LSR-SIFT is beneficial for providing much more initial inliers.
The comparative feature point matching results of RANSAC, WGTM, LLT, and GOR  are presented in the right part of Table II.
It can be observed that the proposed GOR is superior in removing more outliers and preserving sufficient inliers than RANSAC and WGTM, especially for the shear deformations.
RANSAC generally provides the closest recall values to those of GOR, but it degenerates much more seriously in the situations of large proportion of initial outliers.
This is because RANSAC estimates the parameters of the transformation model from a set of correspondences containing outliers. It can produce reasonable results only within certain proportion of outliers.
WGTM has a good performance in removing outliers for images with rigid transformation. However, the performance of WGTM degenerates rapidly when dealing with sheared transformed images.
This is because that shear deformations cause the inconformity of geometrical distance and angles between vectors of corresponding matches.
Accordingly, the angular distances between vectors of corresponding matches utilized in WGTM will be changed with sheared deformations.
Compared with other three algorithms, LLT usually has high value of recall.
However, the performance of LLT degenerates in terms of precision when the initial corresponding set contains large percentage of outliers.
This is because the local structures among neighboring feature points are easily affected by a large amount of outliers.
The local geometrical constraint will be confused by outliers with the similar neighboring feature points and inliers with neighboring outliers.

\begin{table*}[htb]
\centering
  \captionstyle{normal}
  \setlength{\abovecaptionskip}{0pt}
  \setlength{\belowcaptionskip}{10pt}
\caption{THE COMPARISON RESULTS OF FEATURE POINT MATCHING WITH OUTLIER REMOVAL}
\begin{lrbox}{\tablebox}
\begin{tabular}{|c|c|p{0.8cm}|p{0.8cm}|p{0.8cm}|p{0.8cm}|p{0.001cm}|p{0.95cm}|p{0.95cm}|p{0.95cm}|p{0.95cm}|p{0.95cm}|p{0.95cm}|p{0.95cm}|p{0.95cm}|p{0.95cm}|p{0.95cm}|}
\hline
\multirow{2}{*}{Image Pair} &
\multirow{2}{*}{iter} &
\multicolumn{2}{c|}{SIFT} &
\multicolumn{2}{c|}{LSR-SIFT} & &
\multicolumn{2}{c|}{RANSAC} &
\multicolumn{2}{c|}{WGTM} &
\multicolumn{2}{c|}{LLT} &
\multicolumn{2}{c|}{GOR}\\
\cline{3-6}
\cline{8-15}
&& Initial Match & Correct Match & Initial Match & Correct Match & & Recall & Precision & Recall & Precision & Recall & Precision & Recall & Precision\\
\cline{1-6} \cline{8-15}
ImgSet1-1&1&243&175&90&11 && 0.24&0.45&0.36&1.00 &0.91 &0.42 &0.64&1.00 \\
\cline{1-6} \cline{8-15}
ImgSet1-2&1&372&204&178&34 && 0.76&0.70&0.12&0.67&0.97 &0.85 &0.47&1.00  \\
\cline{1-6} \cline{8-15}
\multirow{2}{*}{ImgSet2-1}&1&520&126&201&10 && 0.10&0.04&0.00&0.00&0.60&0.22 &0.20&0.25 \\
\cline{2-6} \cline{8-15}
& 2&170&38&68&13 && 0.92&0.60&0.62&0.23& 0.92& 0.86& 0.77&1.00 \\
\cline{1-6} \cline{8-15}
\multirow{2}{*}{ImgSet2-2} &1&24&7&174&29 && 0.10&0.05&0.21&0.60 &0.83 &0.35 &0.03&0.13  \\
\cline{2-6} \cline{8-15}
&2&19&5&39&17 && 0.59&0.27&0.41&1.00&0.88 &1.00 &0.65&1.00 \\
\cline{1-6} \cline{8-15}
ImgSet2-3&1&136&10&76&19 &&	0.85&0.50&0.53&0.83 &0.79 &0.71 &0.53&1.00\\
\cline{1-6} \cline{8-15}
ImgSet2-4&1&18&3&61&23  &&	0.65&0.36&0.26&1.00 &0.70 &1.00 &0.39&1.00\\
\cline{1-6} \cline{8-15}
ImgSet2-5&1&158&4&90&32	&&0.44&0.45&0.13&1.00&0.66 &0.78 &0.34&1.00 \\
\cline{1-6} \cline{8-15}
ImgSet3-1&1&95&19&288&110 && 0.87&0.79&0.29&0.82 &0.67 &0.59 &0.34&1.00 \\
\cline{1-6} \cline{8-15}
ImgSet3-2&1&60&12&109&48 &&	0.25&0.26&0.15&1.00 &0.60 &0.88 &0.23&1.00\\
\cline{1-6} \cline{8-15}
ImgSet3-3&1&498&54&251&33 && 0.67&0.56&0.12&0.50 &0.61 &0.74 &0.52&1.00\\
\cline{1-6} \cline{8-15}
\hline
\end{tabular}
\end{lrbox}
\scalebox{1}{\usebox{\tablebox}}
\end{table*}

\subsection{Iterative Strategy}
As shown in Table II, feature point matching without iterative strategy is effective to most of image sets, merely except for ImgSet2-1 and ImgSet2-2.
In both of the iterative examples, all of the three matching methods are incapable of excluding outliers with the poor initial inliers at full resolution.
The meaningful matching results with iterative strategy can be respectively achieved when the original images of ImgSet2-1 and ImgSet2-2 are downsampled to 50$\%\times$50$\%$ of their original sizes.

\begin{figure*}[htb]
\centering
 \setlength{\abovecaptionskip}{0pt}
 \setlength{\belowcaptionskip}{0pt}
 \setlength{\intextsep}{8pt plus 3pt minus 2pt}
  \subfigure[]{
    \label{fig:mini:subfig:a}
    \begin{minipage}[c]{0.11\textwidth}
      \centering
      \includegraphics[width=0.8in]{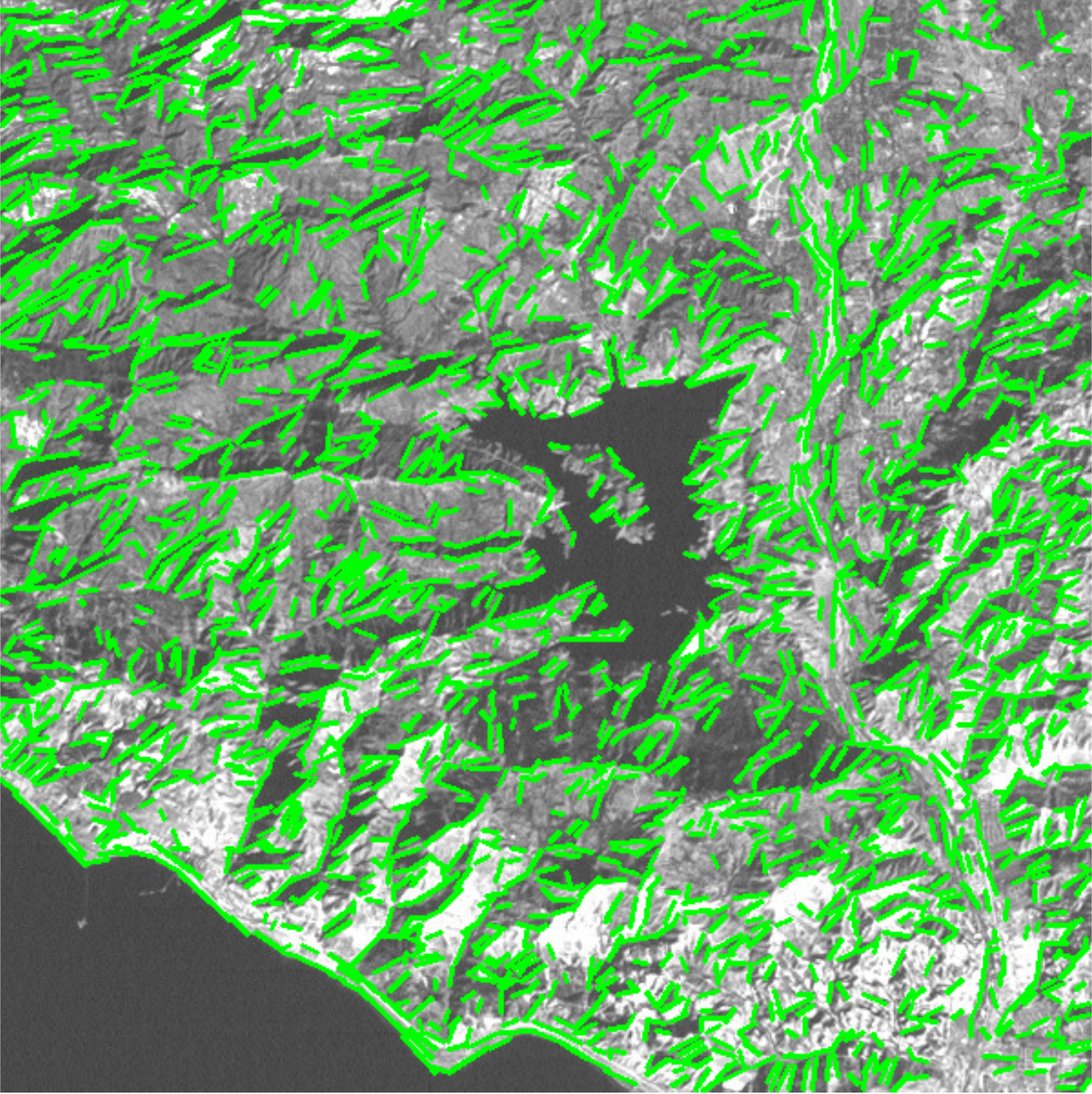}
    \end{minipage}}
  \subfigure[]{
    \label{fig:mini:subfig:b}
    \begin{minipage}[c]{0.11\textwidth}
      \centering
      \includegraphics[width=0.8in]{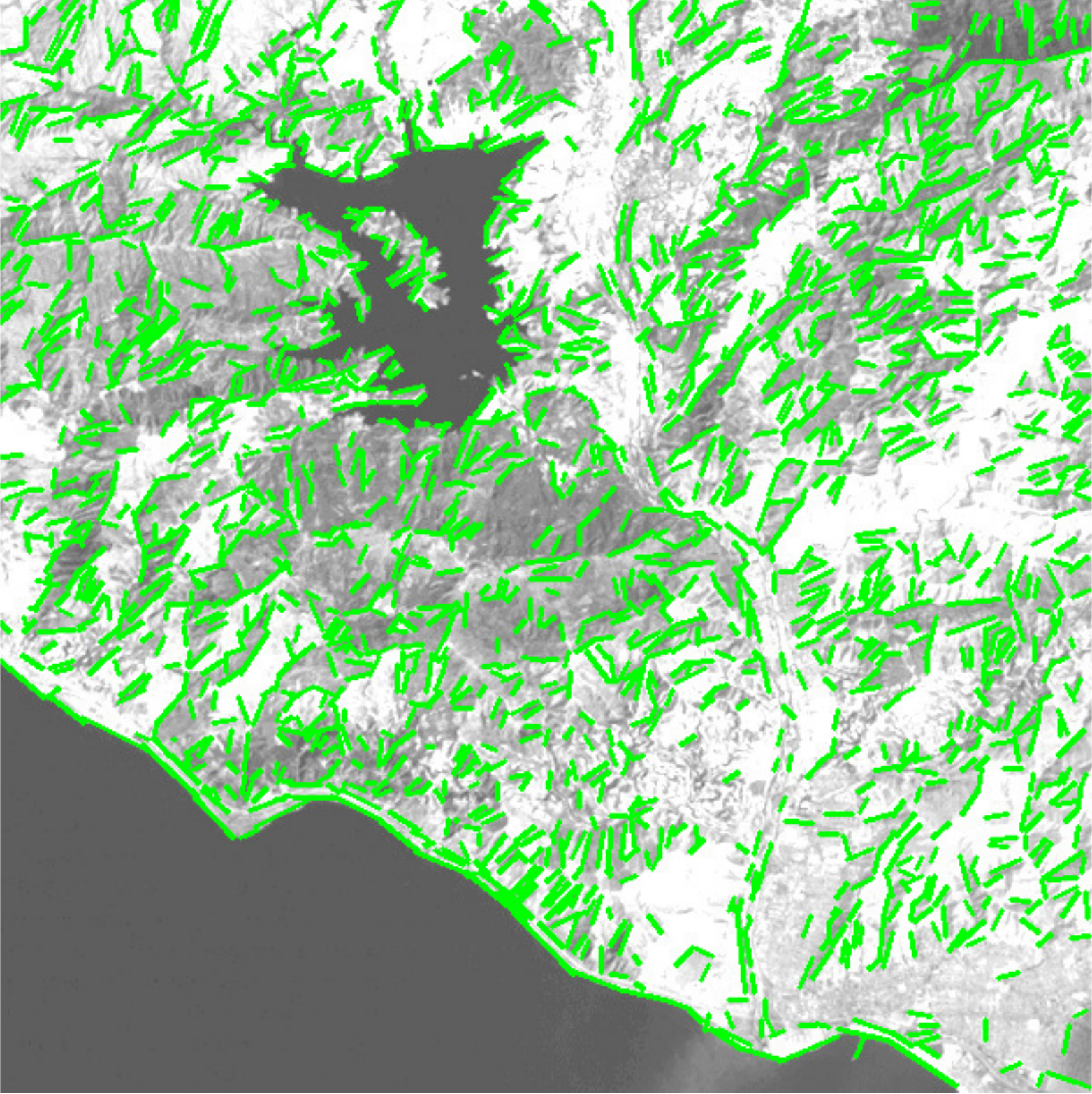}
    \end{minipage}}
  \subfigure[]{
    \label{fig:mini:subfig:a}
    \begin{minipage}[c]{0.11\textwidth}
      \centering
      \includegraphics[width=0.8in]{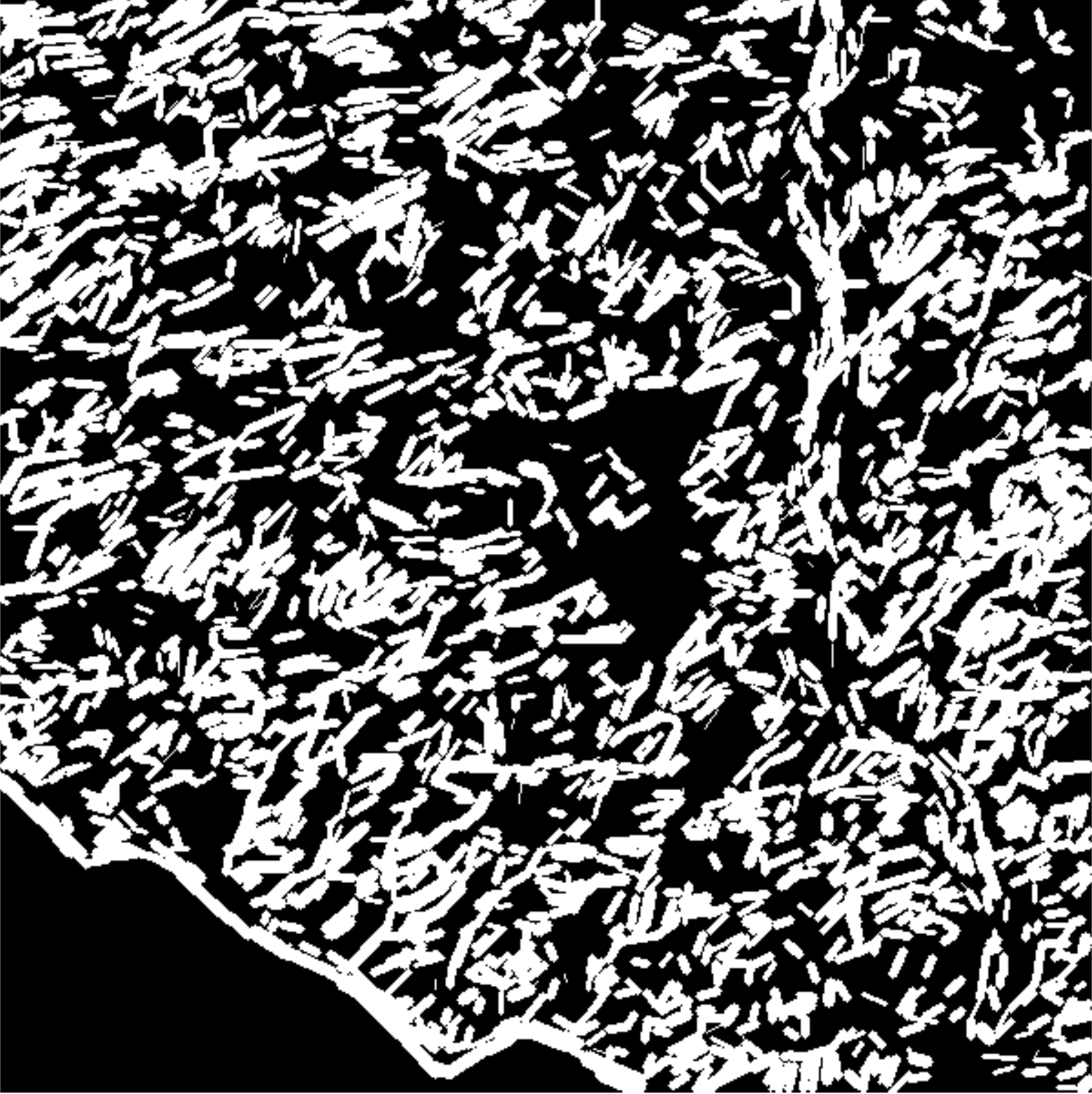}
    \end{minipage}}
  \subfigure[]{
    \label{fig:mini:subfig:b}
    \begin{minipage}[c]{0.11\textwidth}
      \centering
      \includegraphics[width=0.8in]{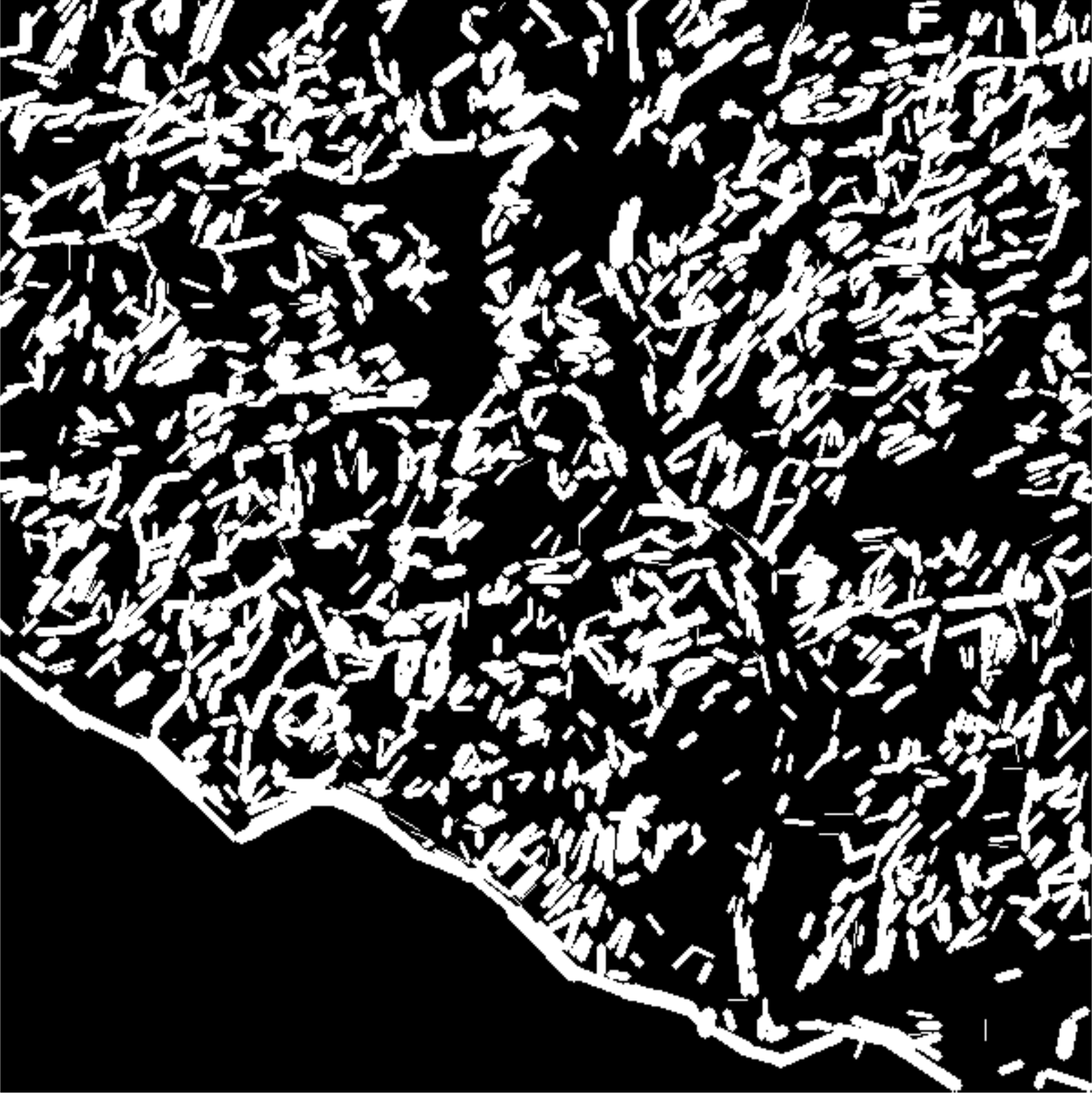}
    \end{minipage}}
  \subfigure[]{
    \label{fig:mini:subfig:a}
    \begin{minipage}[c]{0.22\textwidth}
      \centering
      \includegraphics[width=1.6in]{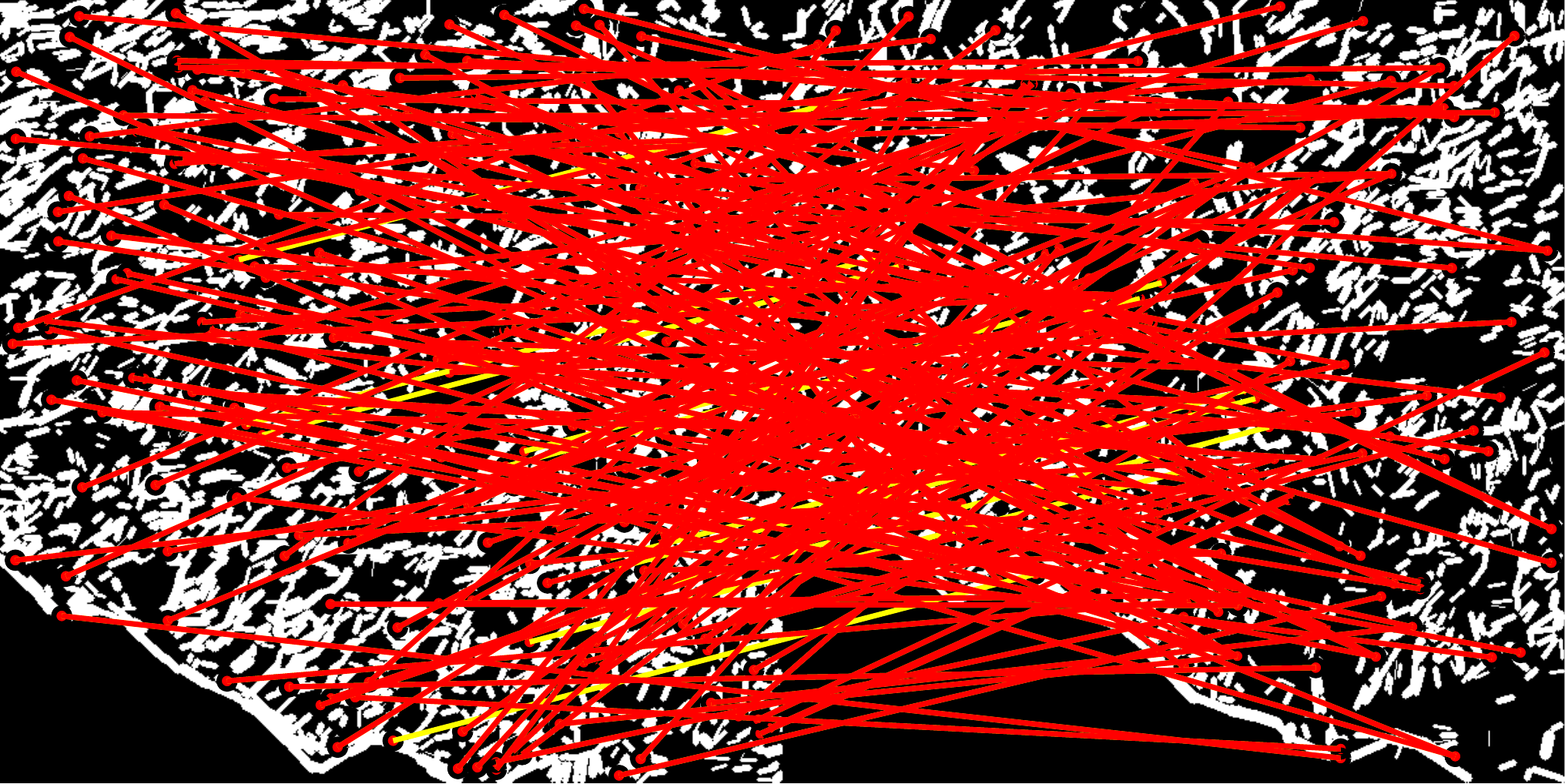}
    \end{minipage}}
  \subfigure[]{
    \label{fig:mini:subfig:a}
    \begin{minipage}[c]{0.22\textwidth}
      \centering
      \includegraphics[width=1.6in]{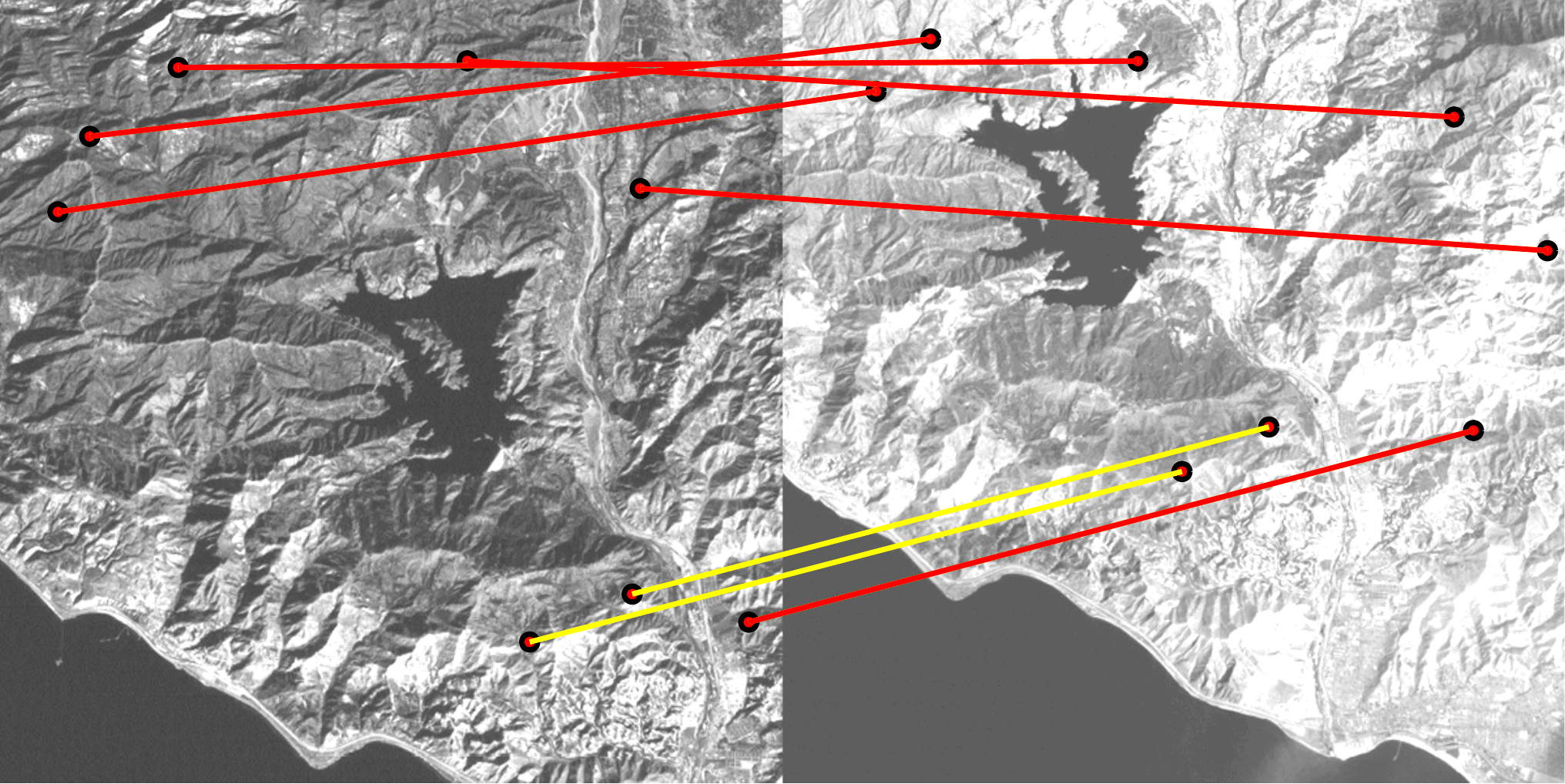}
    \end{minipage}}  \\

   \subfigure[]{
    \label{fig:mini:subfig:a}
    \begin{minipage}[c]{0.11\textwidth}
      \centering
      \includegraphics[width=0.8in]{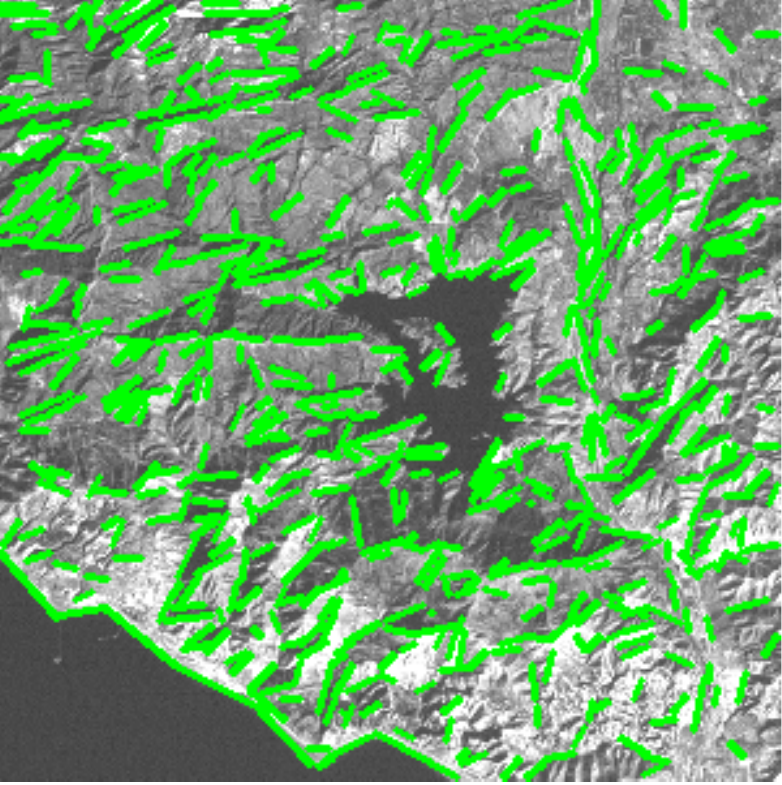}
    \end{minipage}}
  \subfigure[]{
    \label{fig:mini:subfig:b}
    \begin{minipage}[c]{0.11\textwidth}
      \centering
      \includegraphics[width=0.8in]{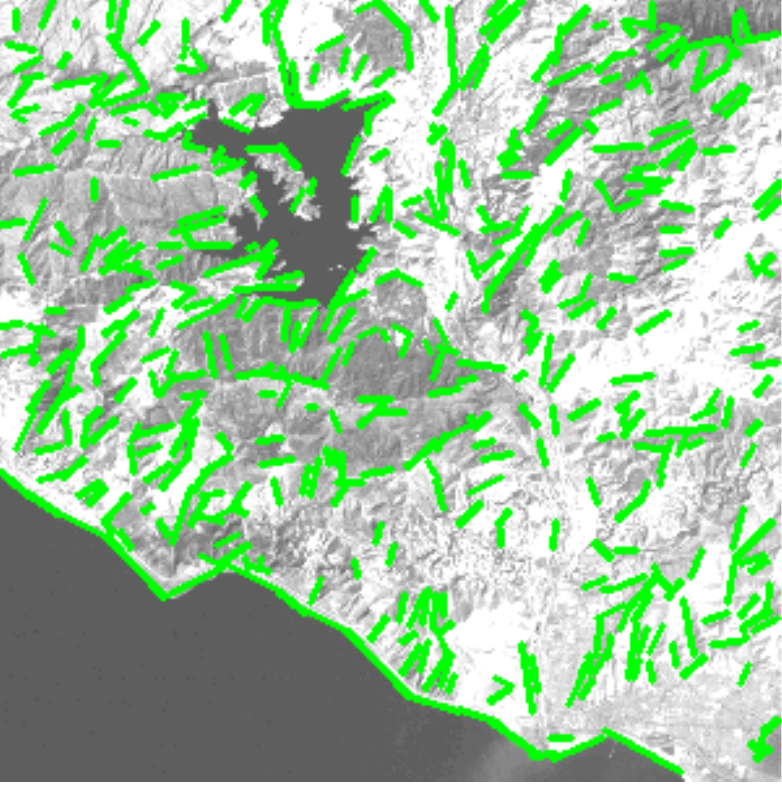}
    \end{minipage}}
  \subfigure[]{
    \label{fig:mini:subfig:a}
    \begin{minipage}[c]{0.11\textwidth}
      \centering
      \includegraphics[width=0.8in]{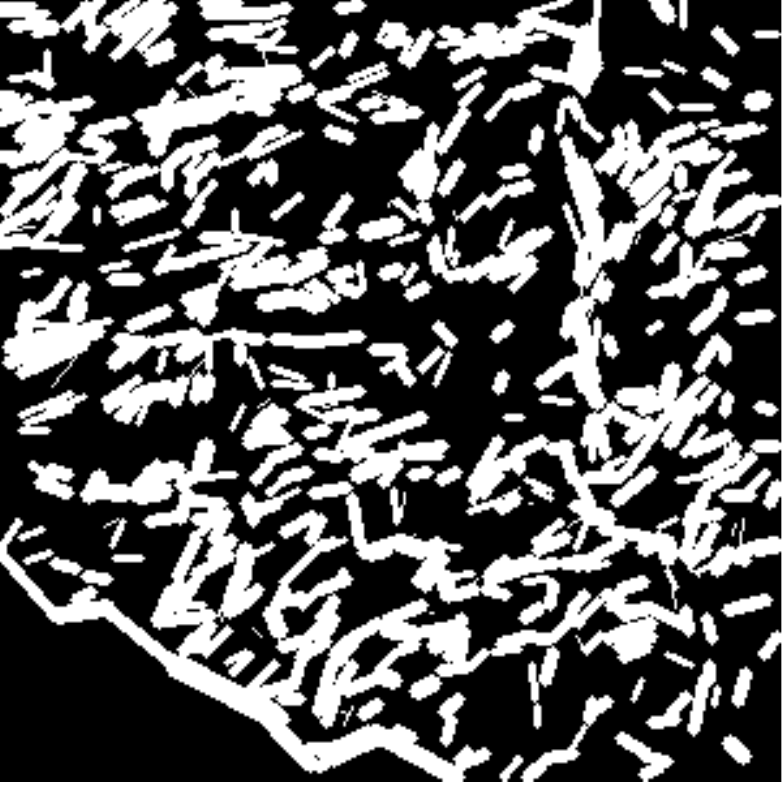}
    \end{minipage}}
  \subfigure[]{
    \label{fig:mini:subfig:b}
    \begin{minipage}[c]{0.11\textwidth}
      \centering
      \includegraphics[width=0.8in]{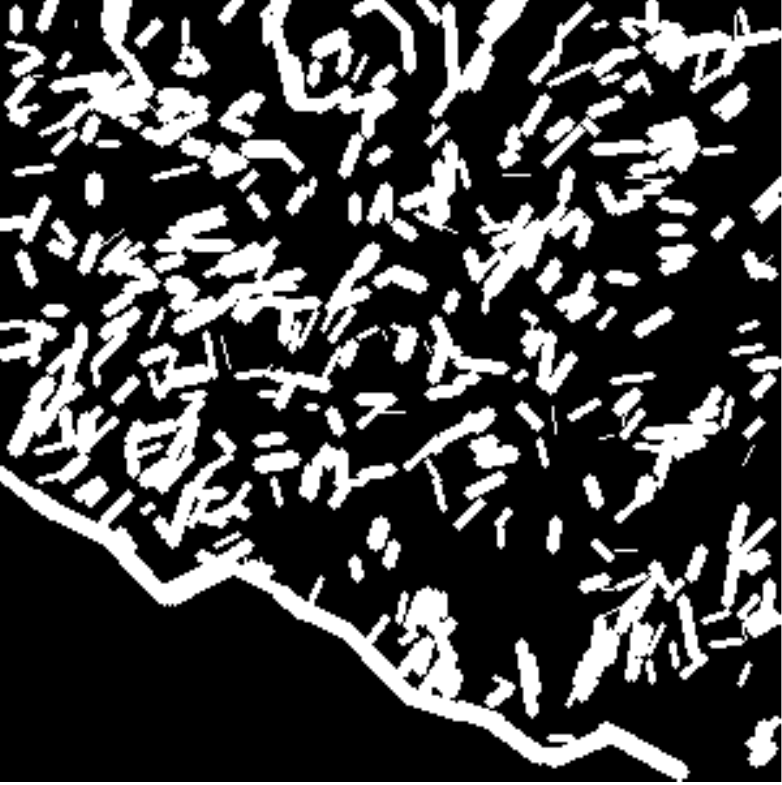}
    \end{minipage}}
  \subfigure[]{
    \label{fig:mini:subfig:b}
    \begin{minipage}[c]{0.22\textwidth}
      \centering
      \includegraphics[width=1.6in]{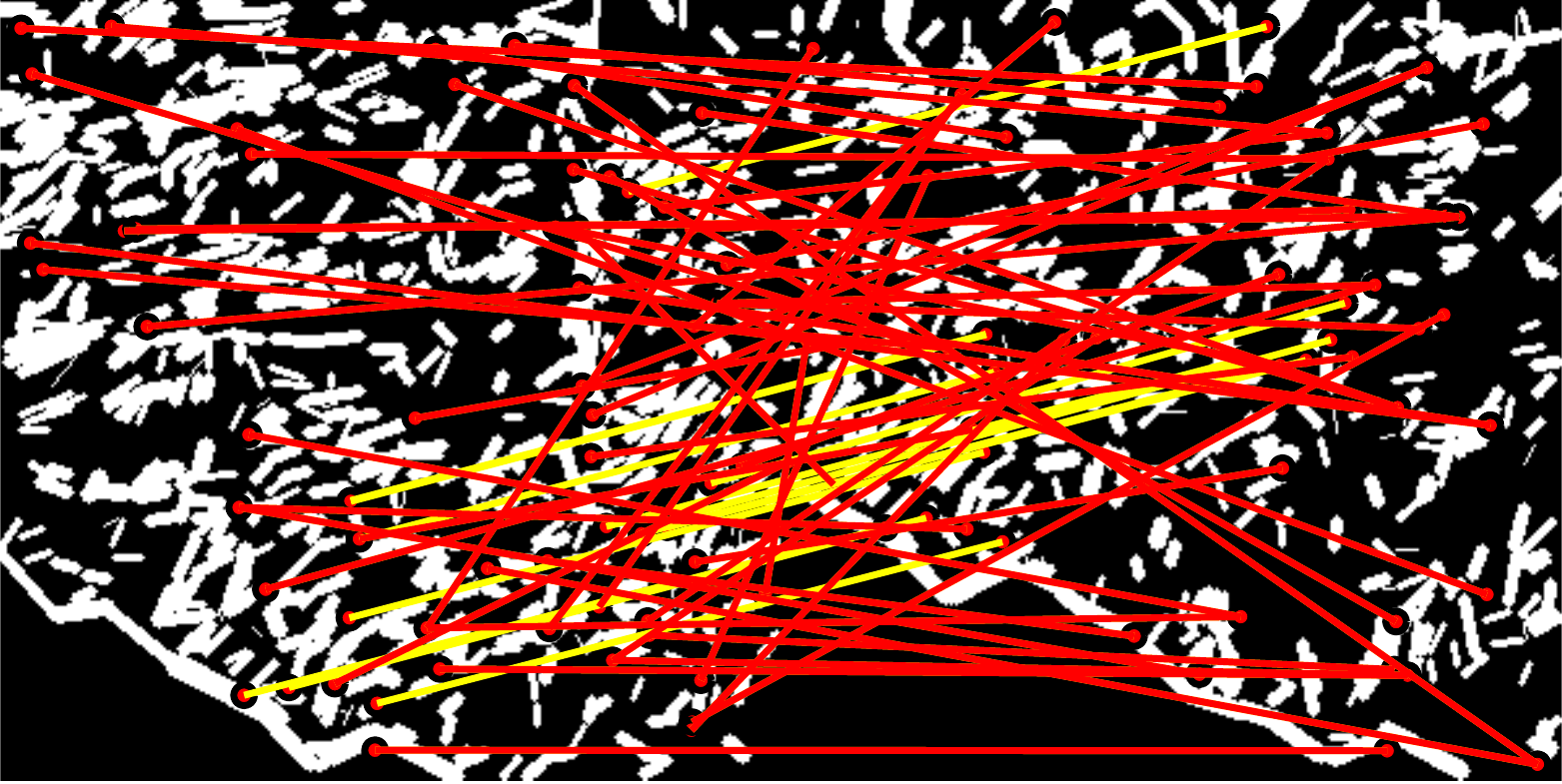}
    \end{minipage}}
  \subfigure[]{
    \label{fig:mini:subfig:b}
    \begin{minipage}[c]{0.22\textwidth}
      \centering
      \includegraphics[width=1.6in]{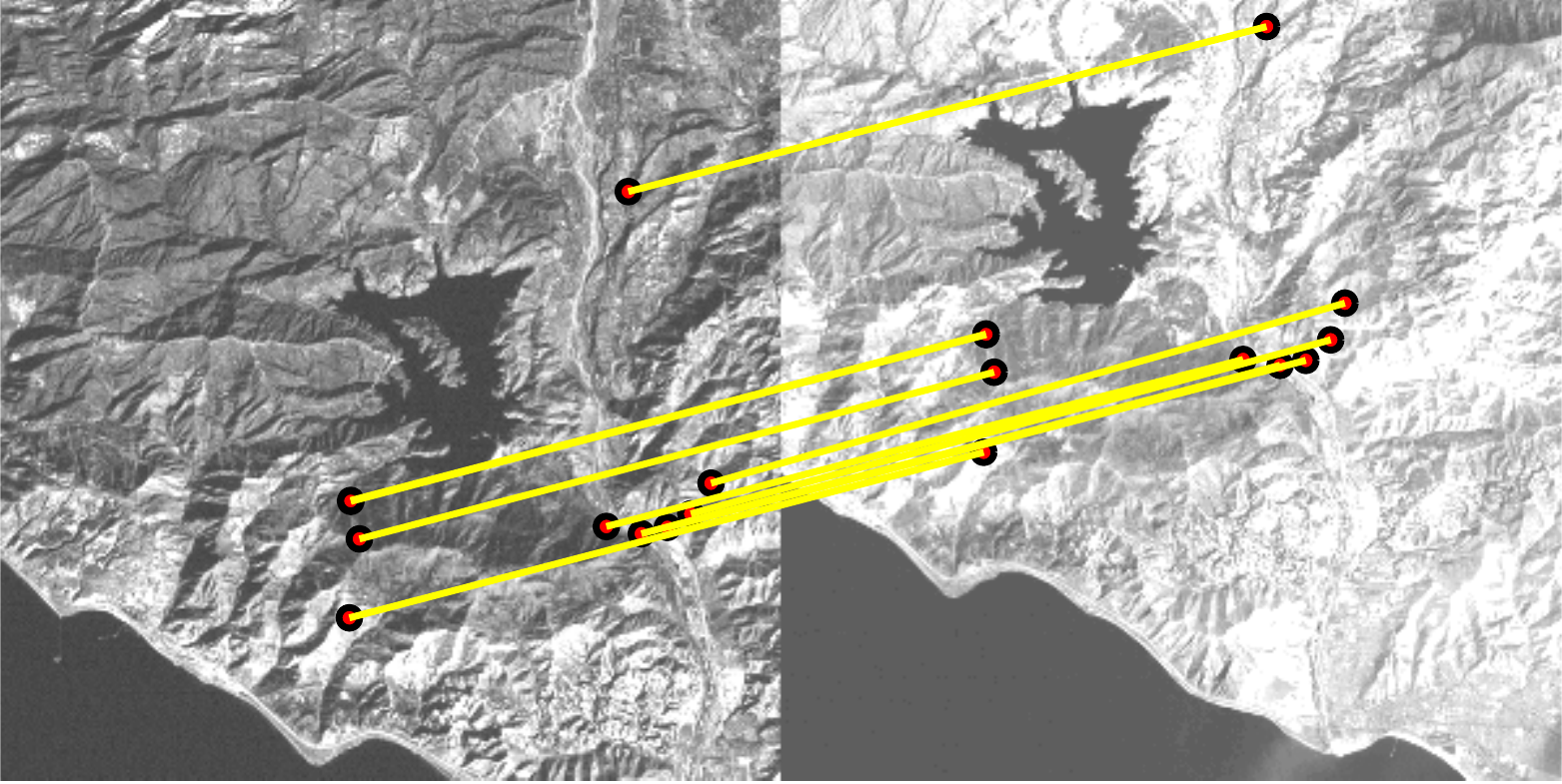}
    \end{minipage}}\\
  \captionstyle{normal}
  \caption{Iterative line segments extraction and matching for ImgSet2-1. (a) Line segments of reference image at the first iteration. (b) Line segments of sensed image at the first iteration. (c) Line-support regions of reference image at the first iteration. (d) Line-support regions of sensed image at the first iteration. (e) Initial correspondences by SIFT at the first iteration. (f) Point correspondences by GOR at the first iteration.
  (g) Line segments of reference image at the second iteration. (h) Line segments of sensed image at the second iteration.  (i) Line-support regions of reference image at the second iteration. (j) Line-support regions of sensed image at the second iteration. (l) Initial correspondences by SIFT at the second iteration.  (l) Point correspondences by GOR at the second iteration.}
  \label{fig-6}
\end{figure*}

The first considered image pair of ImgSet2-1 are obtained from Landsat TM band 5 and band 7 with the same size of 600$\times$600 pixels.
Beyond the spectral difference and displacement between the two images, it can be clearly seen that the similar patterns appeared in the scene.
At the first iteration of full resolution as shown in Fig. \ref{fig-6} (c) and (d), a large amount of isolated line segments are extracted from the original images.
Most of line-support regions are isolated and similar in appearance.
It makes this example a quite difficult to provide sufficient initial inliers at the full resolution.
As shown in Fig. \ref{fig-6} (e), 201 pairs of corresponding matches are initialized by SIFT with only 10 inliers.
At the second iteration of decreasing the original images to 50$\%\times$50$\%$ of their size, both of the reference and sensed image are downsampled to the resolution of 300$\times$300 pixels.
The structures of line-support regions in the second iteration are preserved much more completely.
The number of inliers in the initial set increases to 13, with respect to 68 initial correspondences. Finally, 9 pairs of corresponding matches without any outliers are obtained by the proposed method with two iterations.

\begin{figure*}[htb]
\centering
 \setlength{\abovecaptionskip}{0pt}
 \setlength{\belowcaptionskip}{0pt}
 \setlength{\intextsep}{8pt plus 3pt minus 2pt}
  \subfigure[]{
    \label{fig:mini:subfig:a}
    \begin{minipage}[c]{0.11\textwidth}
      \centering
      \includegraphics[width=0.8in]{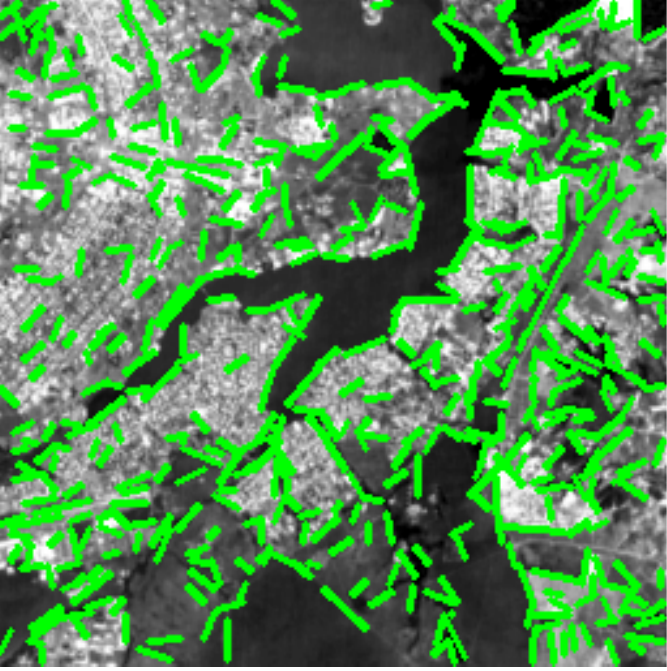}
    \end{minipage}}
  \subfigure[]{
    \label{fig:mini:subfig:b}
    \begin{minipage}[c]{0.11\textwidth}
      \centering
      \includegraphics[width=0.8in]{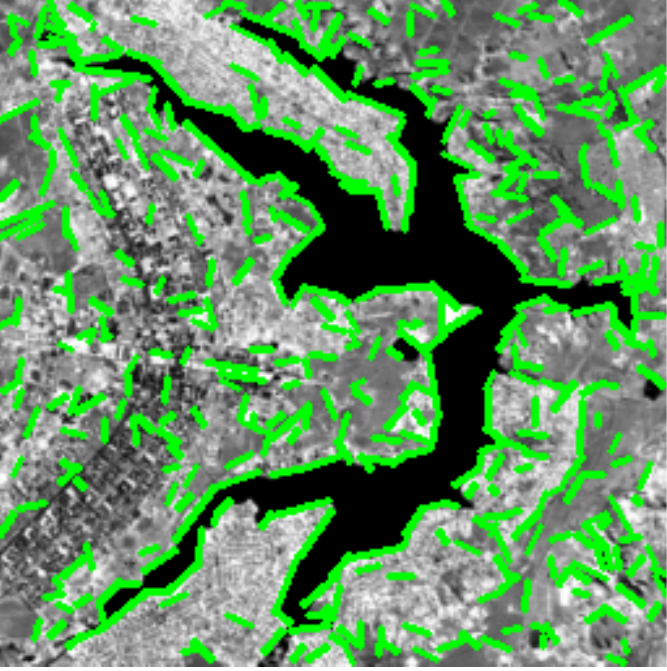}
    \end{minipage}}
  \subfigure[]{
    \label{fig:mini:subfig:a}
    \begin{minipage}[c]{0.11\textwidth}
      \centering
      \includegraphics[width=0.8in]{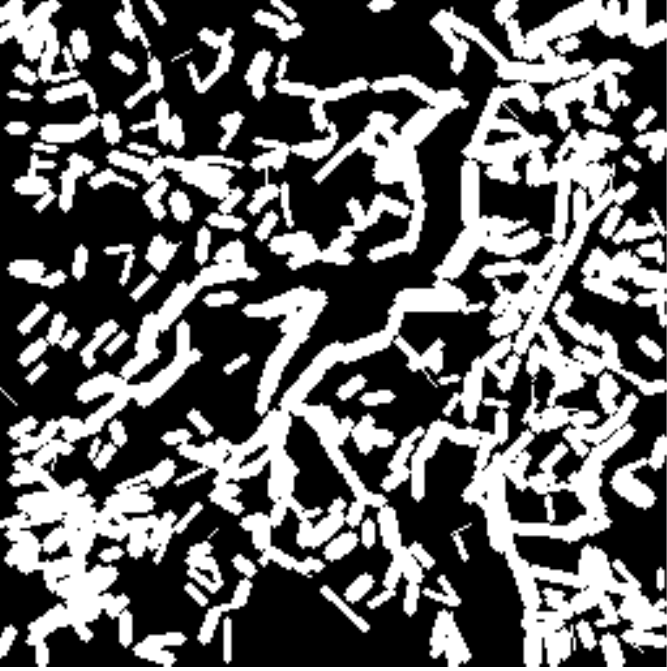}
    \end{minipage}}
  \subfigure[]{
    \label{fig:mini:subfig:b}
    \begin{minipage}[c]{0.11\textwidth}
      \centering
      \includegraphics[width=0.8in]{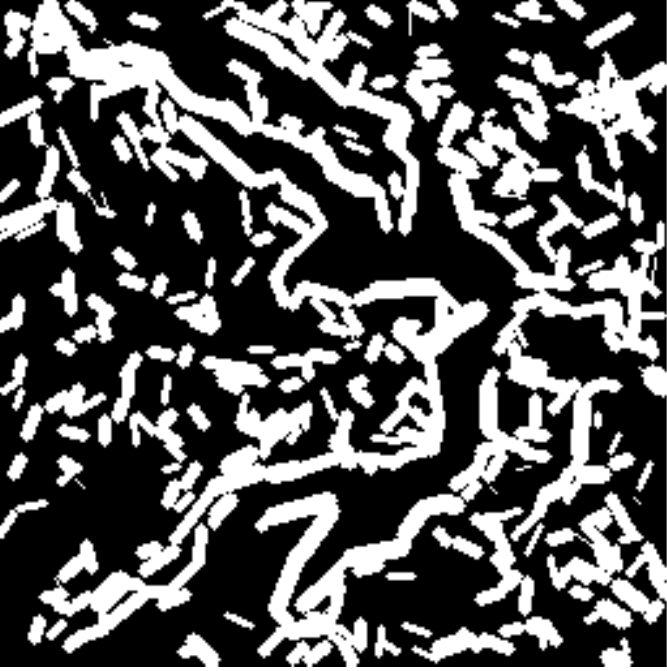}
    \end{minipage}}
  \subfigure[]{
    \label{fig:mini:subfig:a}
    \begin{minipage}[c]{0.22\textwidth}
      \centering
      \includegraphics[width=1.6in]{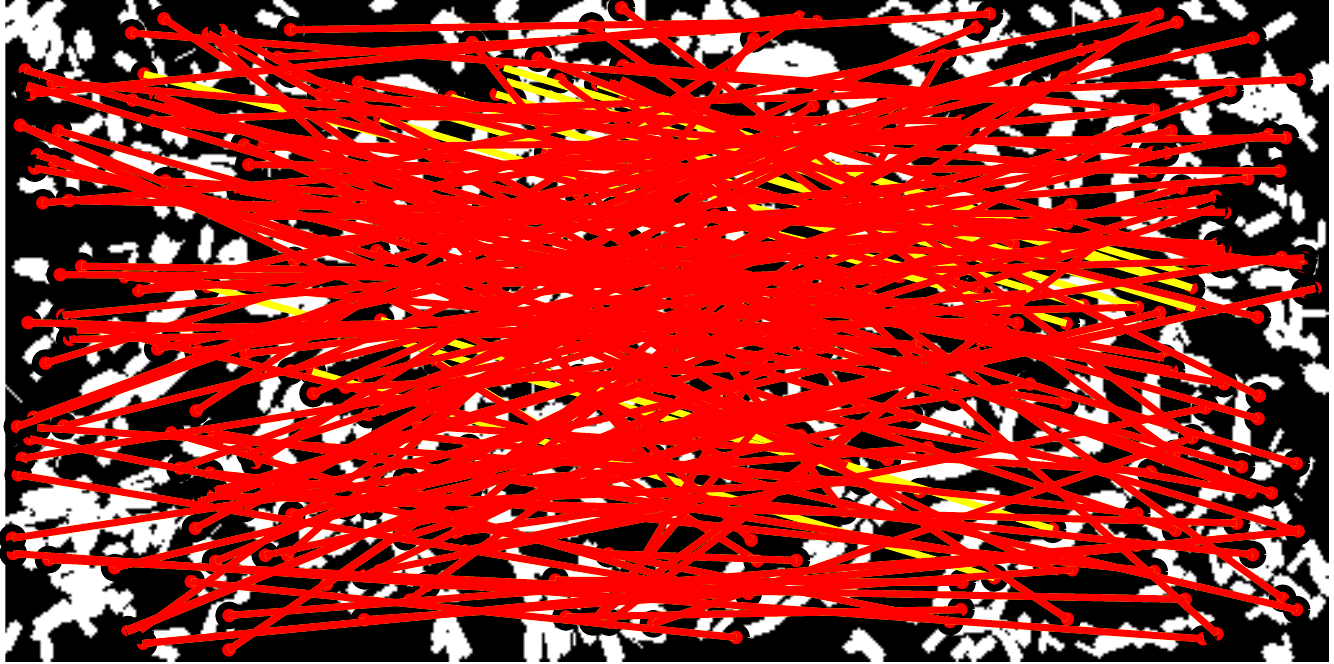}
    \end{minipage}}
  \subfigure[]{
    \label{fig:mini:subfig:a}
    \begin{minipage}[c]{0.22\textwidth}
      \centering
      \includegraphics[width=1.6in]{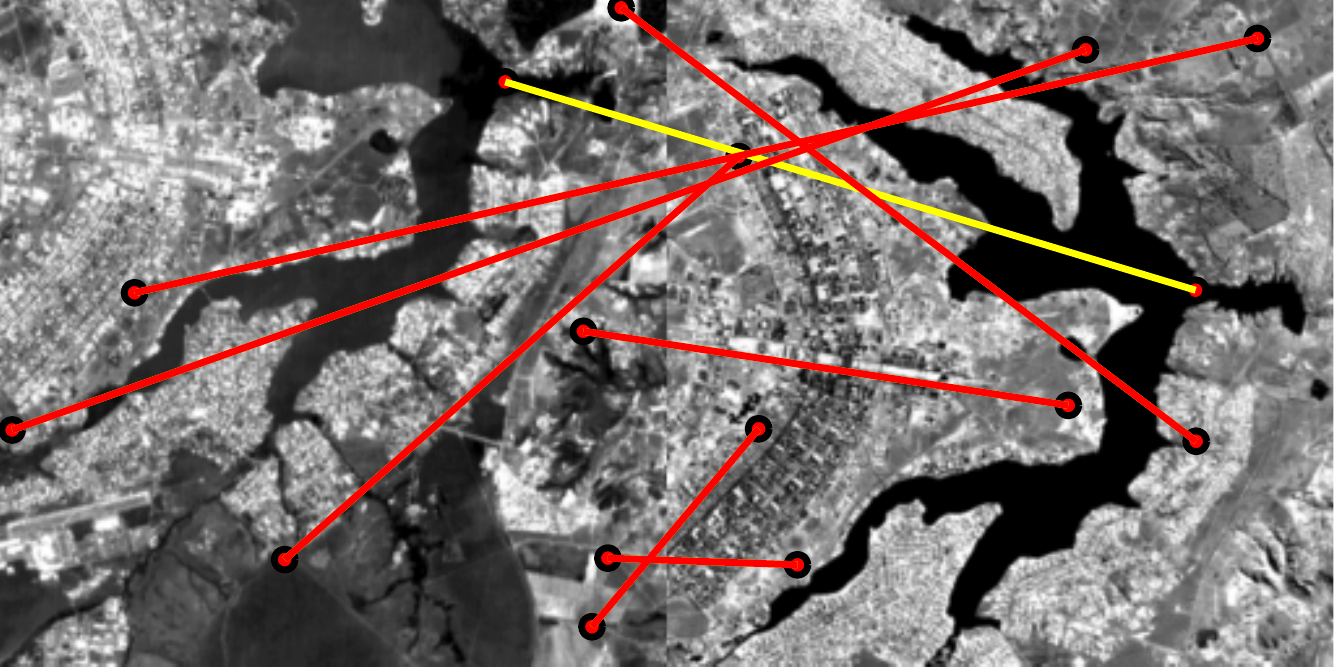}
    \end{minipage}}  \\

   \subfigure[]{
    \label{fig:mini:subfig:a}
    \begin{minipage}[c]{0.11\textwidth}
      \centering
      \includegraphics[width=0.8in]{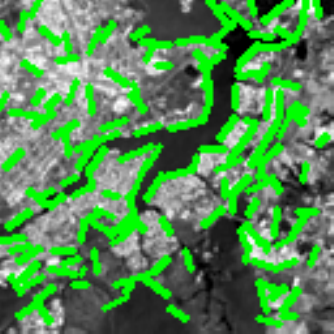}
    \end{minipage}}
  \subfigure[]{
    \label{fig:mini:subfig:b}
    \begin{minipage}[c]{0.11\textwidth}
      \centering
      \includegraphics[width=0.8in]{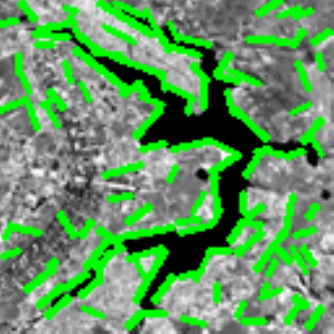}
    \end{minipage}}
  \subfigure[]{
    \label{fig:mini:subfig:a}
    \begin{minipage}[c]{0.11\textwidth}
      \centering
      \includegraphics[width=0.8in]{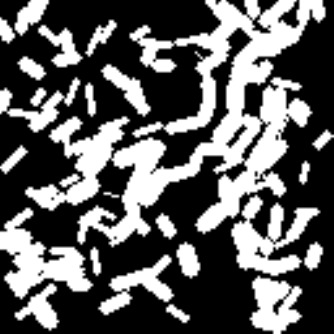}
    \end{minipage}}
  \subfigure[]{
    \label{fig:mini:subfig:b}
    \begin{minipage}[c]{0.11\textwidth}
      \centering
      \includegraphics[width=0.8in]{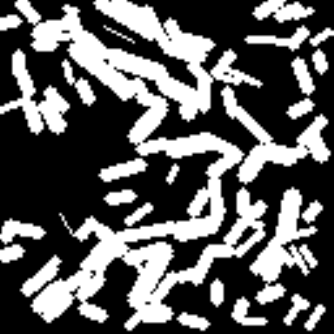}
    \end{minipage}}
  \subfigure[]{
    \label{fig:mini:subfig:b}
    \begin{minipage}[c]{0.22\textwidth}
      \centering
      \includegraphics[width=1.6in]{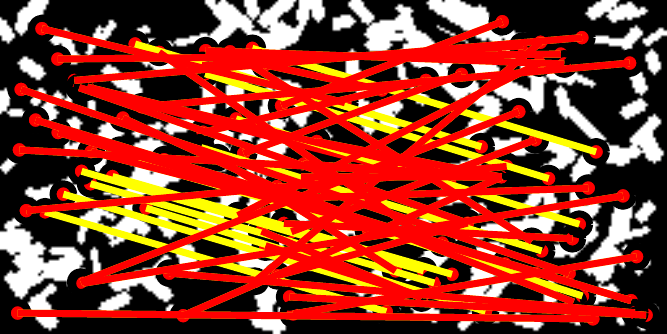}
    \end{minipage}}
  \subfigure[]{
    \label{fig:mini:subfig:b}
    \begin{minipage}[c]{0.22\textwidth}
      \centering
      \includegraphics[width=1.6in]{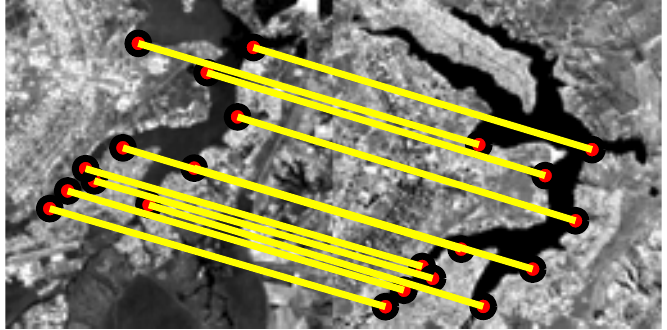}
    \end{minipage}}\\
  \captionstyle{normal}
  \caption{Iterative line segments extraction and matching for ImgSet2-2. (a) Line segments of reference image at the first iteration. (b) Line segments of sensed image at the first iteration. (c) Line-support regions of reference image at the first iteration. (d) Line-support regions of sensed image at the first iteration. (e) Initial correspondences by SIFT at the first iteration. (f) Point correspondences by GOR at the first iteration.
  (g) Line segments of reference image at the second iteration. (h) Line segments of sensed image at the second iteration.  (i) Line-support regions of reference image at the second iteration. (j) Line-support regions of sensed image at the second iteration. (l) Initial correspondences by SIFT at the second iteration.  (l) Point correspondences by GOR at the second iteration.}
  \label{fig-b}
\end{figure*}

The second considered image pair of ImgSet2-2 are composed by the two images with the same size of 256$\times$256 pixels, respectively from SPOT band 3 (0.78-0.89$\mu$m) and Landsat TM band 4 (0.76-0.90$\mu$m) over an urban area in Brasilia, Brazil.
Although the water body are completely preserved in the first iteration, a large amount of isolated line segments are extracted due to the presence of small size of sparkling features around the water body as shown in Fig. \ref{fig-b} (c) and (d).
At the second iteration, the salient structure of the water body are still preserved, while isolated segments extracted from the downsampled images significantly decrease.
It leads to the drastically reduced outliers initialized by SIFT in Fig. \ref{fig-b} (k).
The residual matches
Finally, 11 pairs of corresponding matches without any outliers are obtained by the proposed method with two iterations.

\subsection{Registration for Multispectral Remote Sensing Images}
\begin{figure*}[htb]
\centering
 \setlength{\abovecaptionskip}{0pt}
 \setlength{\belowcaptionskip}{0pt}
 \setlength{\intextsep}{8pt plus 3pt minus 2pt}
  \subfigure[]{
    \label{fig:mini:subfig:a}
    \begin{minipage}[c]{0.18\textwidth}
      \centering
      \includegraphics[width=1.5in]{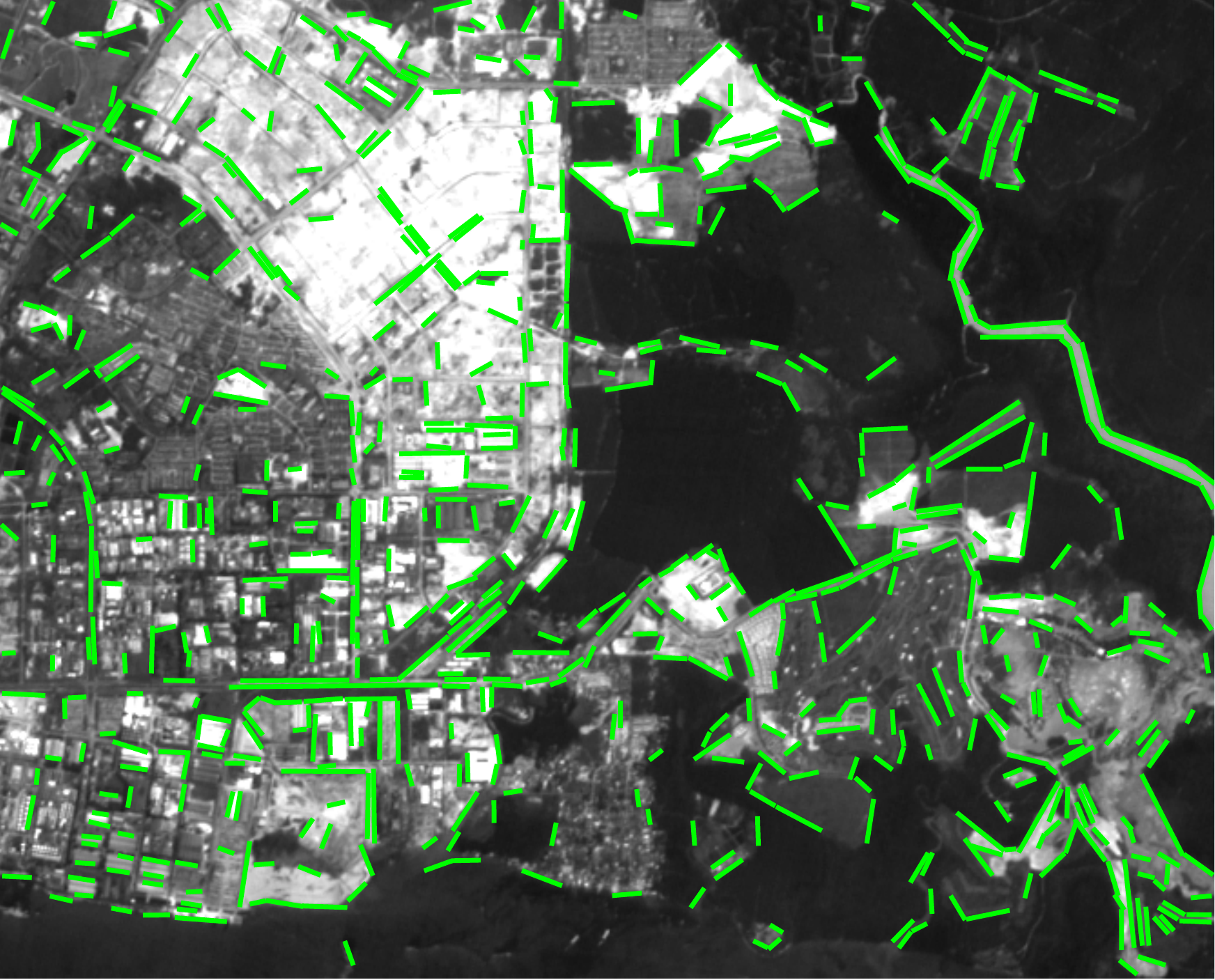}
    \end{minipage}}
  \subfigure[]{
    \label{fig:mini:subfig:b}
    \begin{minipage}[c]{0.18\textwidth}
      \centering
      \includegraphics[width=0.75in]{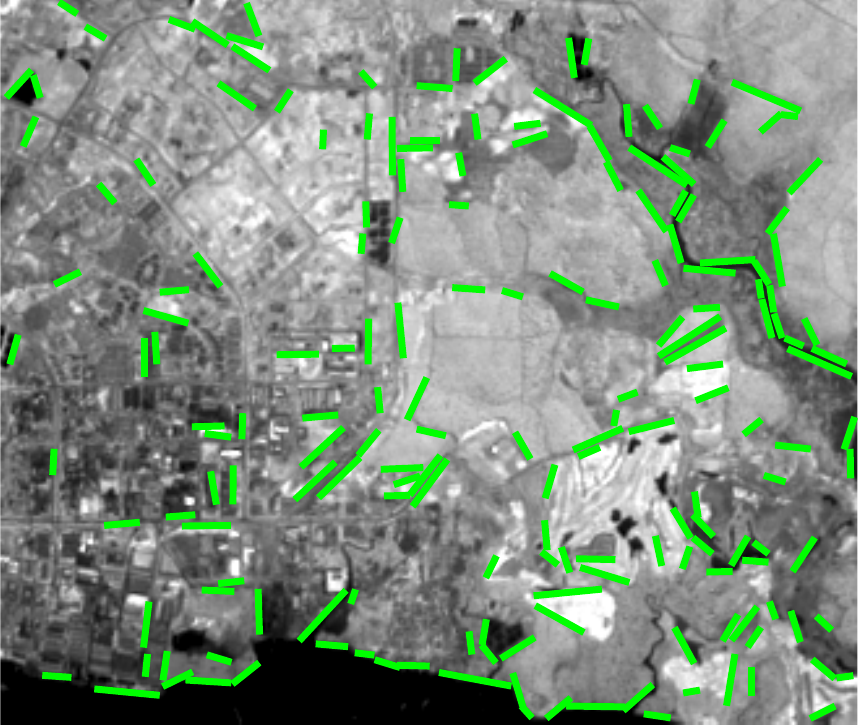}
    \end{minipage}}
  \subfigure[]{
    \label{fig:mini:subfig:a}
    \begin{minipage}[c]{0.18\textwidth}
      \centering
      \includegraphics[width=1.5in]{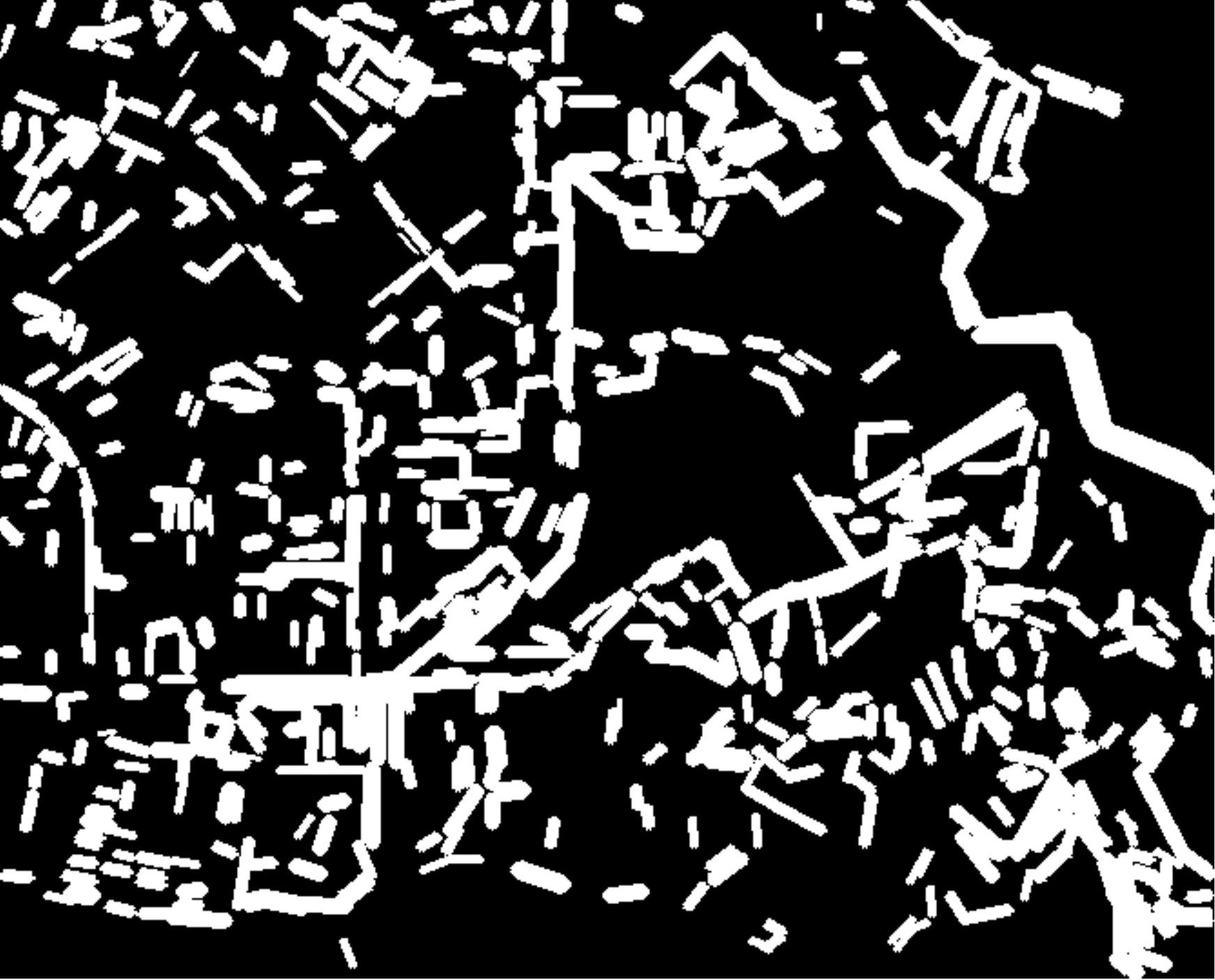}
    \end{minipage}}
  \subfigure[]{
    \label{fig:mini:subfig:b}
    \begin{minipage}[c]{0.18\textwidth}
      \centering
      \includegraphics[width=0.75in]{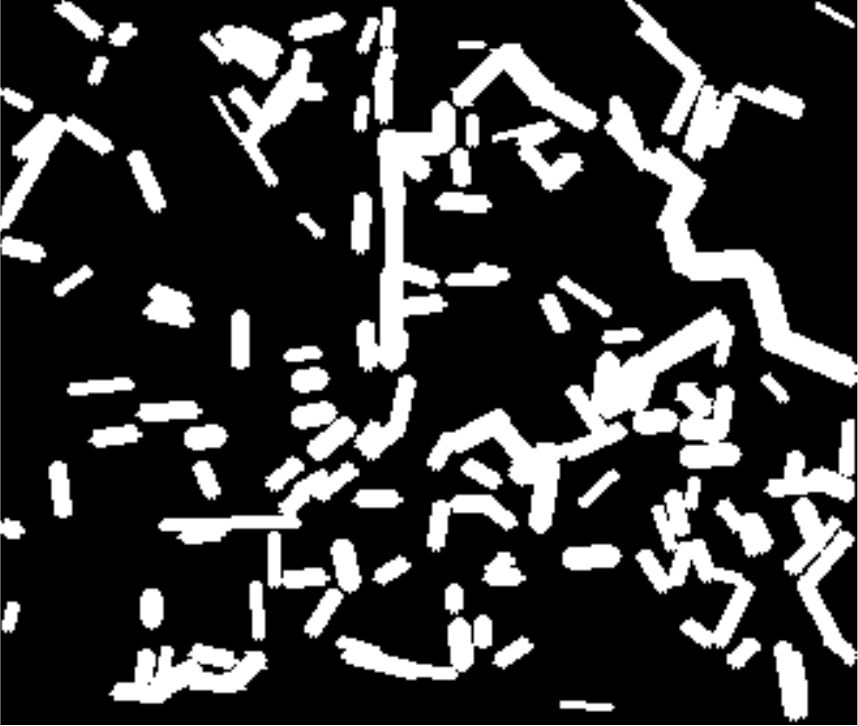}
    \end{minipage}}\\
  \subfigure[]{
    \label{fig:mini:subfig:a}
    \begin{minipage}[c]{0.4\textwidth}
      \centering
      \includegraphics[width=2.3in]{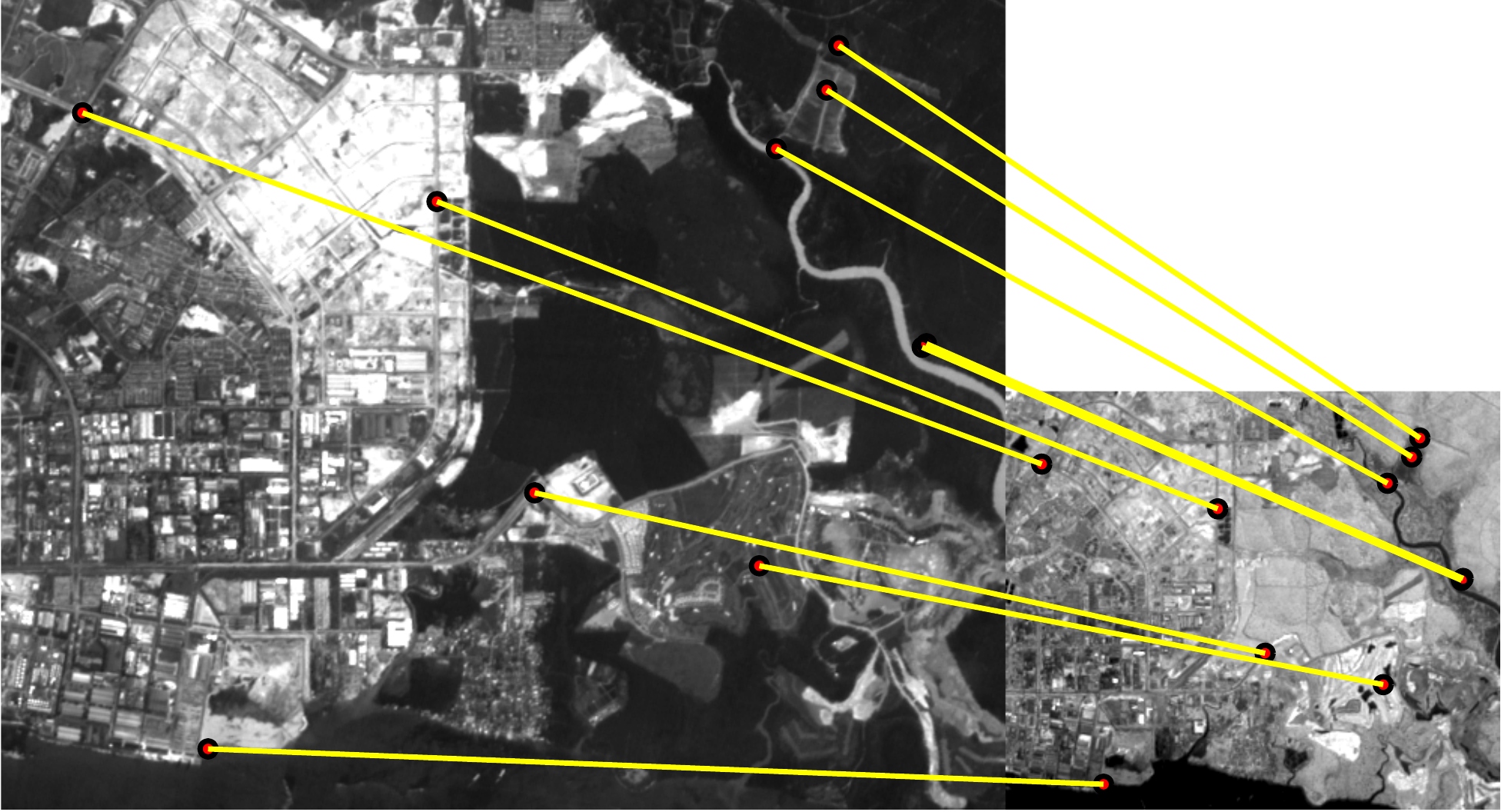}
    \end{minipage}}
  \subfigure[]{
    \label{fig:mini:subfig:b}
    \begin{minipage}[c]{0.4\textwidth}
      \centering
      \includegraphics[width=1.5in]{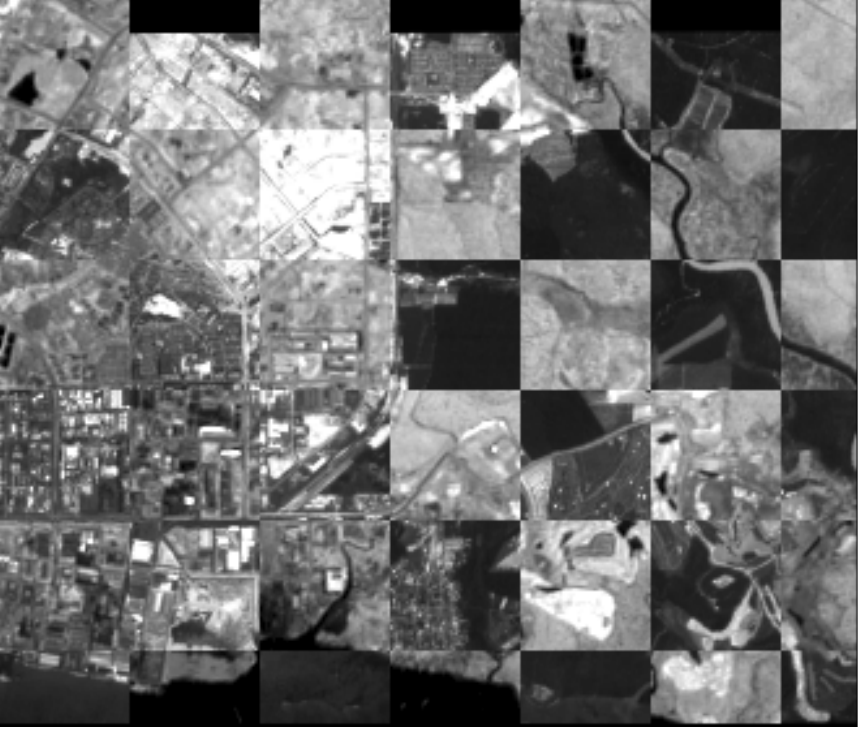}
    \end{minipage}}\\
  \captionstyle{normal}
  \caption{Image registration for ImgSet2-3. (a) Line segments of reference image. (b) Line segments of sensed image. (c) Line-support regions of reference image. (d) Line-support regions of sensed image. (e) Point correspondences by GOR. (f) Checkerboard mosaiced image.}
  \label{fig-8}
\end{figure*}

The image pair of ImgSet2-3 is composed by a segment with a size of 669$\times$539 pixels from a panchromatic SPOT image (with a spatial resolution of 10 m) and a segment with a size of 329$\times$278 pixels from a SPOT image of XS3 (near IR) band (with a spatial resolution of 20 m).
Beyond the scaling deformation between these images, the significant difference exists in the same scene. For example, water mass appears bright in the panchromatic image, but appears dark in the XS3 band.
The extracted line segments are superimposed on the reference and sensed images in Fig. \ref{fig-8} (a) and (b).
The main structures from multispectral contents are preserved in the corresponding line-support regions as shown in Fig. \ref{fig-8} (c) and (d).
9 correct matches are preserved in Fig. \ref{fig-8} (e) with GOR.
The coefficients of affine transformation model are estimated by Least Square Method (LSM) \cite{J_SU_1991_ITPAMI} using the residual corresponding matches.
The checkerboard mosaiced image in Fig. \ref{fig-8} (f) shows that the  features of the two multispectral images such as the river are precisely overlapped.

\begin{figure*}[htb]
\centering
 \setlength{\abovecaptionskip}{0pt}
 \setlength{\belowcaptionskip}{0pt}
 \setlength{\intextsep}{8pt plus 3pt minus 2pt}
  \subfigure[]{
    \label{fig:mini:subfig:a}
    \begin{minipage}[c]{0.182\textwidth}
      \centering
      \includegraphics[width=1.25in]{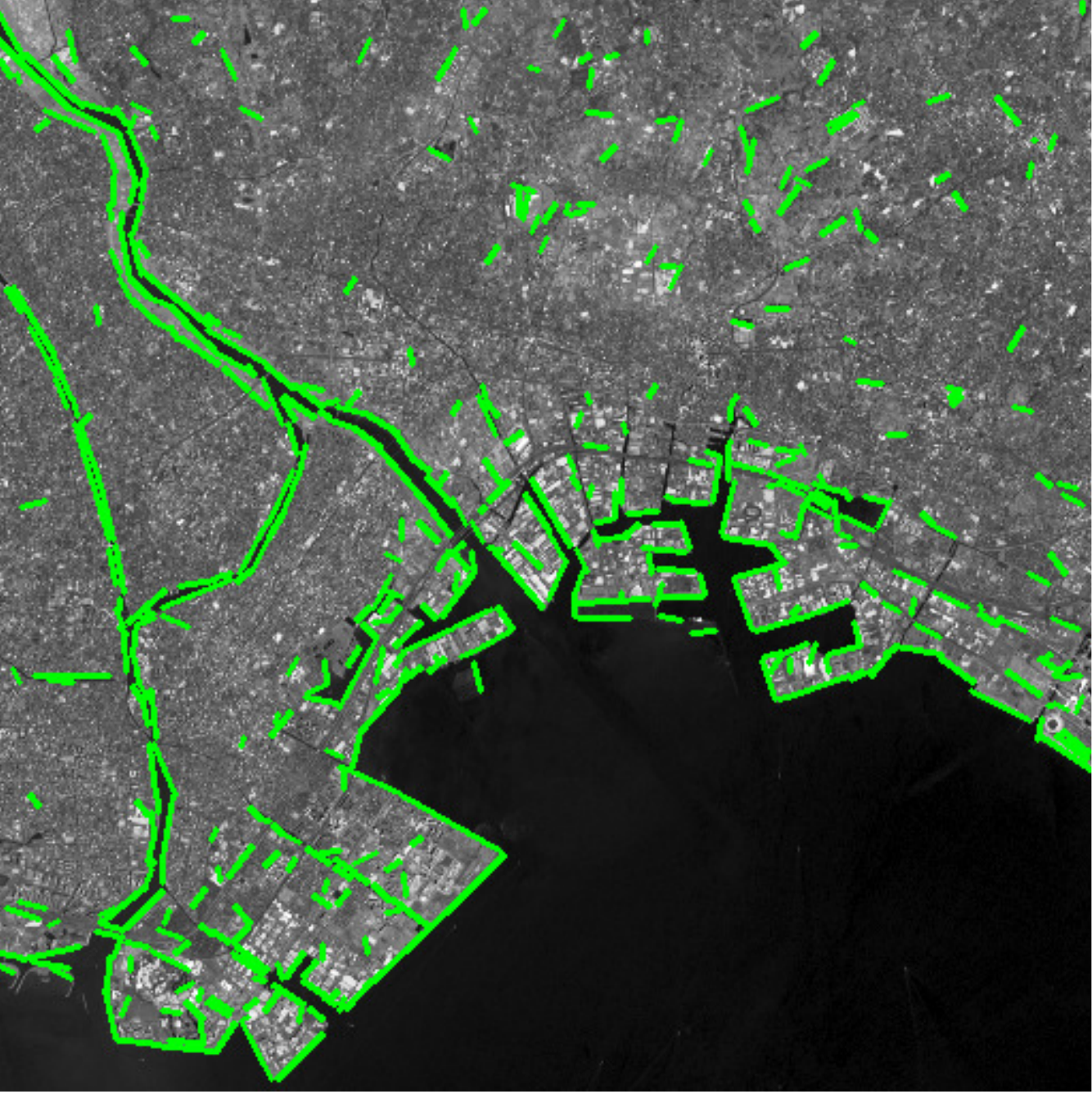}
    \end{minipage}}
  \subfigure[]{
    \label{fig:mini:subfig:b}
    \begin{minipage}[c]{0.182\textwidth}
      \centering
      \includegraphics[width=1.25in]{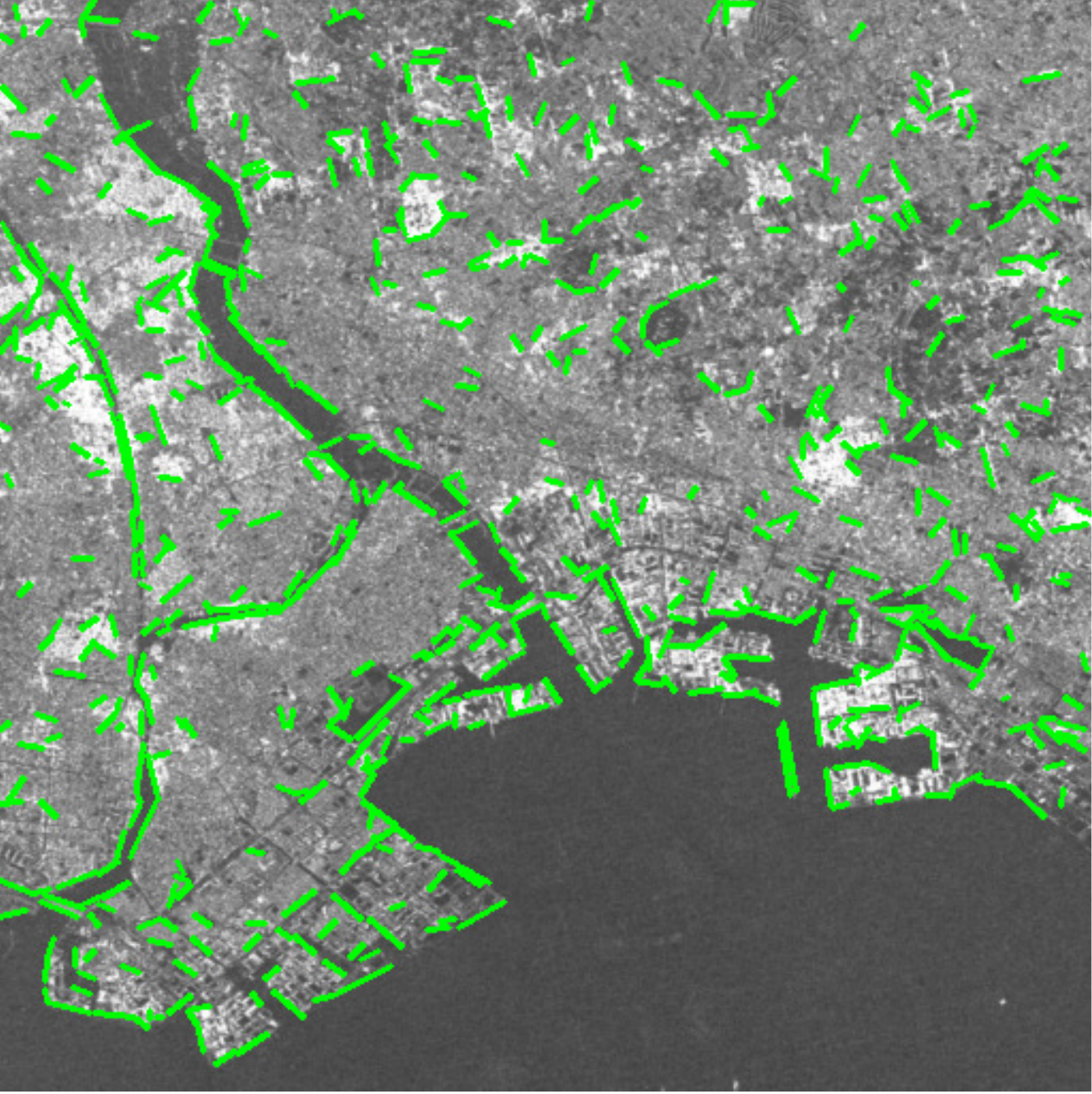}
    \end{minipage}}
  \subfigure[]{
    \label{fig:mini:subfig:a}
    \begin{minipage}[c]{0.182\textwidth}
      \centering
      \includegraphics[width=1.25in]{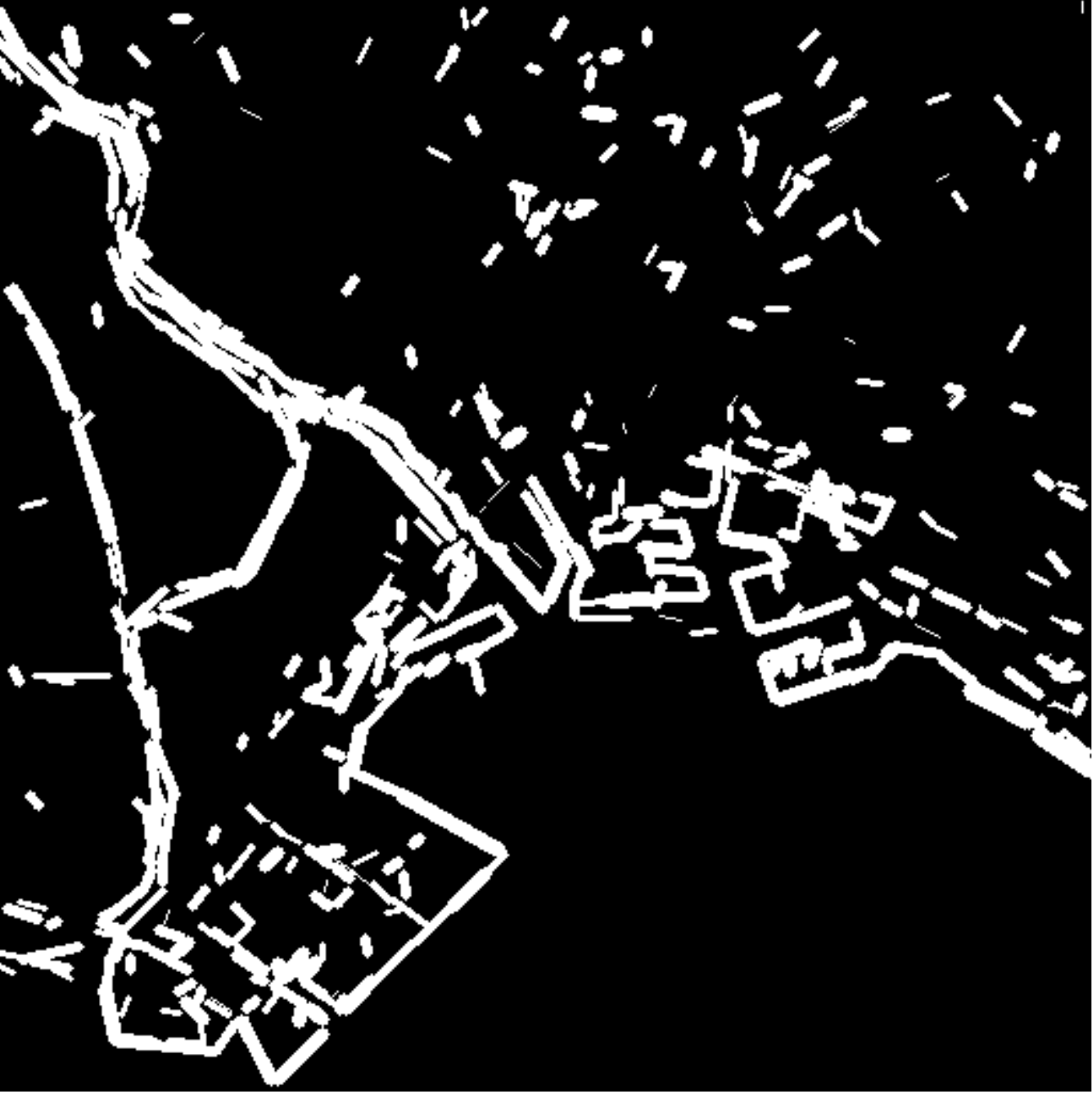}
    \end{minipage}}
  \subfigure[]{
    \label{fig:mini:subfig:b}
    \begin{minipage}[c]{0.182\textwidth}
      \centering
      \includegraphics[width=1.25in]{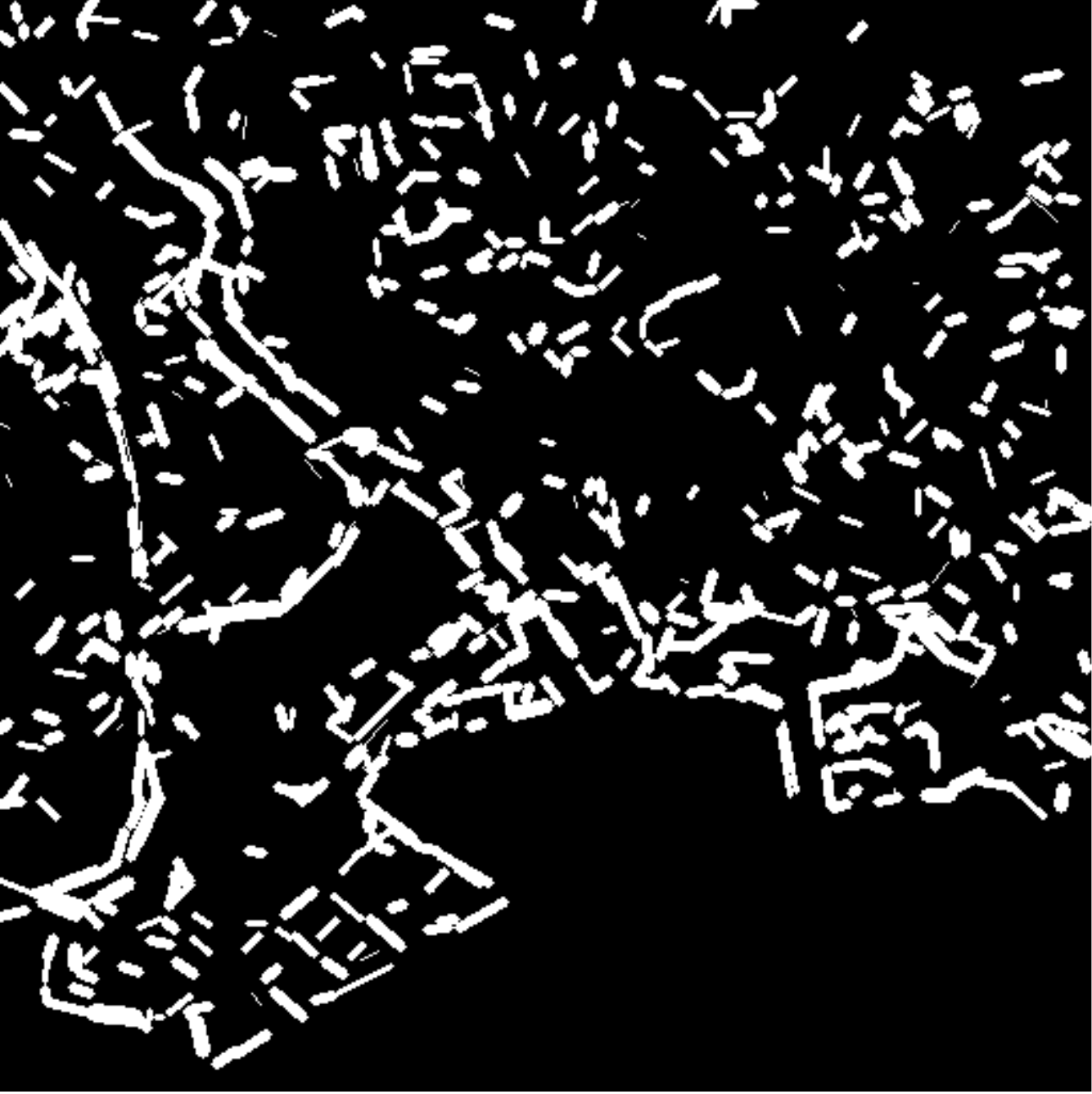}
    \end{minipage}}\\
  \subfigure[]{
    \label{fig:mini:subfig:a}
    \begin{minipage}[c]{0.4\textwidth}
      \centering
      \includegraphics[width=2.5in]{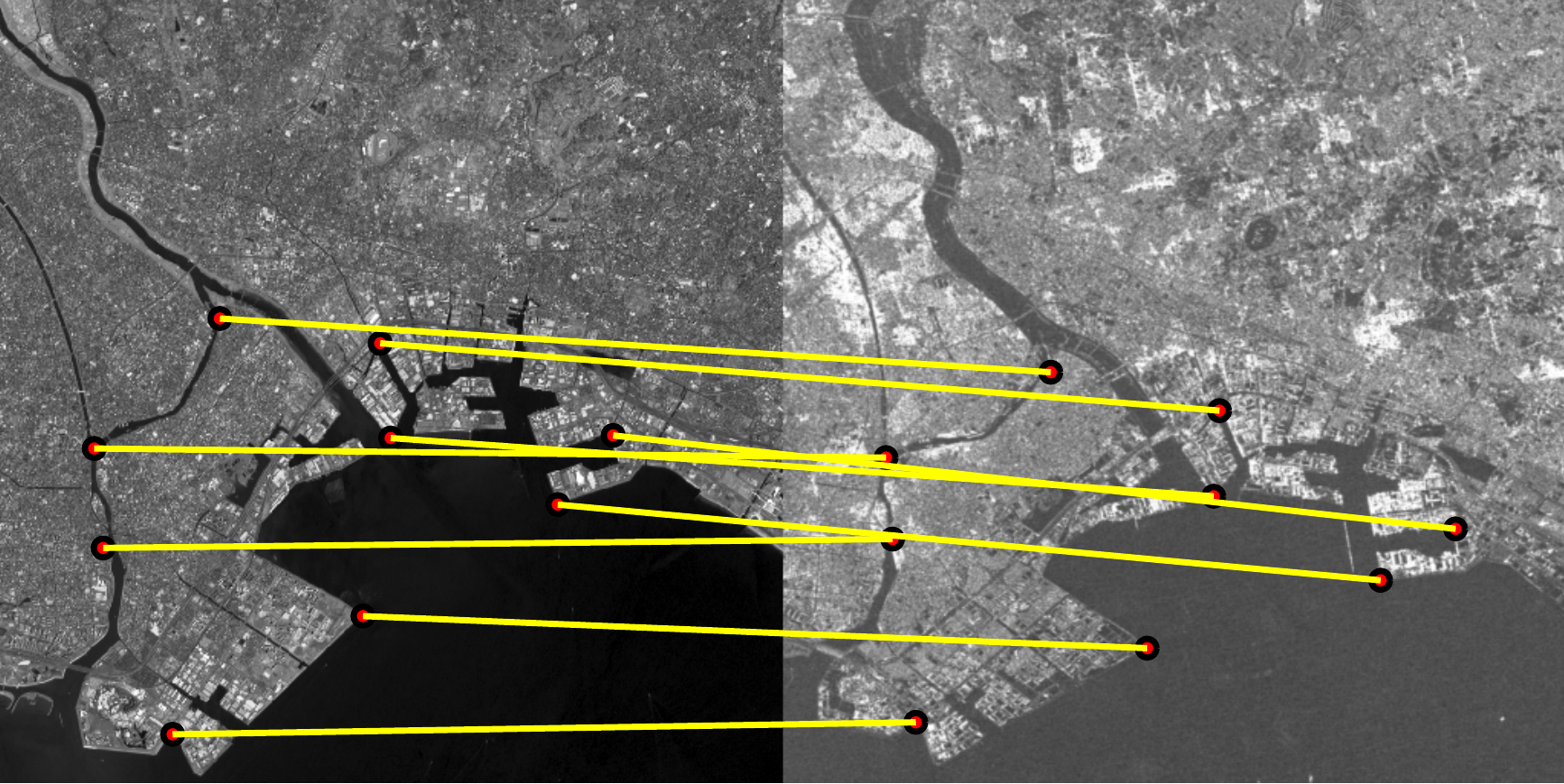}
    \end{minipage}}
  \subfigure[]{
    \label{fig:mini:subfig:b}
    \begin{minipage}[c]{0.4\textwidth}
      \centering
      \includegraphics[width=1.22in]{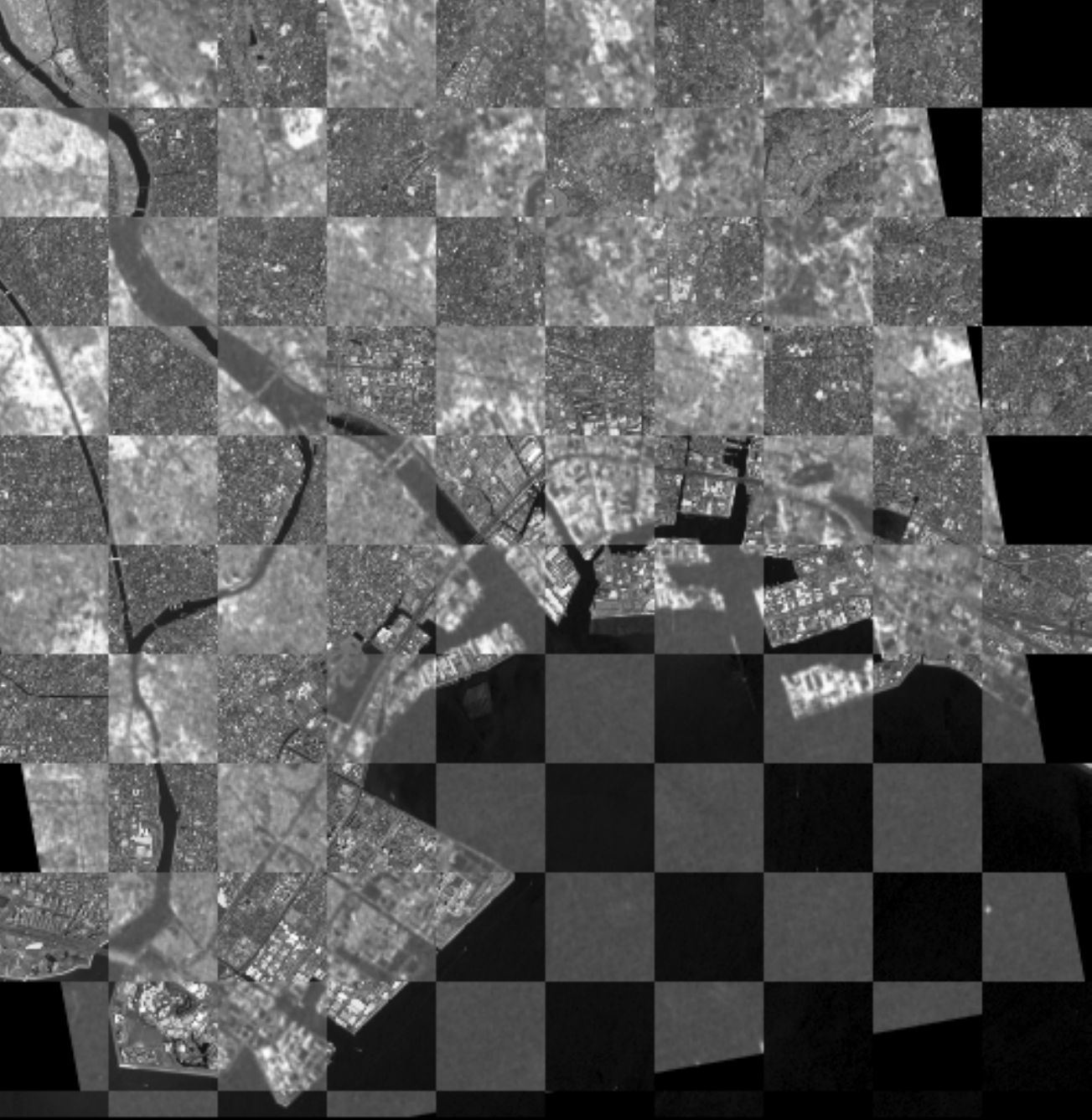}
    \end{minipage}}\\
  \captionstyle{normal}
  \caption{Image registration for ImgSet2-4. (a) Line segments of reference image. (b) Line segments of sensed image. (c) Line-support regions of reference image. (d) Line-support regions of sensed image. (e) Point correspondences by GOR. (f) Checkerboard mosaiced image.}
  \label{fig-9}
\end{figure*}

The image pair of ImgSet2-4 consists of two images with the same size of 512$\times$512 pixels taken by the sensor of ASTER L1B band 1 and PALSAR fine mode, covering the bay of Tokyo, Japan. The substantial disparity according to the visual appearance can be observed between the optical and SAR images.
The SAR image from PALSAR shown in Fig. \ref{fig-9} (b) is inevitably contaminated by the speckle noise and scatter signals from the earth surface, which brings more challenges to multispectral image registration. As illustrated in Fig. \ref{fig-9} (c) and (d), the line-support regions from both of the images perserve similar shapes. It can be indicated from the registered image that image features such as the coasts of the bay are precisely overlapped.

\begin{figure*}[htb]
\centering
 \setlength{\abovecaptionskip}{0pt}
 \setlength{\belowcaptionskip}{0pt}
 \setlength{\intextsep}{8pt plus 3pt minus 2pt}
  \subfigure[]{
    \label{fig:mini:subfig:a}
    \begin{minipage}[c]{0.182\textwidth}
      \centering
      \includegraphics[width=1.25in]{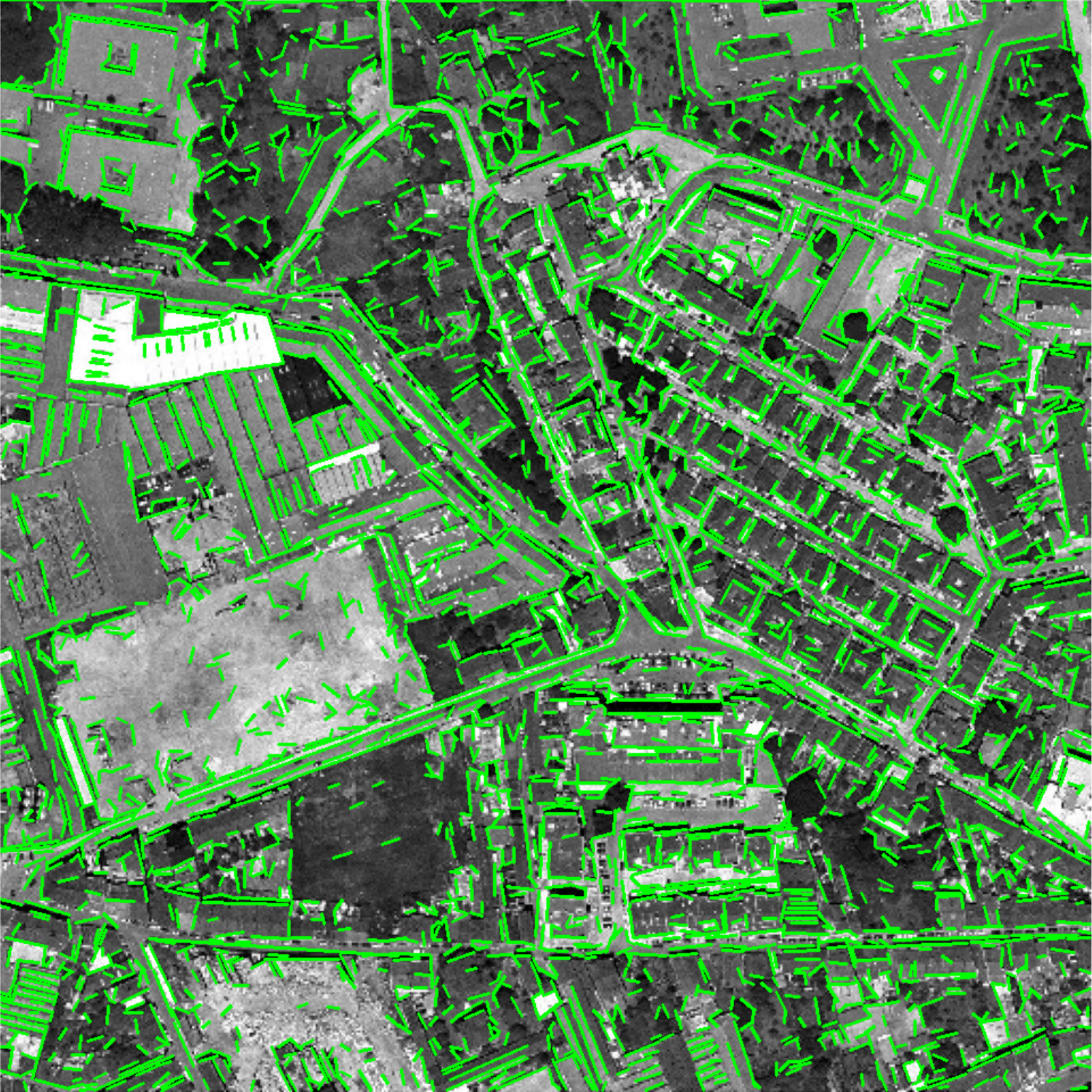}
    \end{minipage}}
  \subfigure[]{
    \label{fig:mini:subfig:b}
    \begin{minipage}[c]{0.182\textwidth}
      \centering
      \includegraphics[width=1.25in]{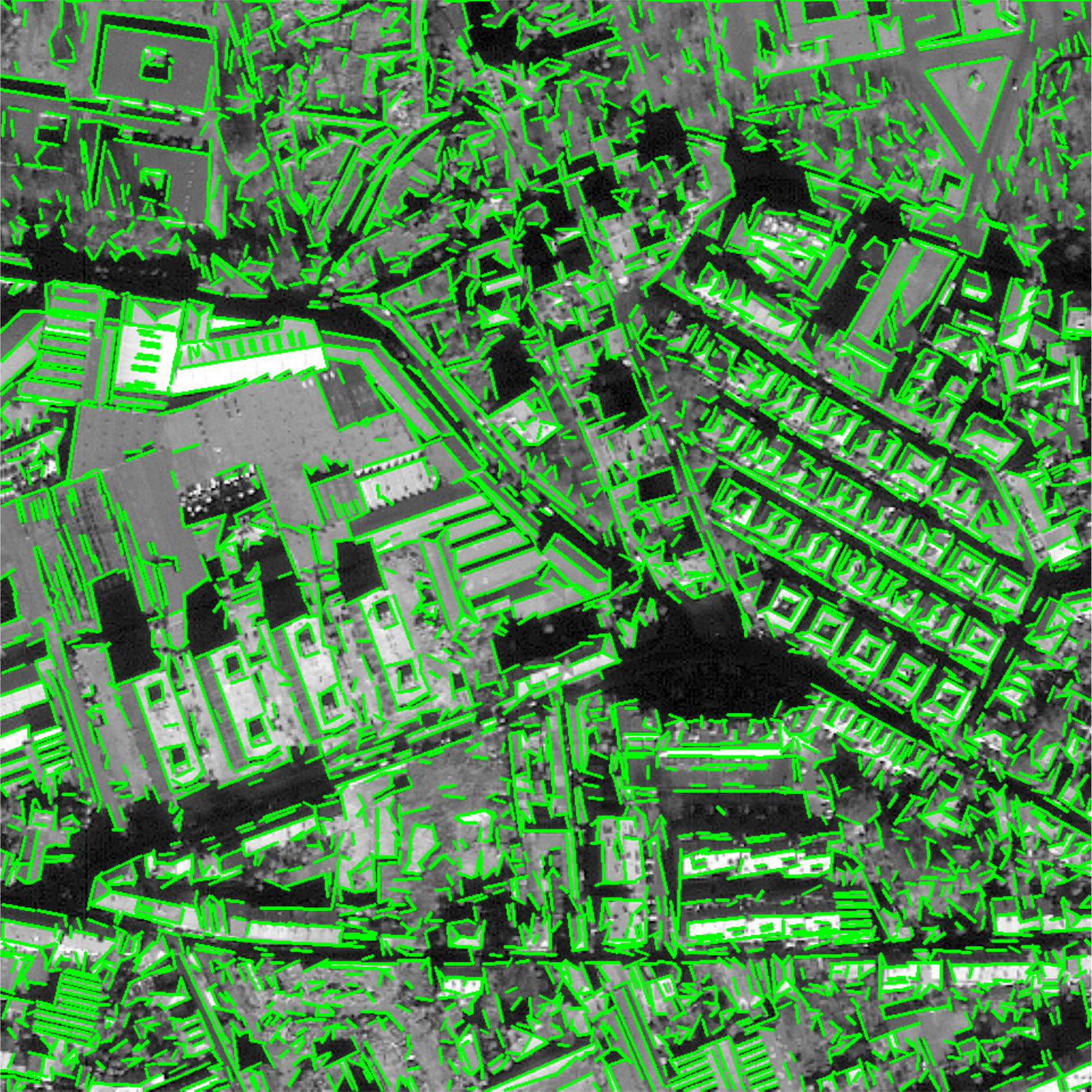}
    \end{minipage}}
  \subfigure[]{
    \label{fig:mini:subfig:a}
    \begin{minipage}[c]{0.182\textwidth}
      \centering
      \includegraphics[width=1.25in]{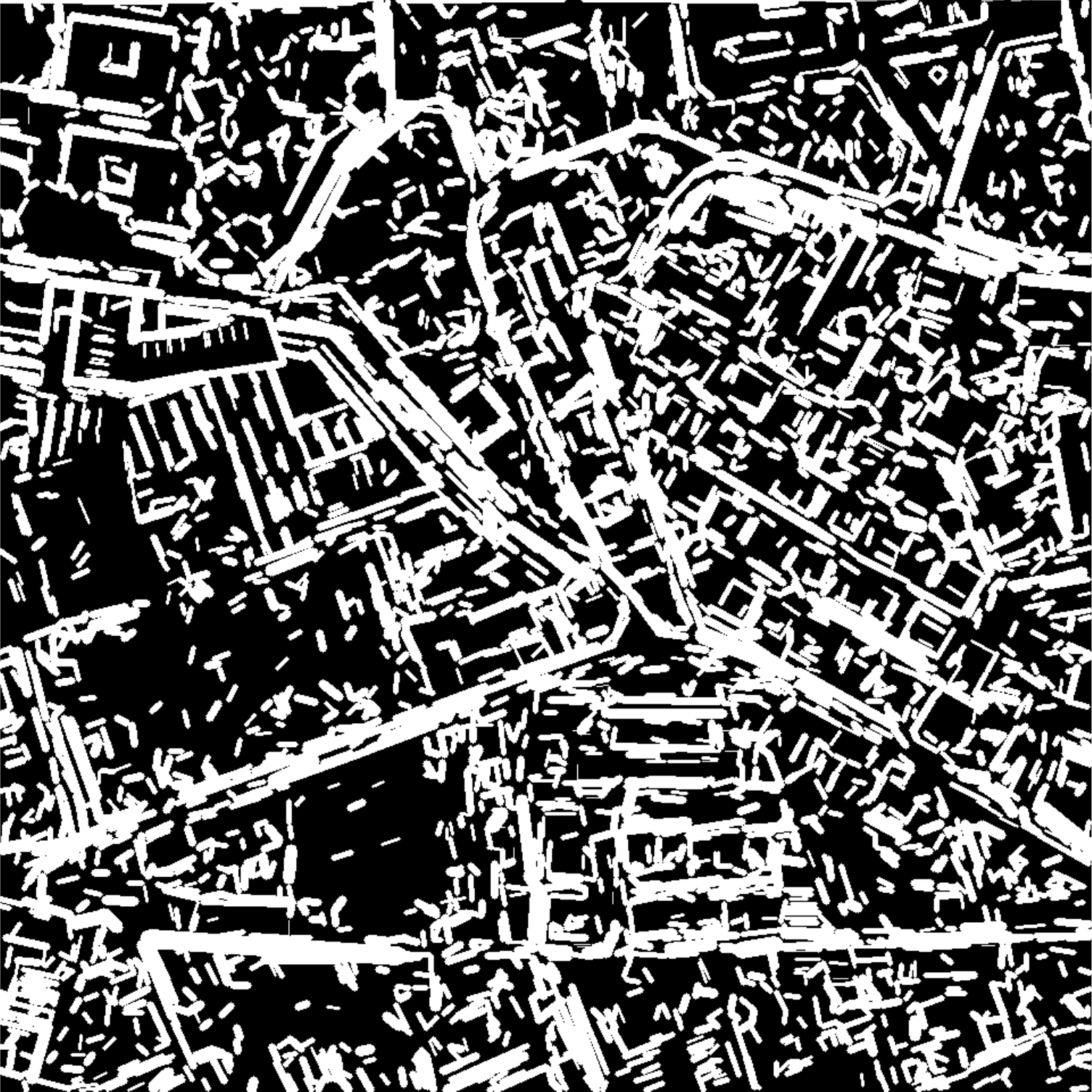}
    \end{minipage}}
  \subfigure[]{
    \label{fig:mini:subfig:b}
    \begin{minipage}[c]{0.182\textwidth}
      \centering
      \includegraphics[width=1.25in]{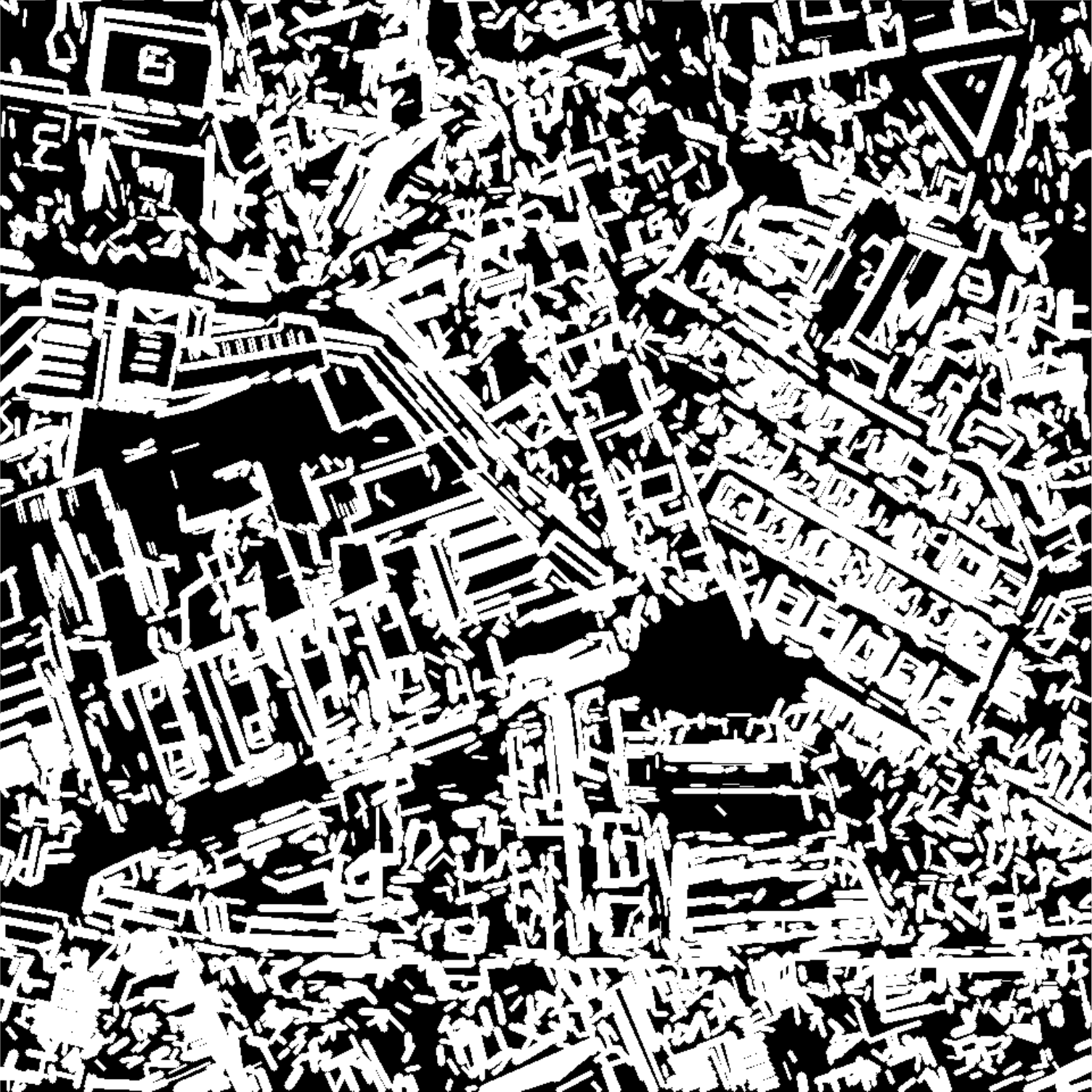}
    \end{minipage}}\\
  \subfigure[]{
    \label{fig:mini:subfig:a}
    \begin{minipage}[c]{0.4\textwidth}
      \centering
      \includegraphics[width=2.5in]{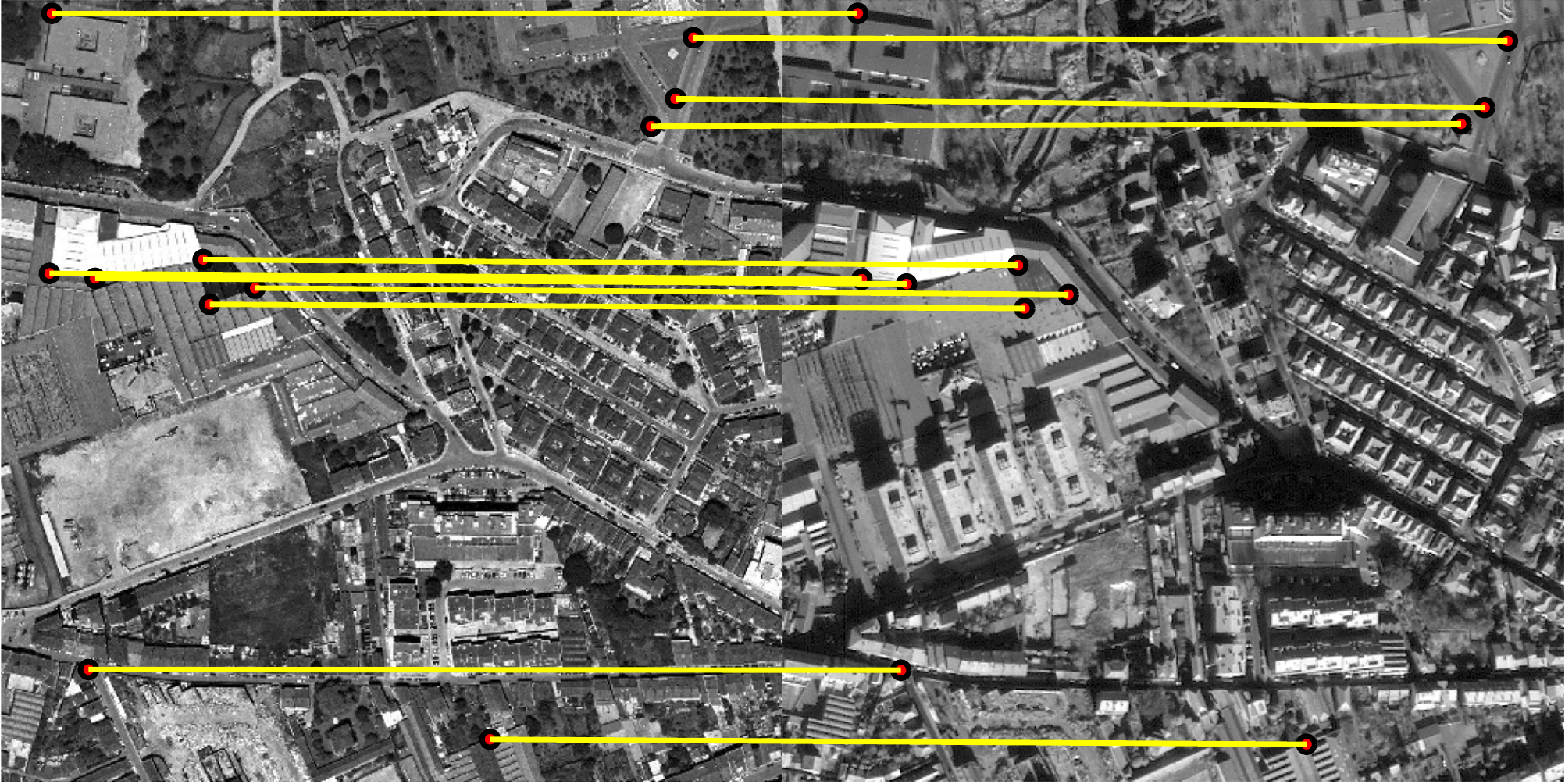}
    \end{minipage}}
  \subfigure[]{
    \label{fig:mini:subfig:b}
    \begin{minipage}[c]{0.4\textwidth}
      \centering
      \includegraphics[width=1.24in]{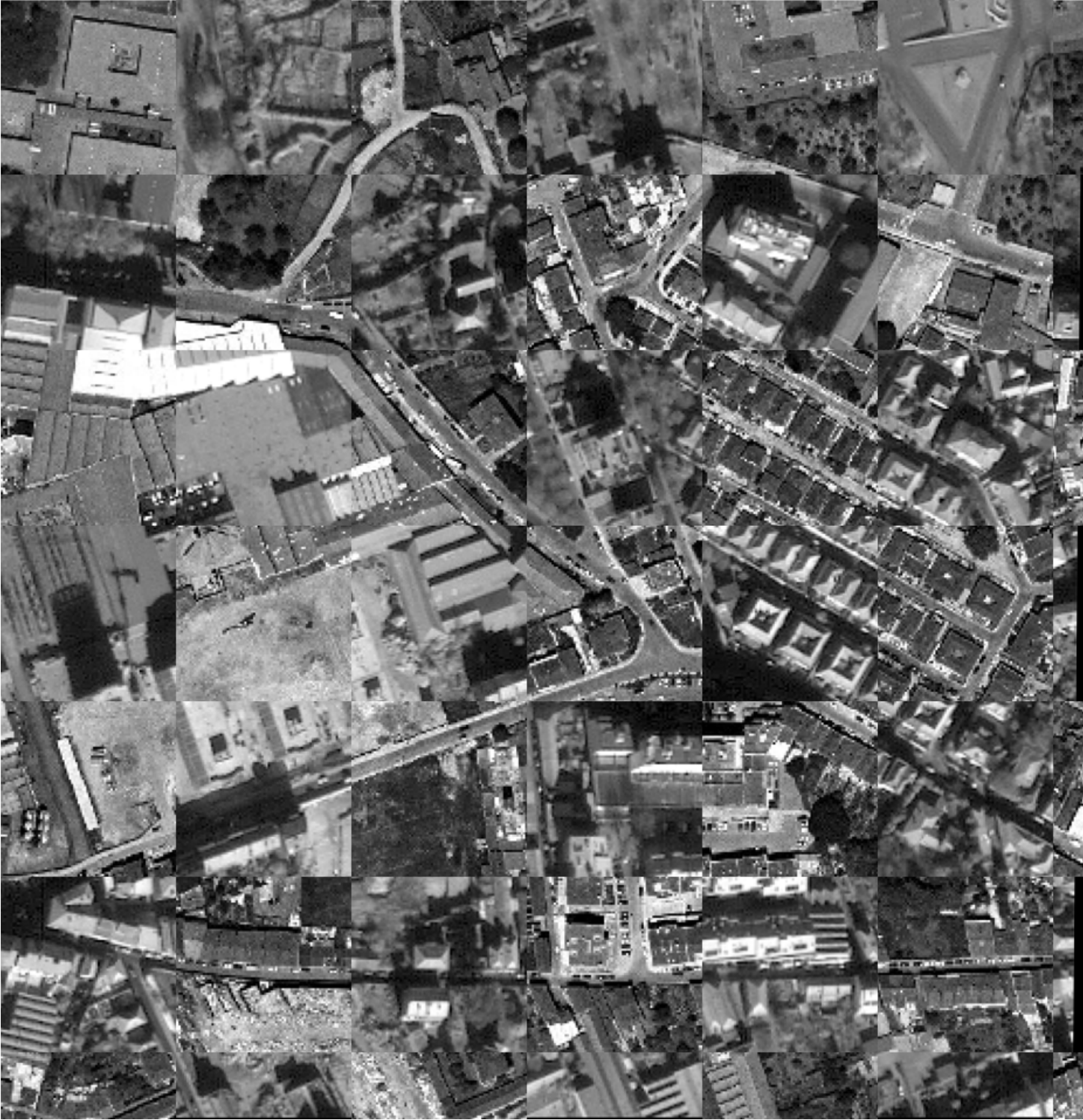}
    \end{minipage}}\\
  \captionstyle{normal}
  \caption{Image registration for ImgSet2-5. (a) Line segments of reference image. (b) Line segments of sensed image. (c) Line-support regions of reference image. (d) Line-support regions of sensed image. (e) Point correspondences by GOR. (f) Checkerboard mosaiced image.}
  \label{fig-10}
\end{figure*}

The image pair of ImgSet2-5 consists of an orthophotograph image with the green band and a panchromatic image covering part of the city of Porto with the same size of 512$\times$512 pixels.
Significant changes exist between these images because of spectral and one year temporal difference. For example, several new buildings appears in the sensed image, which cannot be found in the reference image.
These temporal differences make the situations more complicated.
As shown in Fig. \ref{fig-10} (c) and (d), most of structures of the urban roads are preserved in the segmented line-support regions.
10 corresponding inliers are matched by GOR, and the checkerboard mosaiced image are shown in Fig. \ref{fig-10} (f).


\vspace{-0.5cm}

\subsection{Registration for Images with Inconsistent Annotation}
ImgSet3-1 comprises two navigation maps generated by google map for Android mobile system with the same location but different orientations. The map rotates globally in clockwise, except for the bus icons and texts of street names.
The icons and texts stay nearly horizontal for a better visualization.

ImgSet3-2 consists of two google map images with computer-generated graphics by searching for the place of ``Oriental Pearl TV Tower of Shanghai". The sensed map image is obtained by zooming in the reference map for a large scale. Noted that most of the interest icons and Chinese characters remain in the same size in the scaled map. Moreover, several new icons and texts of street names appear in the scaled map.

ImgSet3-3 consists of the maps overlaid with real satellite images generated by google map. The maps covers the part of the city of Erlangen, Germany. The reference map is the top view of google map with the spatial resolution of 1000 m. The sensed map is obtained by rotating the reference map $90^ \circ$ in clockwise, and scaling to the spatial resolution of 500 m.
The icons and texts are not exactly transformed along with the map.

\begin{figure*}[htb]
\centering
 \setlength{\abovecaptionskip}{0pt}
 \setlength{\belowcaptionskip}{0pt}
 \setlength{\intextsep}{8pt plus 3pt minus 2pt}
  \subfigure[]{
    \label{fig:mini:subfig:a}
    \begin{minipage}[c]{0.182\textwidth}
      \centering
      \includegraphics[width=1.25in]{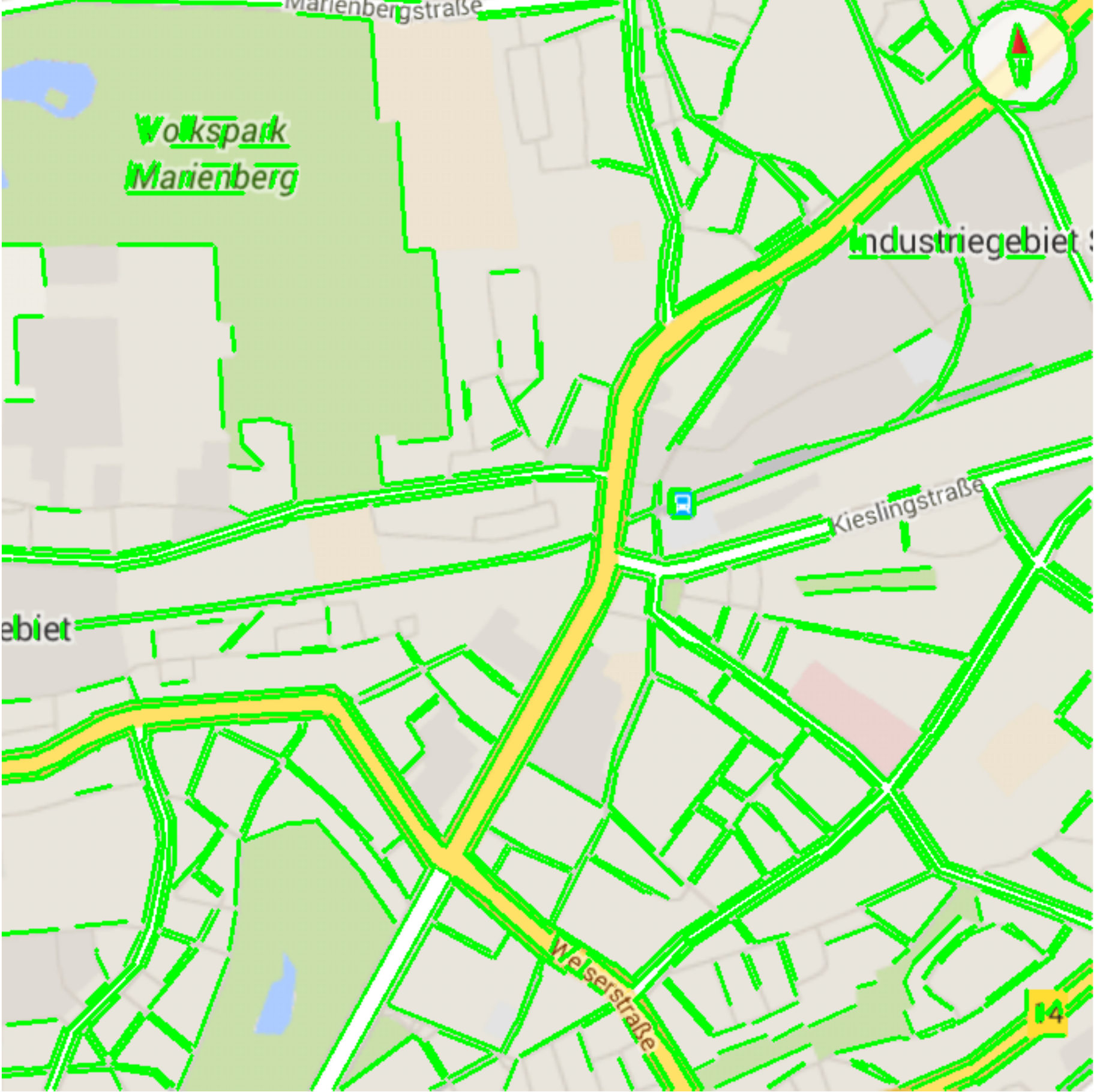}
    \end{minipage}}
  \subfigure[]{
    \label{fig:mini:subfig:b}
    \begin{minipage}[c]{0.182\textwidth}
      \centering
      \includegraphics[width=1.25in]{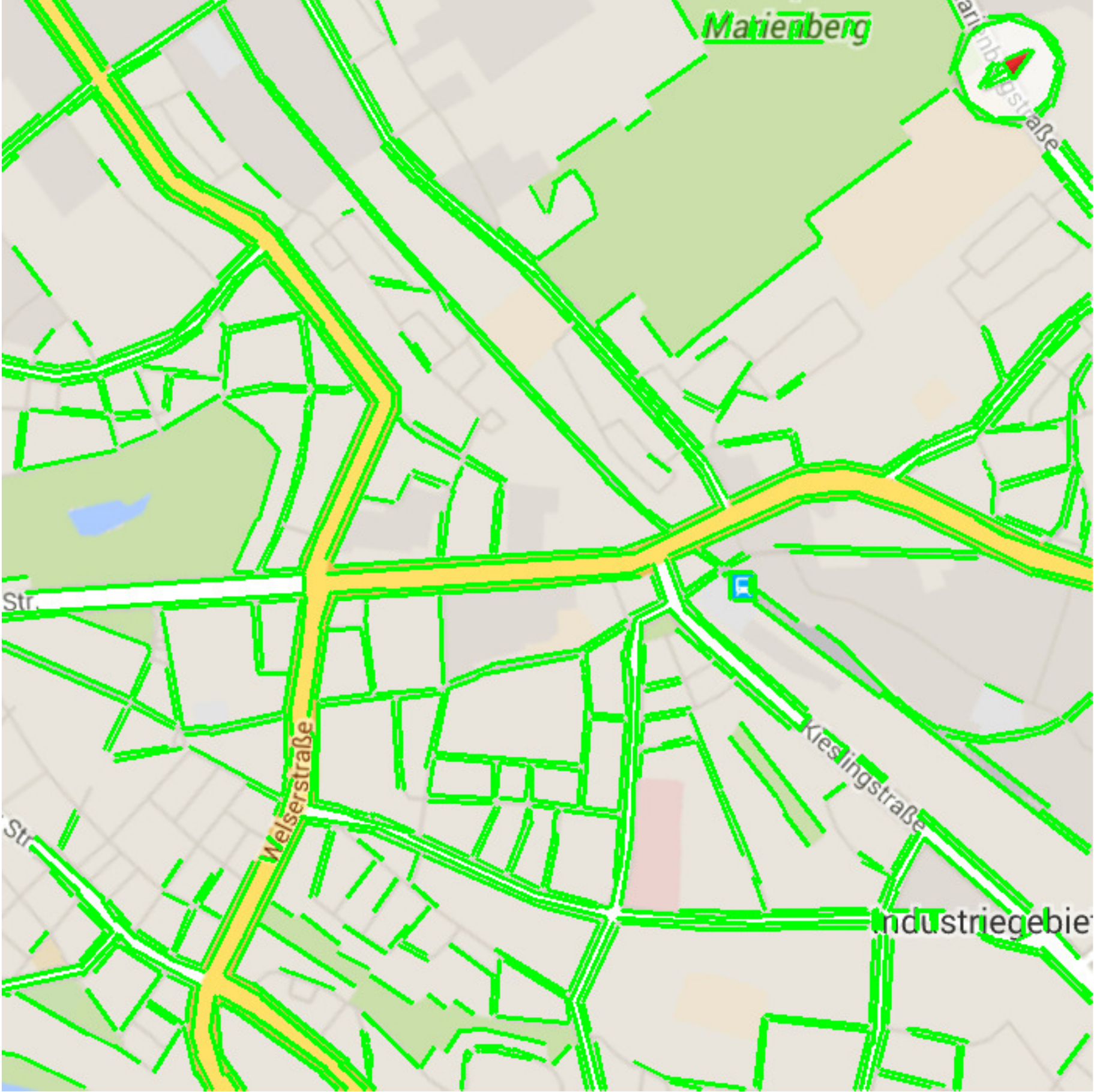}
    \end{minipage}}
  \subfigure[]{
    \label{fig:mini:subfig:a}
    \begin{minipage}[c]{0.182\textwidth}
      \centering
      \includegraphics[width=1.25in]{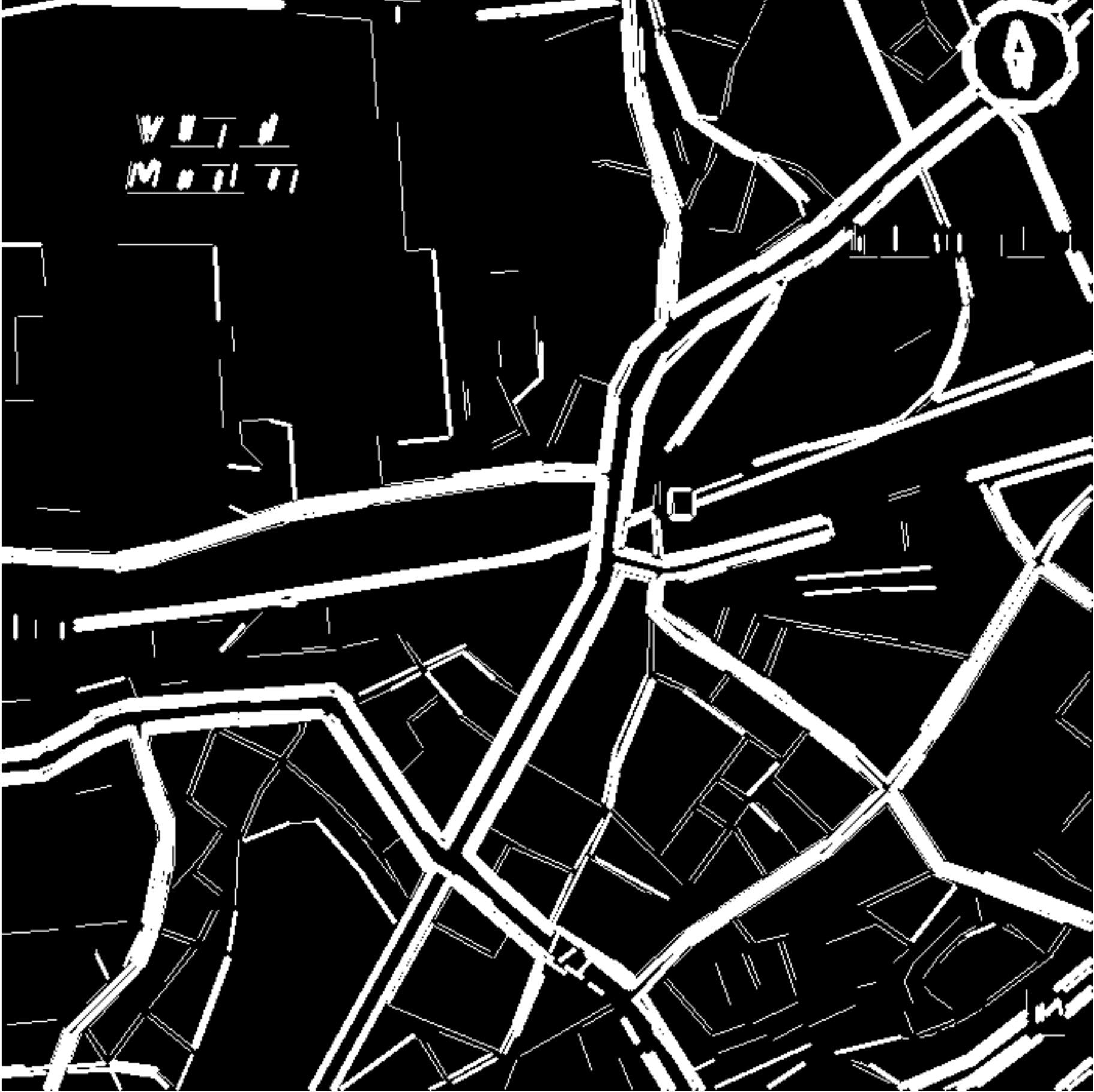}
    \end{minipage}}
  \subfigure[]{
    \label{fig:mini:subfig:b}
    \begin{minipage}[c]{0.182\textwidth}
      \centering
      \includegraphics[width=1.25in]{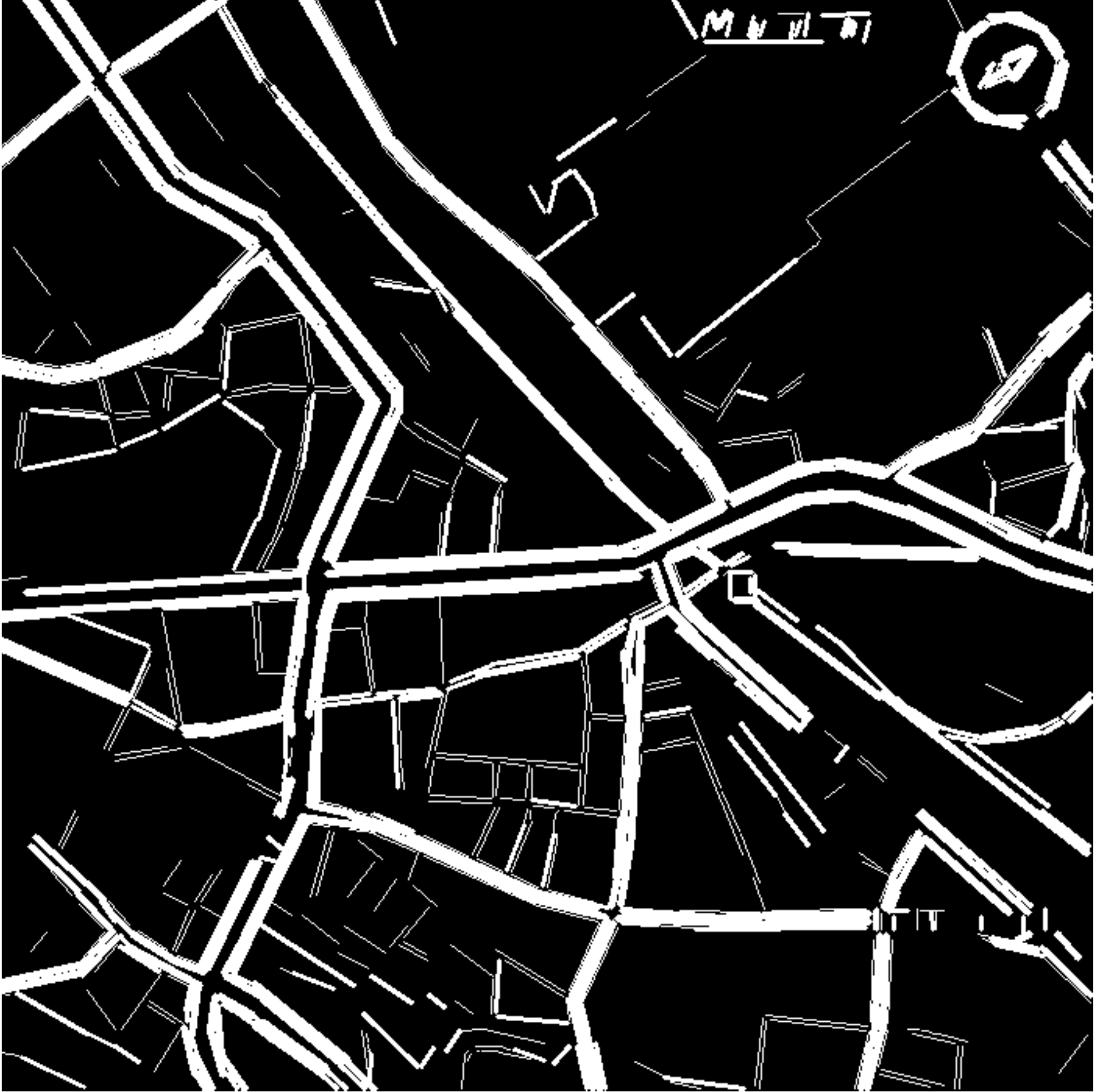}
    \end{minipage}}\\
  \subfigure[]{
    \label{fig:mini:subfig:a}
    \begin{minipage}[c]{0.4\textwidth}
      \centering
      \includegraphics[width=2.5in]{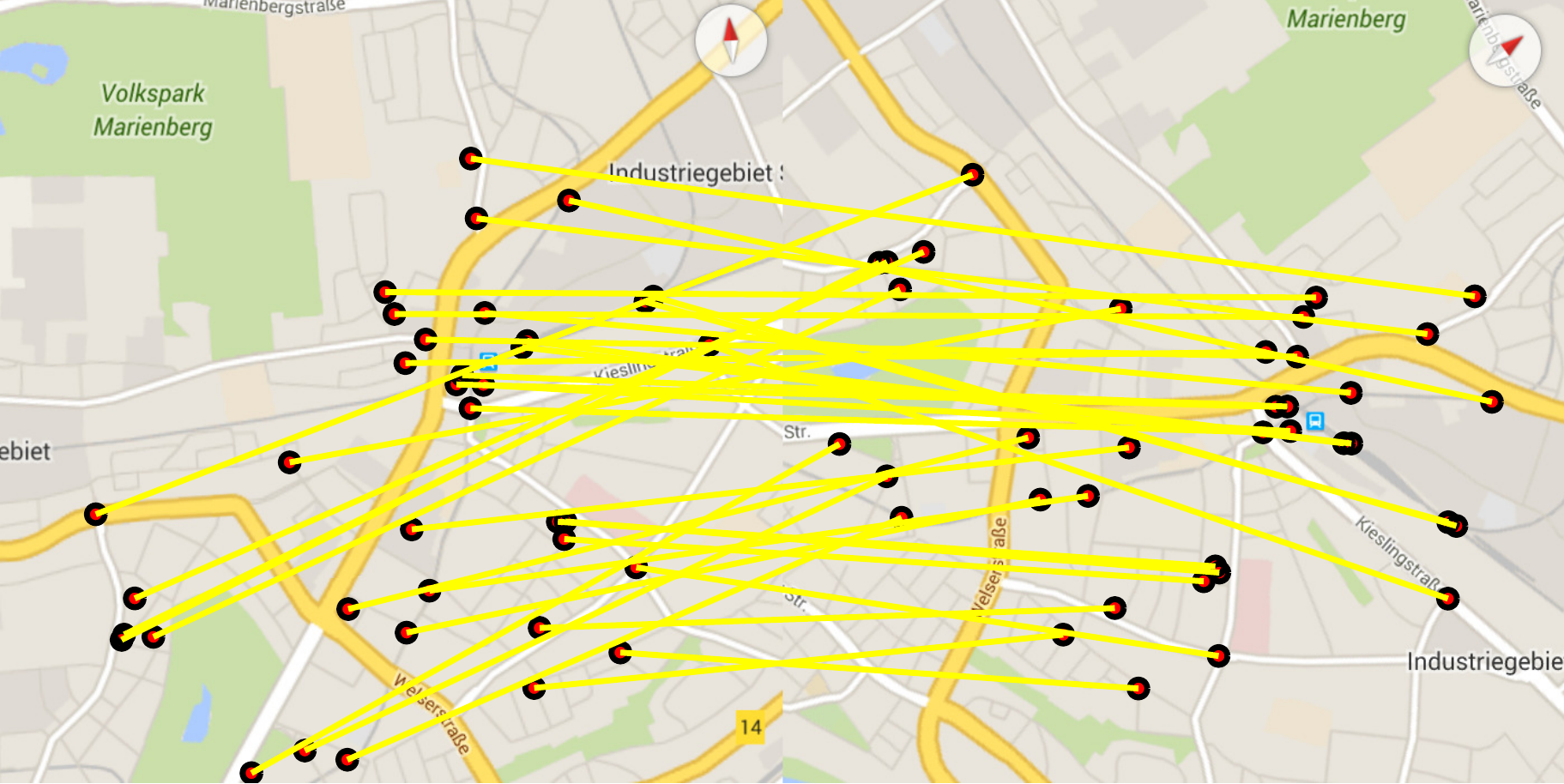}
    \end{minipage}}
  \subfigure[]{
    \label{fig:mini:subfig:b}
    \begin{minipage}[c]{0.4\textwidth}
      \centering
      \includegraphics[width=1.24in]{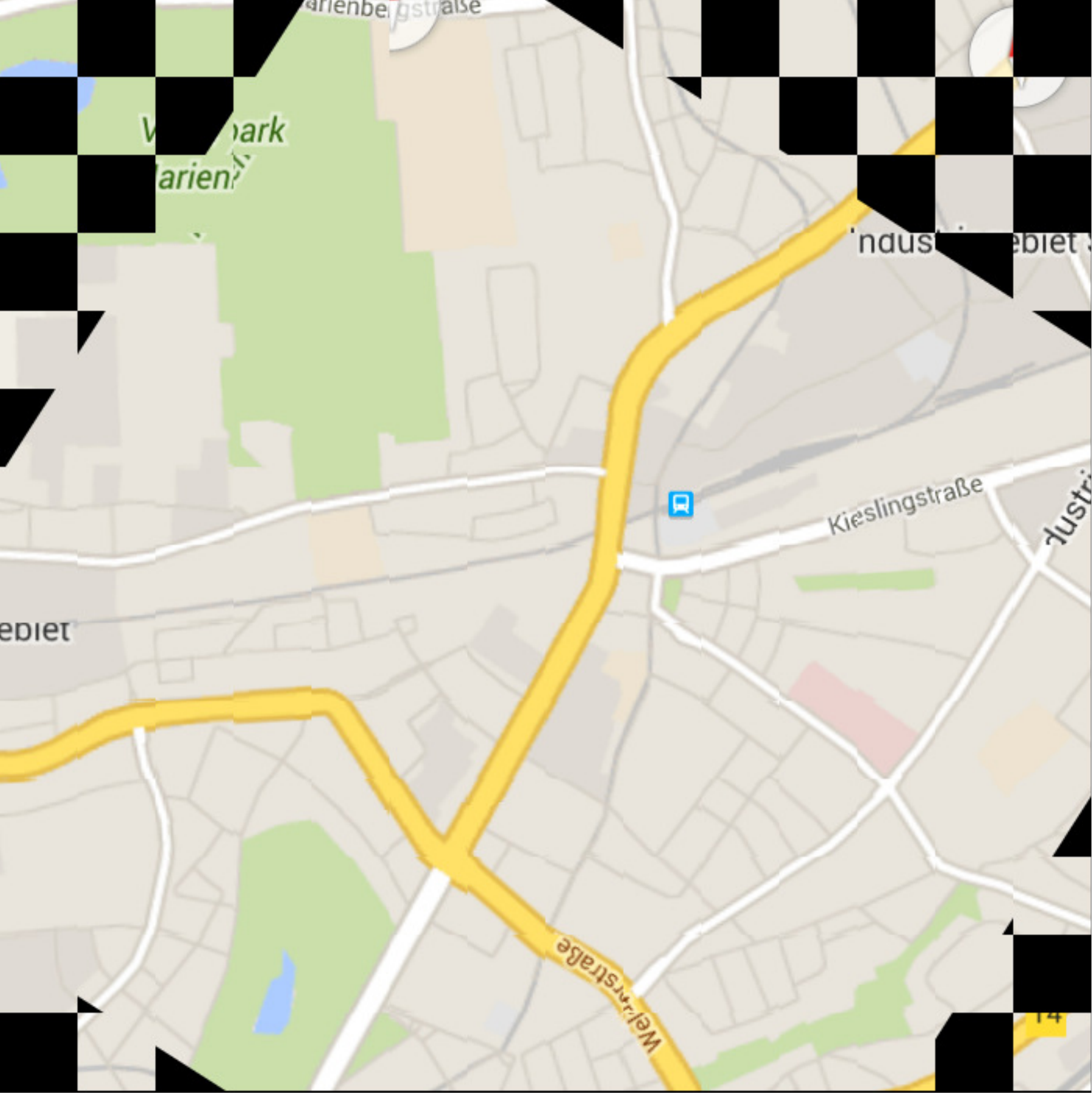}
    \end{minipage}}\\
  \captionstyle{normal}
  \caption{Image registration for ImgSet3-1. (a) Line segments of reference image. (b) Line segments of sensed image. (c) Line-support regions of reference image. (d) Line-support regions of sensed image. (e) Point correspondences by GOR. (f) Checkerboard mosaiced image.}
  \label{fig-11}
\end{figure*}

\begin{figure*}[htb]
\centering
 \setlength{\abovecaptionskip}{0pt}
 \setlength{\belowcaptionskip}{0pt}
 \setlength{\intextsep}{8pt plus 3pt minus 2pt}
  \subfigure[]{
    \label{fig:mini:subfig:a}
    \begin{minipage}[c]{0.182\textwidth}
      \centering
      \includegraphics[width=1.25in]{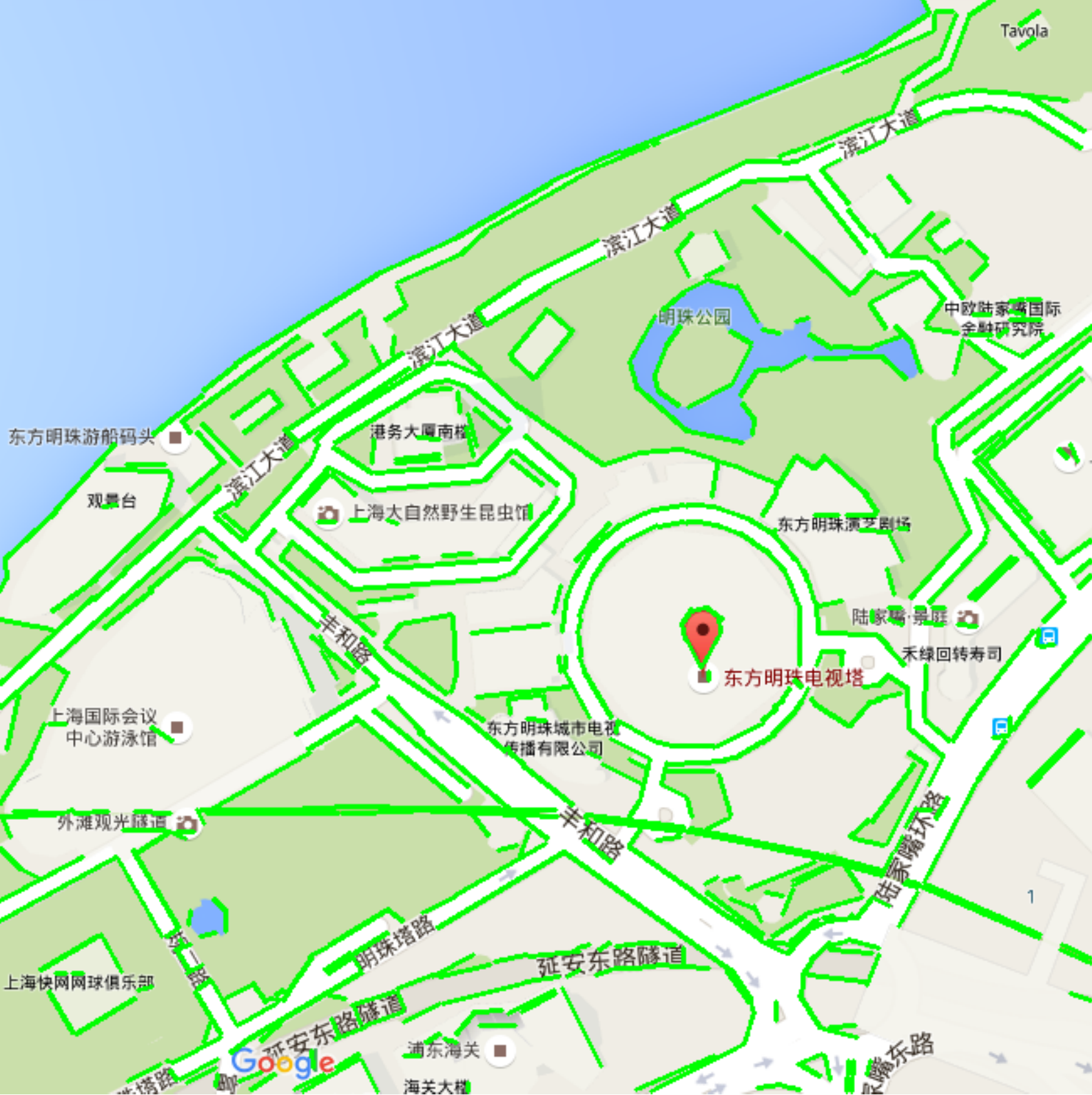}
    \end{minipage}}
  \subfigure[]{
    \label{fig:mini:subfig:b}
    \begin{minipage}[c]{0.182\textwidth}
      \centering
      \includegraphics[width=1.25in]{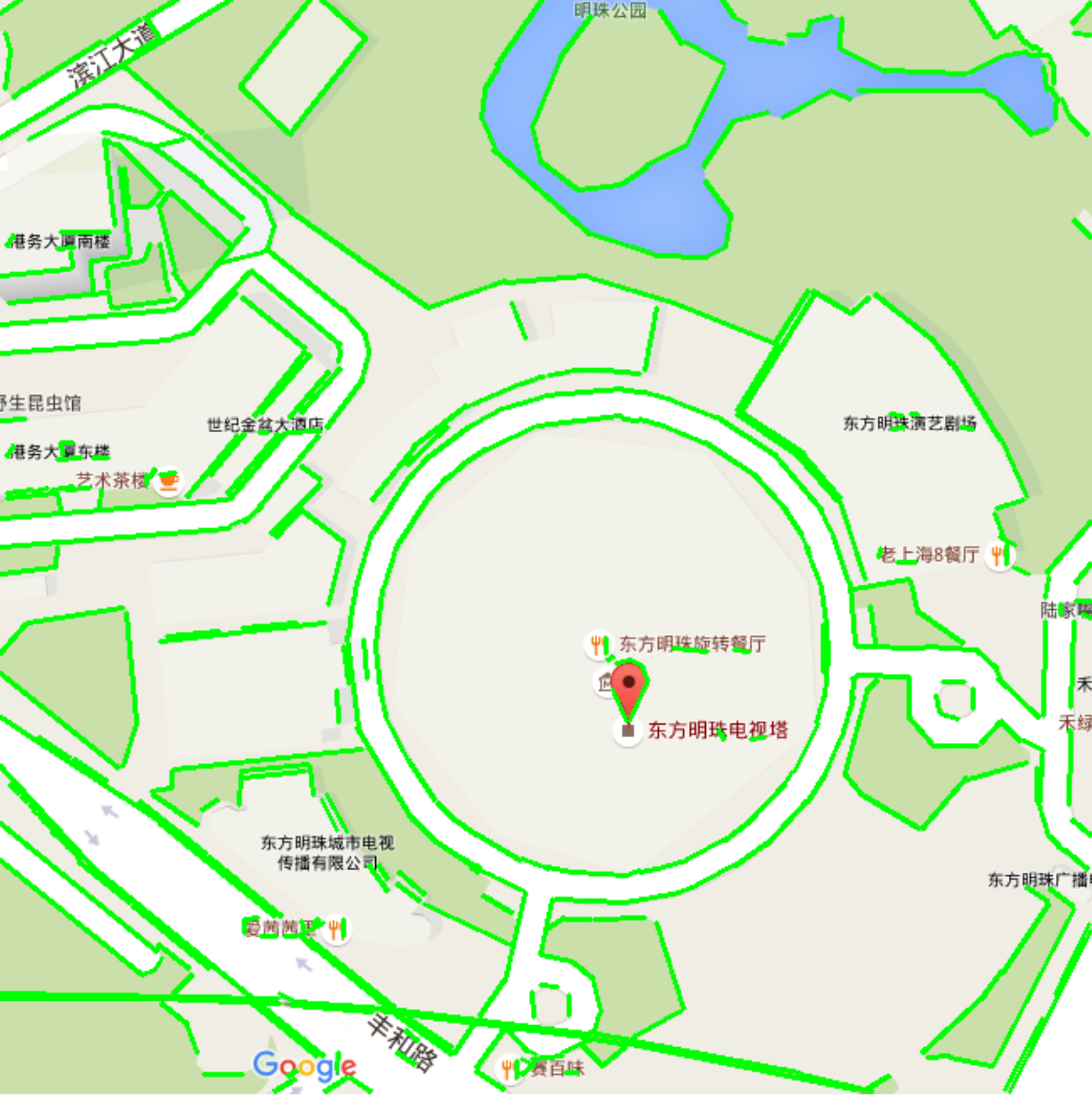}
    \end{minipage}}
  \subfigure[]{
    \label{fig:mini:subfig:a}
    \begin{minipage}[c]{0.182\textwidth}
      \centering
      \includegraphics[width=1.25in]{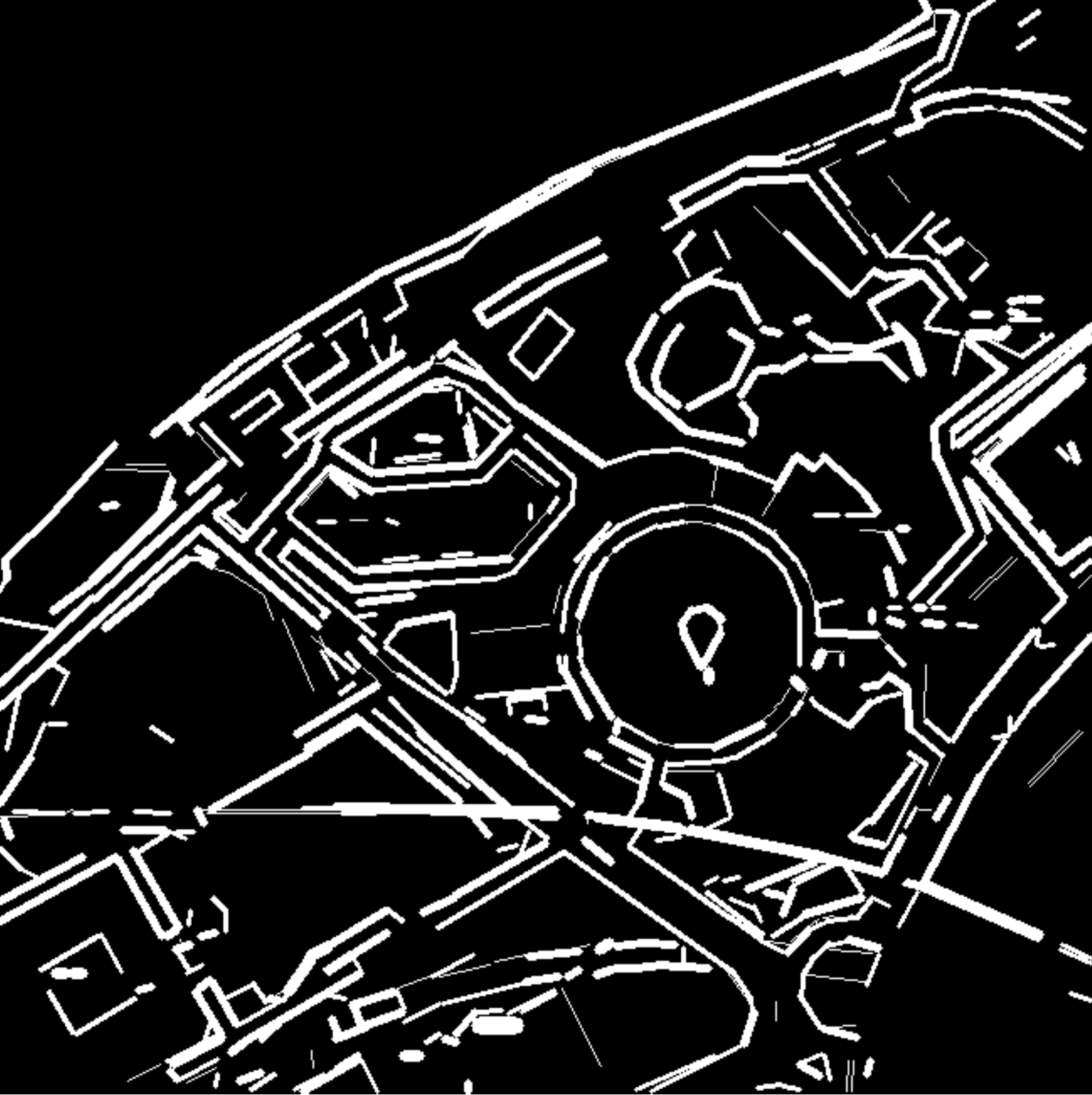}
    \end{minipage}}
  \subfigure[]{
    \label{fig:mini:subfig:b}
    \begin{minipage}[c]{0.182\textwidth}
      \centering
      \includegraphics[width=1.25in]{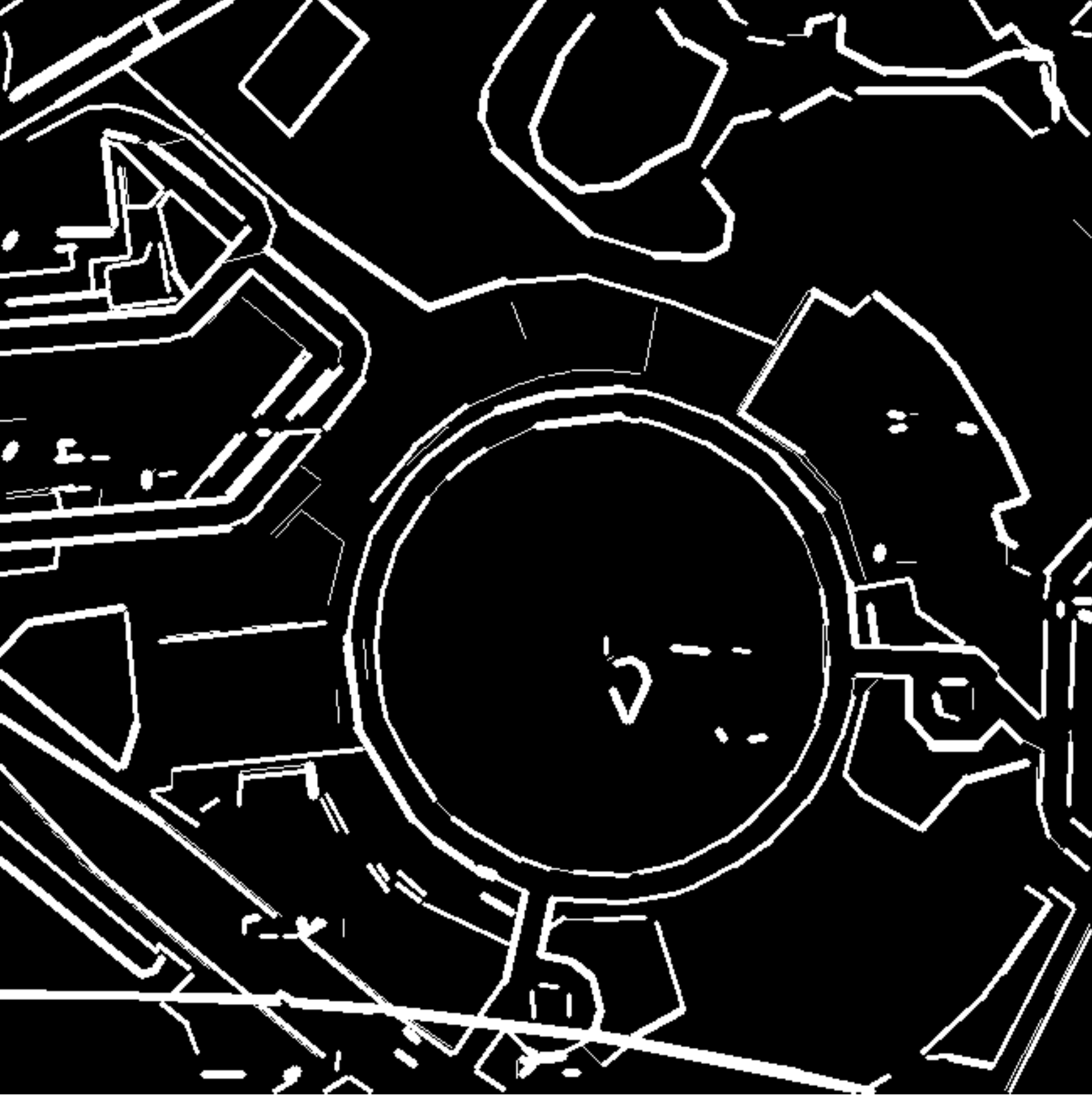}
    \end{minipage}}\\
  \subfigure[]{
    \label{fig:mini:subfig:a}
    \begin{minipage}[c]{0.4\textwidth}
      \centering
      \includegraphics[width=2.5in]{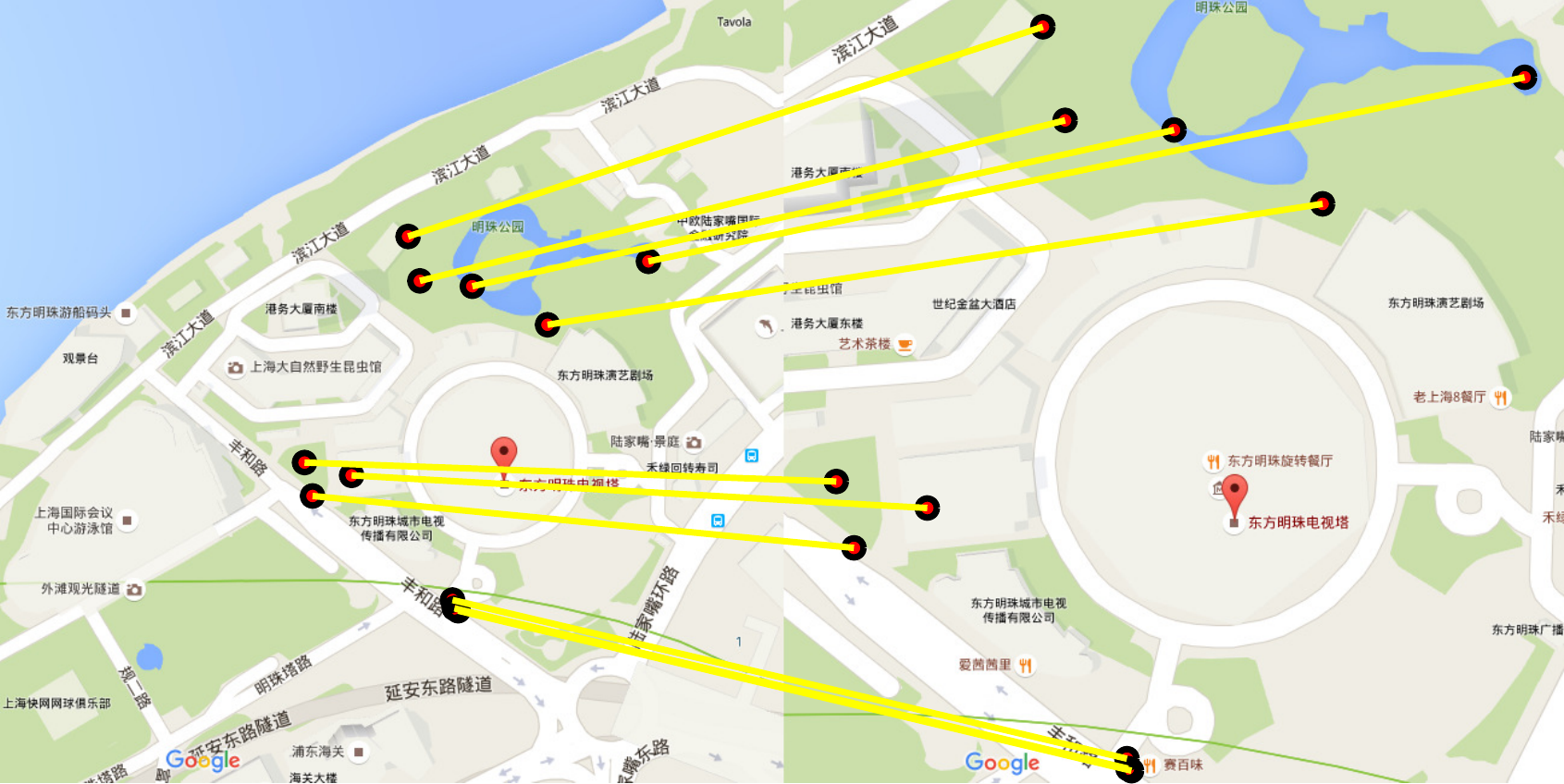}
    \end{minipage}}
  \subfigure[]{
    \label{fig:mini:subfig:b}
    \begin{minipage}[c]{0.4\textwidth}
      \centering
      \includegraphics[width=1.34in]{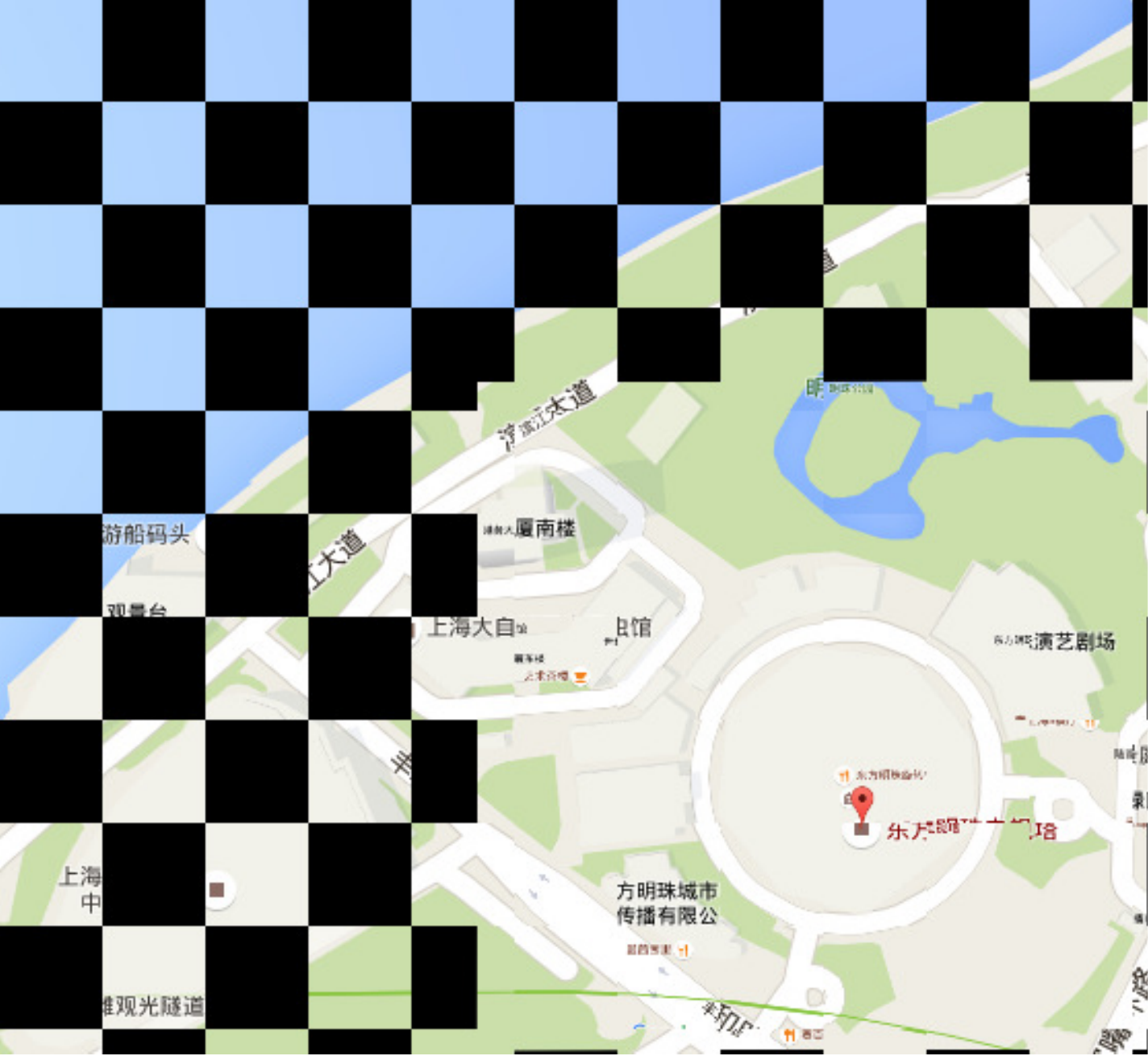}
    \end{minipage}}\\
  \captionstyle{normal}
  \caption{Image registration for ImgSet3-2. (a) Line segments of reference image. (b) Line segments of sensed image. (c) Line-support regions of reference image. (d) Line-support regions of sensed image. (e) Point correspondences by GOR. (f) Checkerboard mosaiced image.}
  \label{fig-12}
\end{figure*}

\begin{figure*}[htb]
\centering
 \setlength{\abovecaptionskip}{0pt}
 \setlength{\belowcaptionskip}{0pt}
 \setlength{\intextsep}{8pt plus 3pt minus 2pt}
  \subfigure[]{
    \label{fig:mini:subfig:a}
    \begin{minipage}[c]{0.182\textwidth}
      \centering
      \includegraphics[width=1.25in]{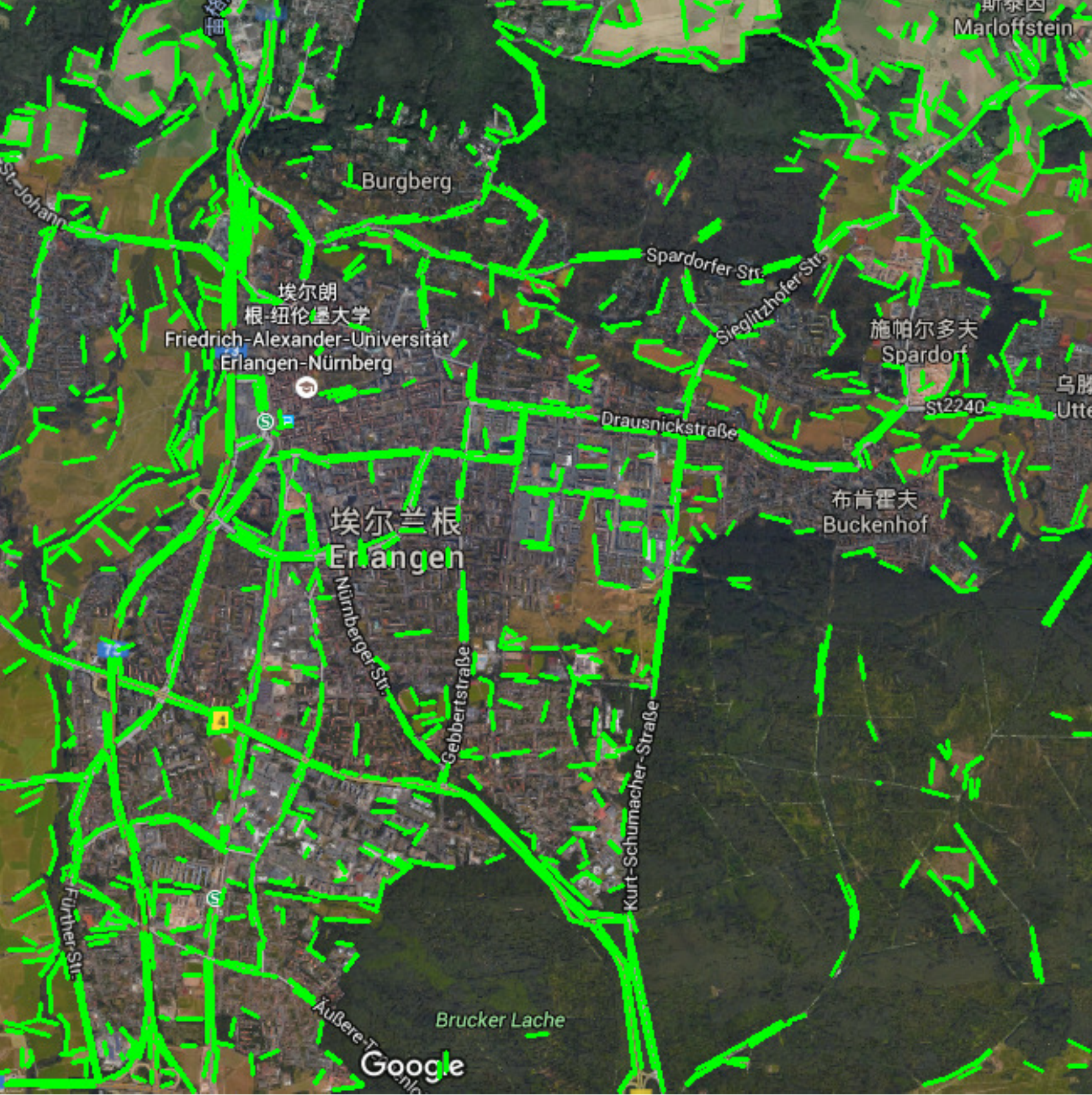}
    \end{minipage}}
  \subfigure[]{
    \label{fig:mini:subfig:b}
    \begin{minipage}[c]{0.182\textwidth}
      \centering
      \includegraphics[width=1.25in]{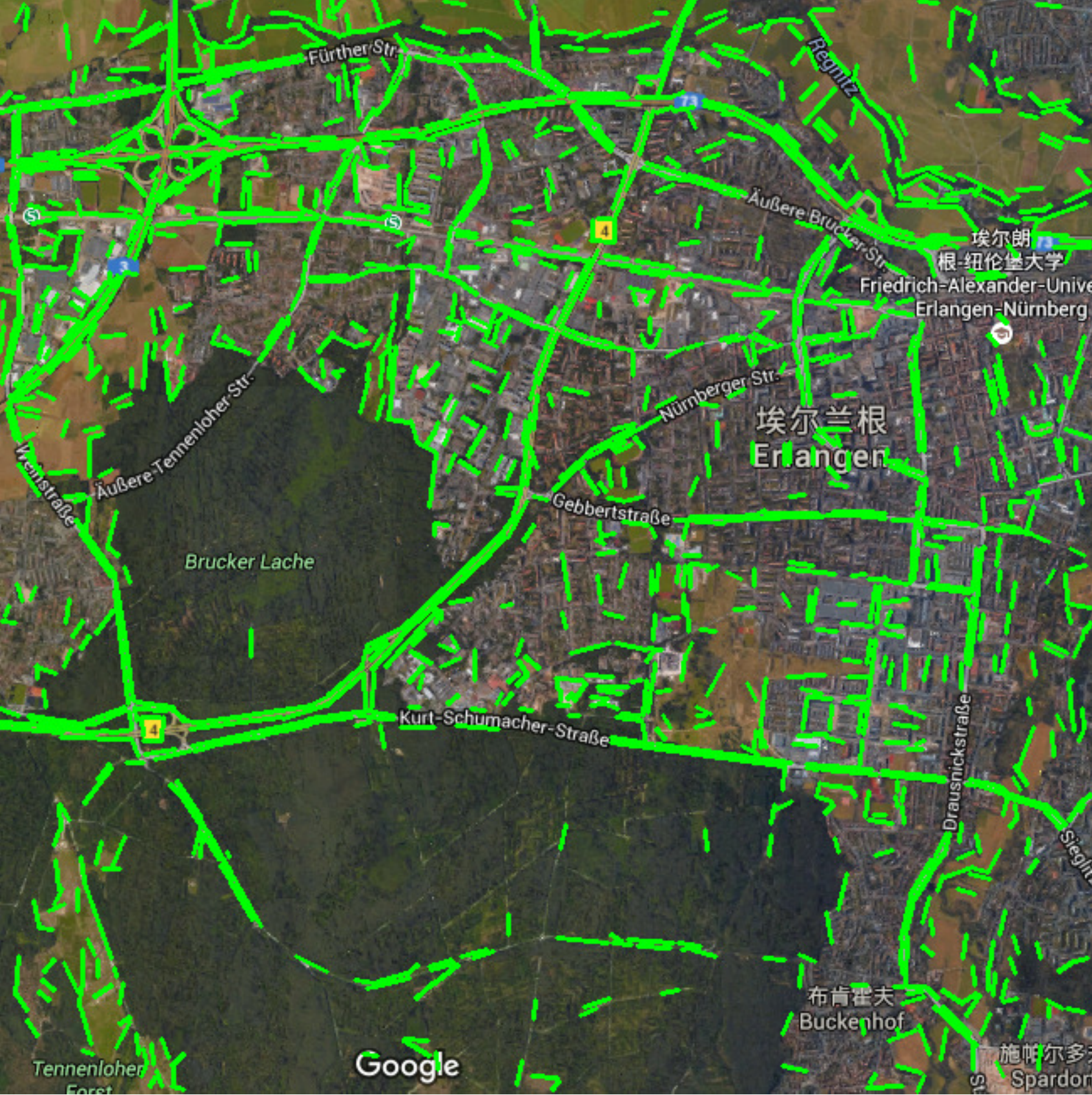}
    \end{minipage}}
  \subfigure[]{
    \label{fig:mini:subfig:a}
    \begin{minipage}[c]{0.182\textwidth}
      \centering
      \includegraphics[width=1.25in]{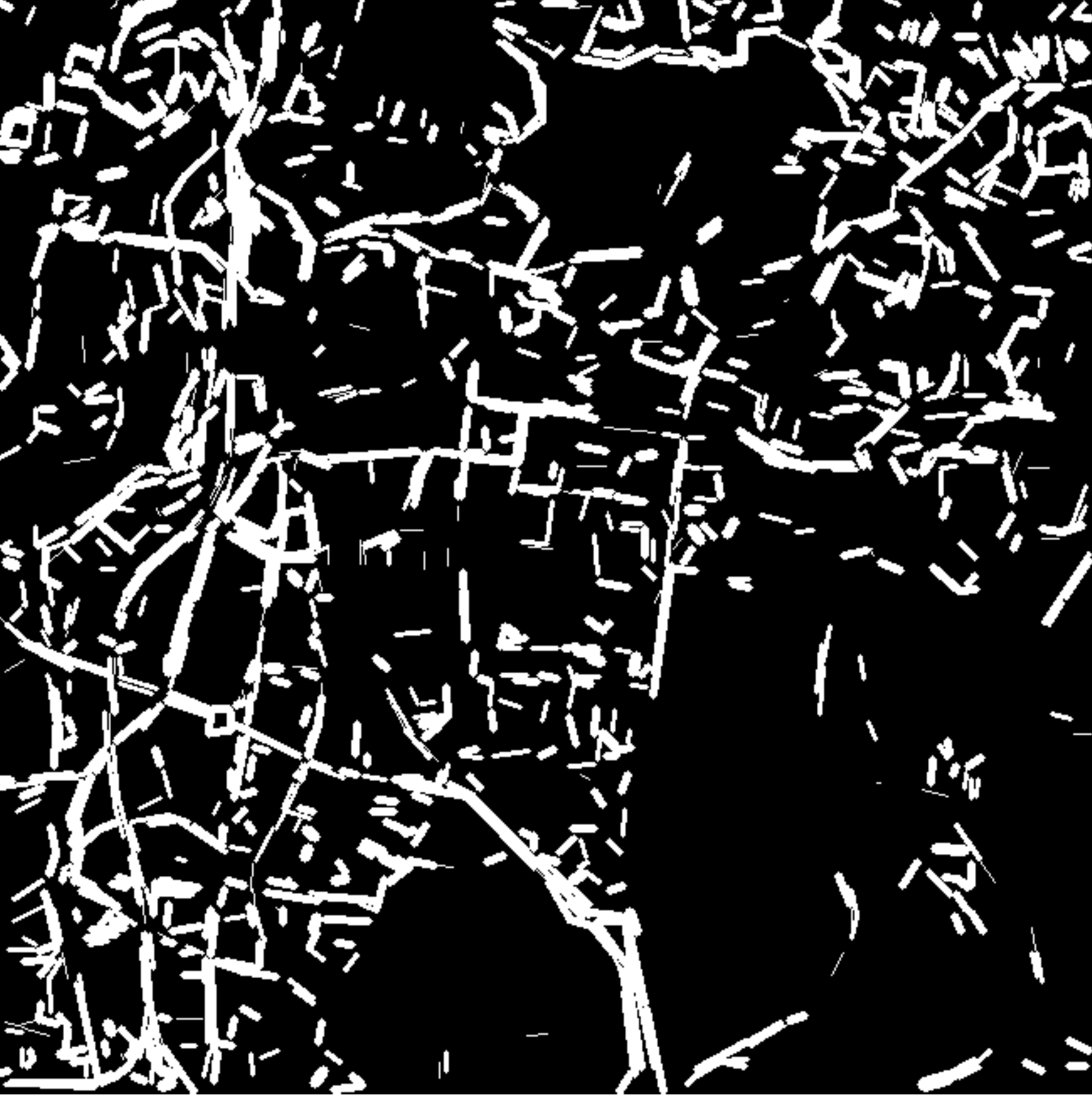}
    \end{minipage}}
  \subfigure[]{
    \label{fig:mini:subfig:b}
    \begin{minipage}[c]{0.182\textwidth}
      \centering
      \includegraphics[width=1.25in]{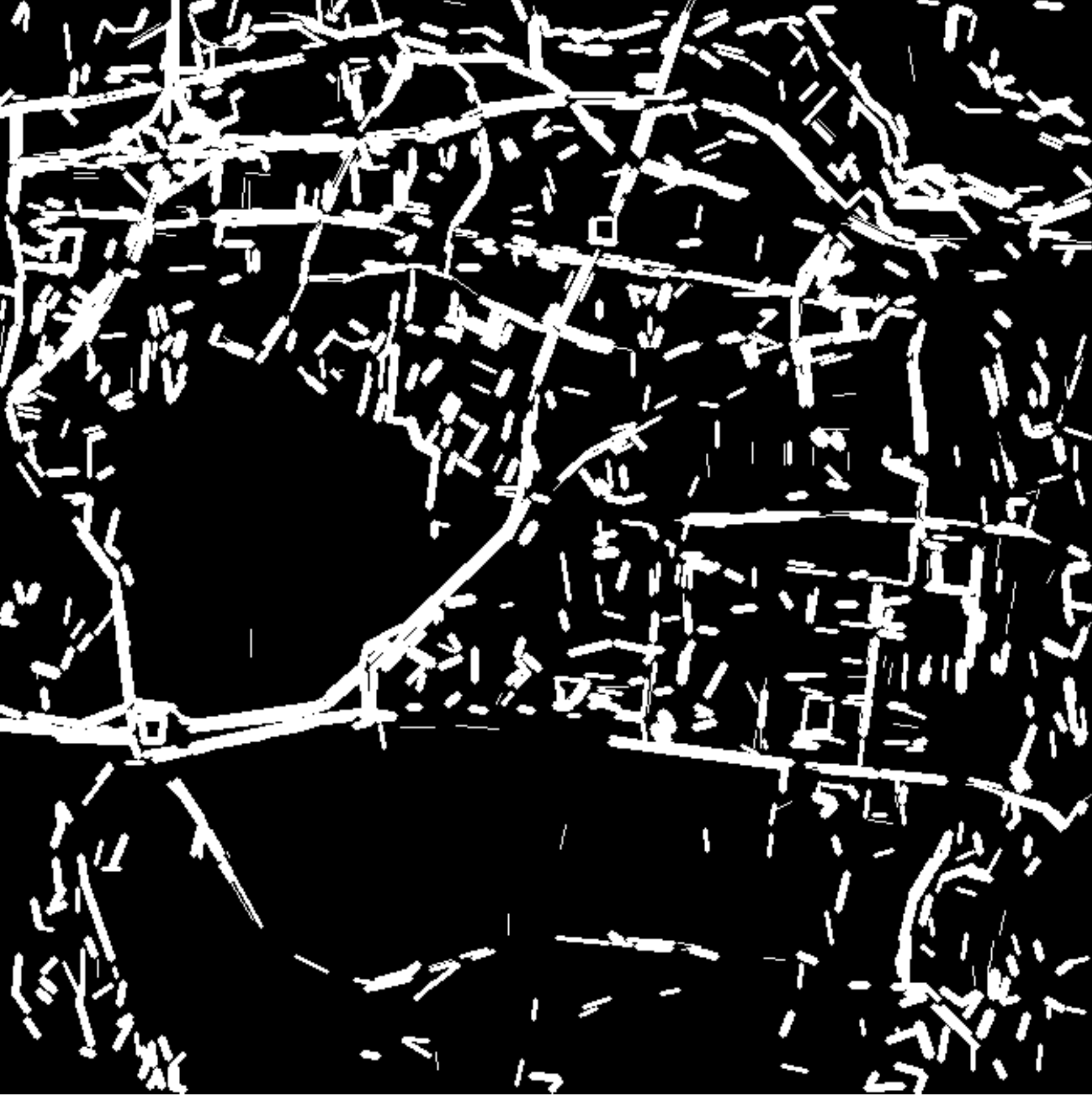}
    \end{minipage}}\\
  \subfigure[]{
    \label{fig:mini:subfig:a}
    \begin{minipage}[c]{0.4\textwidth}
      \centering
      \includegraphics[width=2.5in]{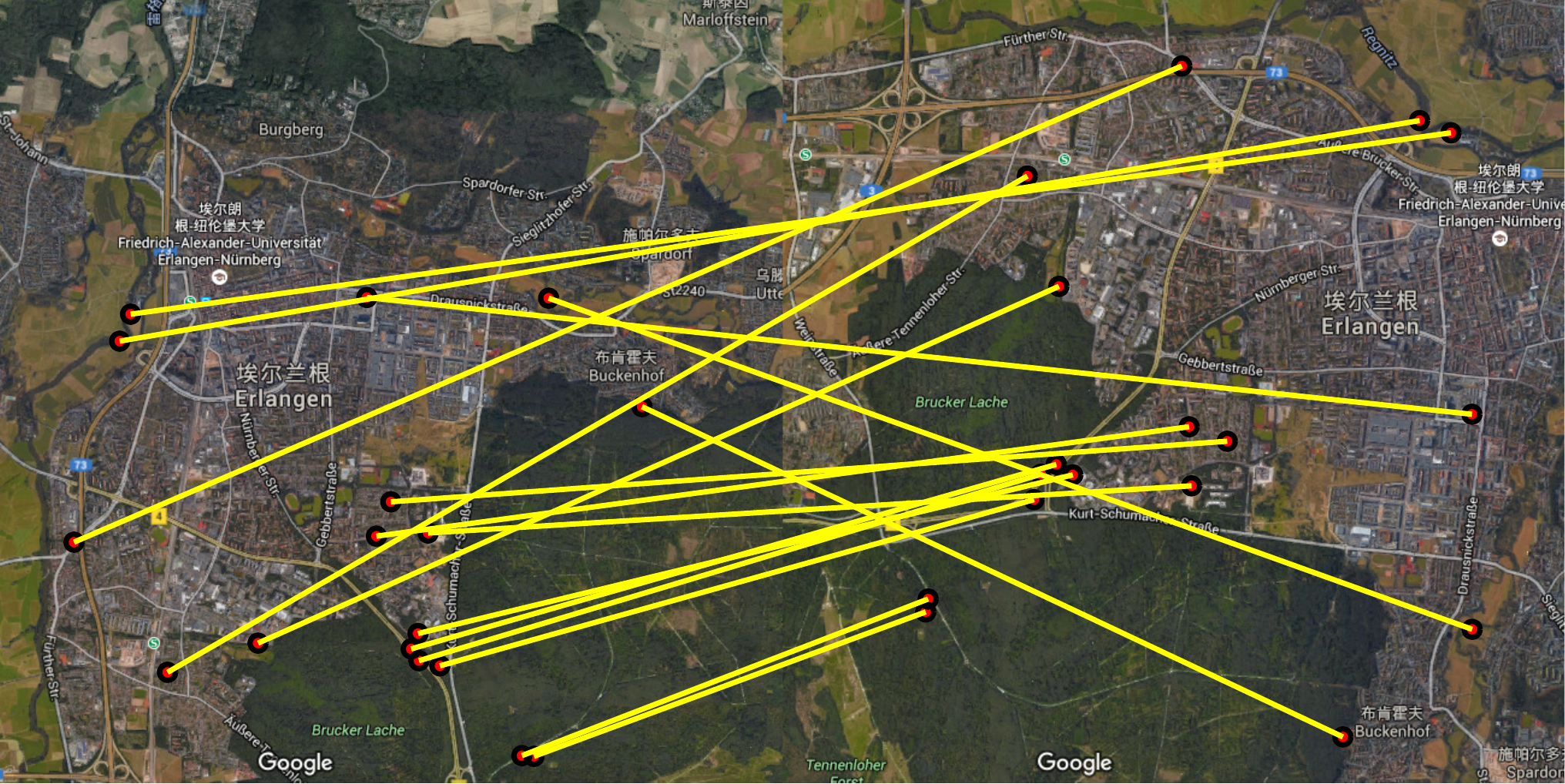}
    \end{minipage}}
  \subfigure[]{
    \label{fig:mini:subfig:b}
    \begin{minipage}[c]{0.4\textwidth}
      \centering
      \includegraphics[width=1.1in]{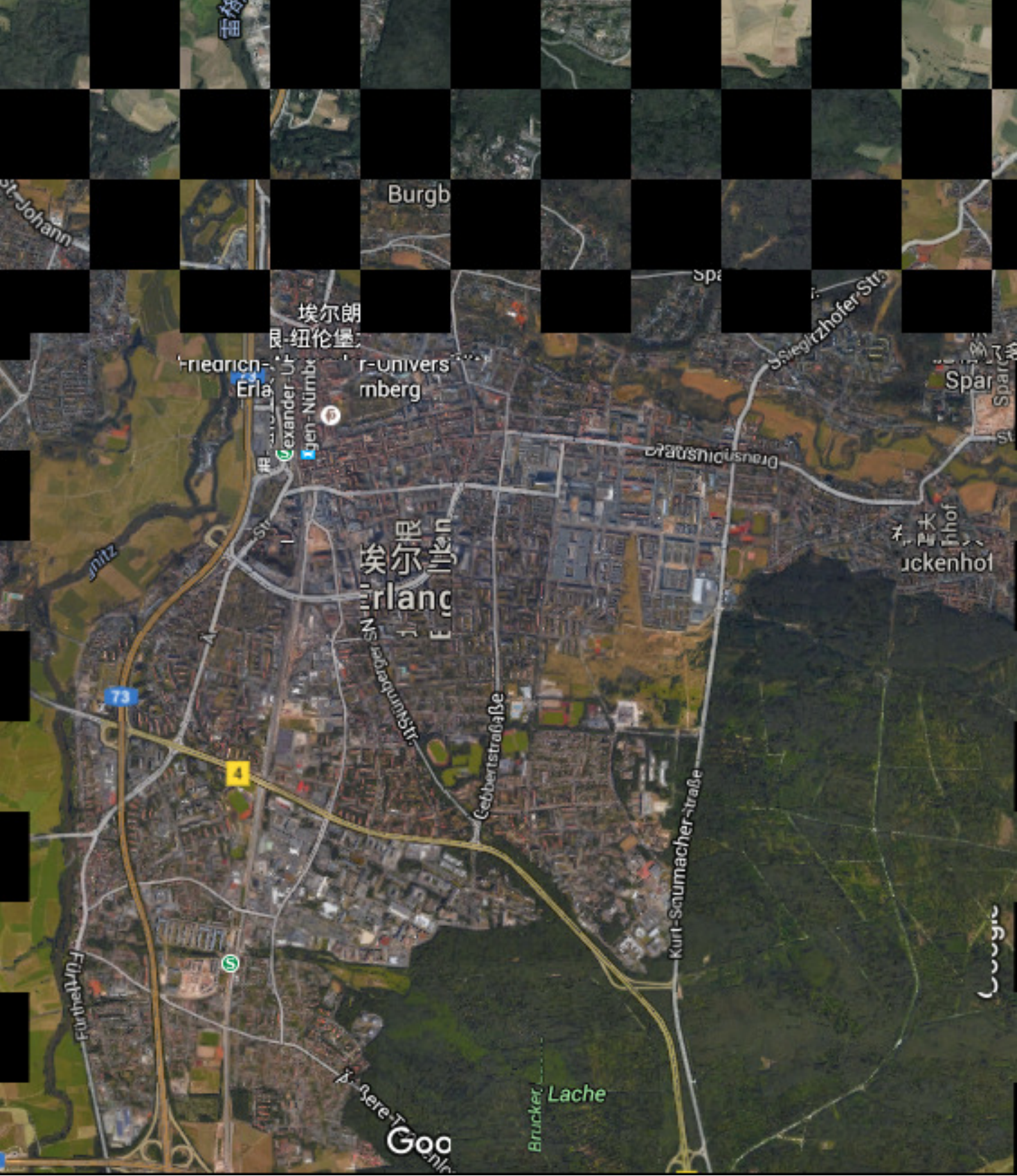}
    \end{minipage}}\\
  \captionstyle{normal}
  \caption{Image registration for ImgSet3-3. (a) Line segments of reference image. (b) Line segments of sensed image. (c) Line-support regions of reference image. (d) Line-support regions of sensed image. (e) Point correspondences by GOR. (f) Checkerboard mosaiced image.}
  \label{fig-13}
\end{figure*}

The examples of the registration process achieved for ImgSet3-1, ImgSet3-2, and ImgSet3-3 are shown in Fig. \ref{fig-11}-\ref{fig-13} respectively.
Distinguished from real images, 2-D map images with computer-generated graphics are mainly composed by linear features.
It is very helpful to detect line segments.
Therefore, most of line graphics in both of the map images are extracted as line segments as shown in Fig. \ref{fig-11} (a)-(b) and Fig. \ref{fig-12} (a)-(b).
The streets can be represented by the line-support regions in Fig. \ref{fig-11} and Fig. \ref{fig-12}.
The map overlaid with real satellite image in ImgSet3-3 is a complicated registration case due to the scale, rotation, and inconsistent text annotations.
Similar to the remote sensing images, line-support regions extracted for ImgSet3-3 are shown in Fig. \ref{fig-13} (c) and (d). For map images in ImgSet3, it is worth to mention that most of the text annotations are excluded in the line-support regions.
Relying on the well-extracted line-support regions, sufficient inliers are obtained by SIFT matching equipped with GOR as shown in Fig. \ref{fig-11}-\ref{fig-13} (e).
By visual inspection of the registered image in Fig. \ref{fig-11}-\ref{fig-13} (f), it appears that the registration is valid and accurate for map images with affine transformation and inconsistent annotations. The registration results presented in Table IV can also verify the effectiveness of the proposed method for registering map images.

\vspace{-0.5cm}

\subsection{Comparisons with Other Automatic Image Registration Methods}
The proposed methodology for automatic image registration is compared with
three automatic registration methods for the previously described and demonstrated image sets.
The first method SIFT-MI in \cite{J_MGG_2014_ITGRS} designs a coarse-to-fine scheme based on SIFT and MI.
The coarse process adopts SIFT approach equipped with the outlier removal, which
removes outliers that scattered away from the cluster of the scale histogram.
The subsequent fine-tuning process is implemented by the maximization of mutual information.
The second method SIFT-WGTM adopts SIFT matching as the initial set of matches, then utilizes Weighted Graph Transformation Matching (WGTM) \cite{J_MI_2012_ITGRS} as outlier rejection.
IS-SIFT \cite{J_HG_2011_ITGRS} is the third considered method, which is the combination of the Otsu's thresholding-based image segmentation and SIFT with bivariate  histogram-based outlier removal.
The performance of these methods is evaluated through the following measures proposed in \cite{J_BZ_2009_IGRSL}:
$N_{red}$ (number of redundant points),
$RMS_{all}$ (rmse considering all residual points together),
$RMS_{LOO}$ (rmse computation of the residual points based on the leave-one-out method), and $BPP(2.0)$ (bad point proportion with norm higher than 2.0).
The obtained registration results are presented in Table III, where the failure cases are marked with `-'.

It can be observed that ALRS-GOR generally outperforms other three registration methods, in particular for remote sensing images with inconsistent spectral contents and map images with inconsistent annotations.
Although SIFT-WGTM has achieved the accuracy results comparable to ALRS-GOR for the images with rigid transformation, their performance is limited for the images of ImgSet1-2 with sheared transformation.
This is because that the angular distance adopted in WGTM is invariant with scaling and rotation deformations, except for sheared transformation.

Both of SIFT-MI and SIFT-WGTM are equipped with SIFT initial matches from the original images.
They cannot deal with more complex situations, such as images with significant spectral differences and inconsistent annotations.
The main reason behind this is that the initial matching by SIFT from inconsistent contents are not able to provide sufficient correct matches.
The performance of the subsequent outlier removal decreases with less initial inliers.
By adopting the Otsu's thresholding-based method as segmentation, IS-SIFT achieves acceptable results in the cases of remote sensing images with small differences in the spectral contents and map images with few annotations.
However, significant differences in image content cannot be excluded by the coarse thresholding-based segmentation. It exhibits in general not precisely enough with the more complex situations, such as optical-SAR image pairs with the presence of noise, or map images with significant inconsistent texts.

Regarding computational efficiency, feature matching is the most time-consuming stage for SIFT-based registration methods. The computational cost depends on the number of detected feature points. Therefore, SIFT-MI and SIFT-WGTM is computationally expensive than IS-SIFT and ALRS-GOR. This is because that they require much higher computation time to detect and match feature points in the original images with more texture details than the segmented images. Compared to IS-SIFT, the proposed method takes more acceptable computational costs with the exceptions of ImgSet2-1 and ImgSet2-2. Re-extraction and re-matching of the proposed method in both of these cases require additional computational costs with multi-resolution strategy.

\vspace{-0.5cm}
\begin{table*}[htb]
\centering
  \captionstyle{normal}
  \setlength{\abovecaptionskip}{0pt}
  \setlength{\belowcaptionskip}{10pt}
\caption{QUANTITATIVE COMPARISONS OF REGISTRATION RESULTS (-: THIS METHOD WAS NOT ABLE TO REGISTER THIS PAIR OF IMAGE.)}
\begin{lrbox}{\tablebox}
\begin{tabular}{|c|c|c|c|c|c|c|c|c|c|c|c|c|c|c|}
\hline
 Pair & AIR & $N_{red}$	& $RMS_{all}$ & $RMS_{LOO}$ & $BPP(2.0)$  & Time(sec) & &
 Pair & AIR & $N_{red}$	& $RMS_{all}$ & $RMS_{LOO}$ & $BPP(2.0)$  & Time(sec) \\
\cline{1-7} \cline{9-15}
\multirow{4}{*}{ImgSet1-1} & SIFT-MI &17&0.510&0.493&0.412&39 &  & \multirow{4}{*}{ImgSet2-4} & SIFT-MI &-&-&-&-&- \\
\cline{2-7} \cline{10-15}
& SIFT-WGTM  &32&0.356&0.307&0.313&33   &&& SIFT-WGTM   &-&-&-&-&- \\
\cline{2-7} \cline{10-15}
& IS-SIFT &5&0.593&0.614&0.200&27  &&& IS-SIFT &7& 39.728&27.632&0.429&22 \\
\cline{2-7} \cline{10-15}
& ALRS-GOR    &7&0.466&0.392&0.143&14  &&& ALRS-GOR    &9&0.459&1.091&0.111&17 \\
\cline{1-7} \cline{9-15}
\multirow{4}{*}{ImgSet1-2} & SIFT-MI &-&-&-&-&- &  &
\multirow{4}{*}{ImgSet2-5} & SIFT-MI &-&-&-&-&- \\
\cline{2-7} \cline{10-15}
 &SIFT-WGTM & 21&37.726&26.349&0.476&24   &&&SIFT-WGTM   &-&-&-&-&- \\
\cline{2-7} \cline{10-15}
& IS-SIFT&29&22.041&20.730&0.069&15   &  & &IS-SIFT & 17&0.655&0.620&0.118&37 \\
\cline{2-7} \cline{10-15}
& ALRS-GOR    &16&0.289&0.305&0.000&19   &&& ALRS-GOR  & 11&0.274&0.233&0.000&28 \\
\cline{1-7} \cline{9-15}
\multirow{4}{*}{ImgSet2-1} & SIFT-MI &28&20.147&20.166&20.071&274 &  &
\multirow{4}{*}{ImgSet3-1} & SIFT-MI &-&-&-&-&- \\
\cline{2-7} \cline{10-15}
 &SIFT-WGTM &33&0.205&0.273&0.030&60   &&&SIFT-WGTM   &14&0.281&0.326&0.071&49 \\
\cline{2-7} \cline{10-15}
& IS-SIFT&7&0.392&0.361&0.000&48   &&&IS-SIFT & 26& 0.626&0.654&0.115&32 \\
\cline{2-7} \cline{10-15}
& ALRS-GOR    &10&0.168&0.252&0.000&56   &&& ALRS-GOR  & 37&0.304&0.380&0.054&34 \\
\cline{1-7} \cline{9-15}
\multirow{4}{*}{ImgSet2-2} & SIFT-MI &4&0.493&0.607&0.250&49 &  &
\multirow{4}{*}{ImgSet3-2} & SIFT-MI &-&-&-&-&- \\
\cline{2-7} \cline{10-15}
 &SIFT-WGTM &5&0.354&0.362&0.000&42   &&&SIFT-WGTM   &-&-&-&-&- \\
\cline{2-7} \cline{10-15}
& IS-SIFT&9&0.308&0.214&0.222&37   &&&IS-SIFT &-&-&-&-&-  \\
\cline{2-7} \cline{10-15}
& ALRS-GOR    &11&0.275&0.209&0.000&54   &&& ALRS-GOR  & 11&0.162&0.187&0.000&21 \\
\cline{1-7} \cline{9-15}
\multirow{4}{*}{ImgSet2-3} & SIFT-MI &-&-&-&-&- &  &
\multirow{4}{*}{ImgSet3-3} & SIFT-MI &-&-&-&-&- \\
\cline{2-7} \cline{10-15}
 &SIFT-WGTM &-&-&-&-&-   &&&SIFT-WGTM   &-&-&-&-&- \\
\cline{2-7} \cline{10-15}
& IS-SIFT&14&0.702&1.680&0.143&32   &&&IS-SIFT &-&-&-&-&-  \\
\cline{2-7} \cline{10-15}
& ALRS-GOR    &10&0.318&0.475&0.100&29   &&& ALRS-GOR  & 17&0.493&0.417&0.118&40 \\
\hline
\end{tabular}
\end{lrbox}
\scalebox{0.65}{\usebox{\tablebox}}
\end{table*}

\subsection{Sensitivity to the Stopping Condition Threshold ($\epsilon$)}
The proposed ALRS-GOR relies on the stopping threshold $\epsilon$ to terminate the iterative process.  The value of $\epsilon$ depends on the required accuracy of registration.
The following experimental results in this section  provide a brief analysis of the sensitivity of the algorithm to the stopping condition threshold $\epsilon$.

\begin{figure*}[htb]
\centering
 \setlength{\abovecaptionskip}{0pt}
 \setlength{\belowcaptionskip}{0pt}
 \setlength{\intextsep}{8pt plus 3pt minus 2pt}
  \subfigure[]{
    \label{fig:mini:subfig:a}
    \begin{minipage}[c]{0.3\textwidth}
      \centering
      \includegraphics[width=2in]{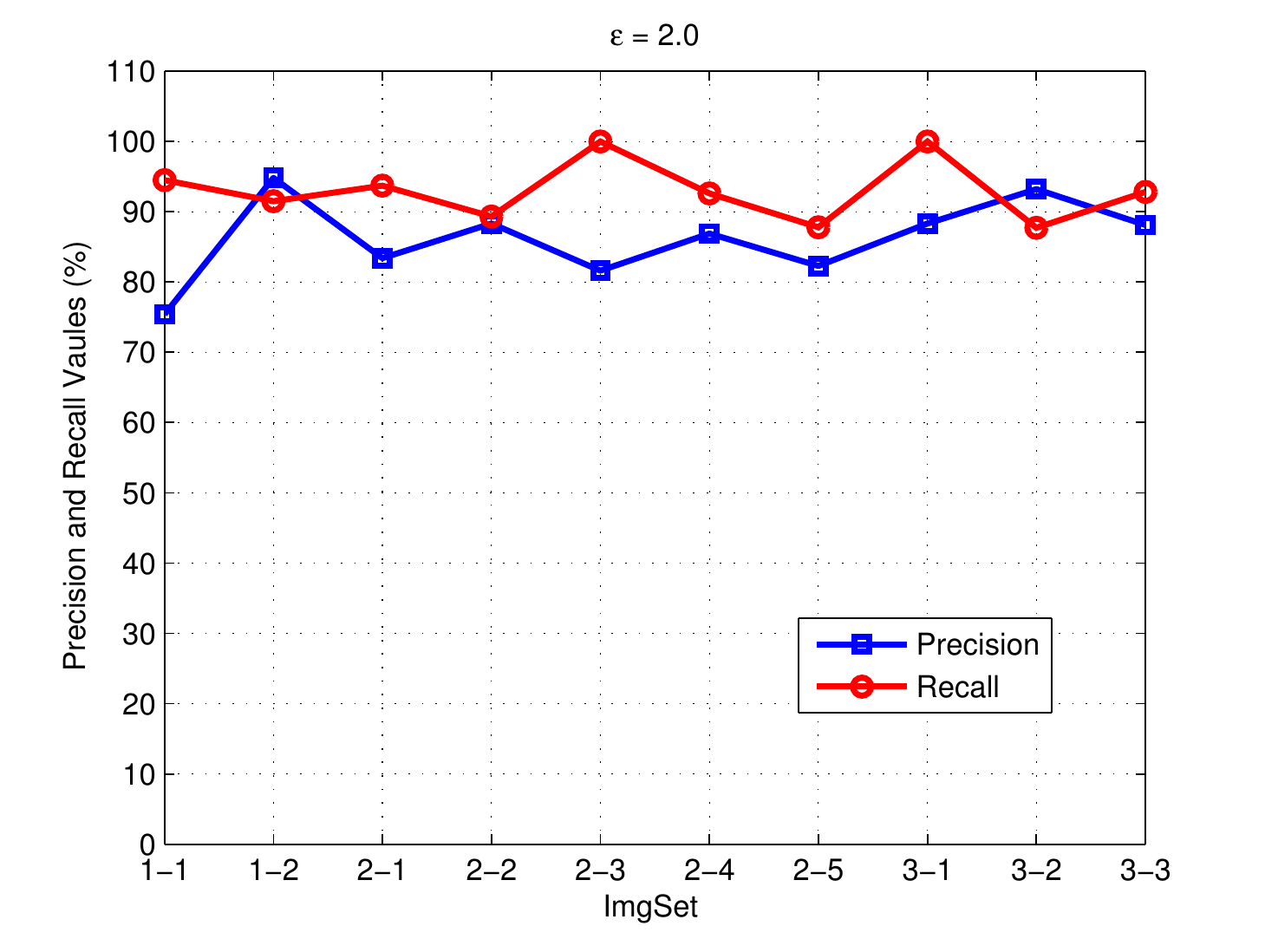}
    \end{minipage}}
  \subfigure[]{
    \label{fig:mini:subfig:b}
    \begin{minipage}[c]{0.3\textwidth}
      \centering
      \includegraphics[width=2in]{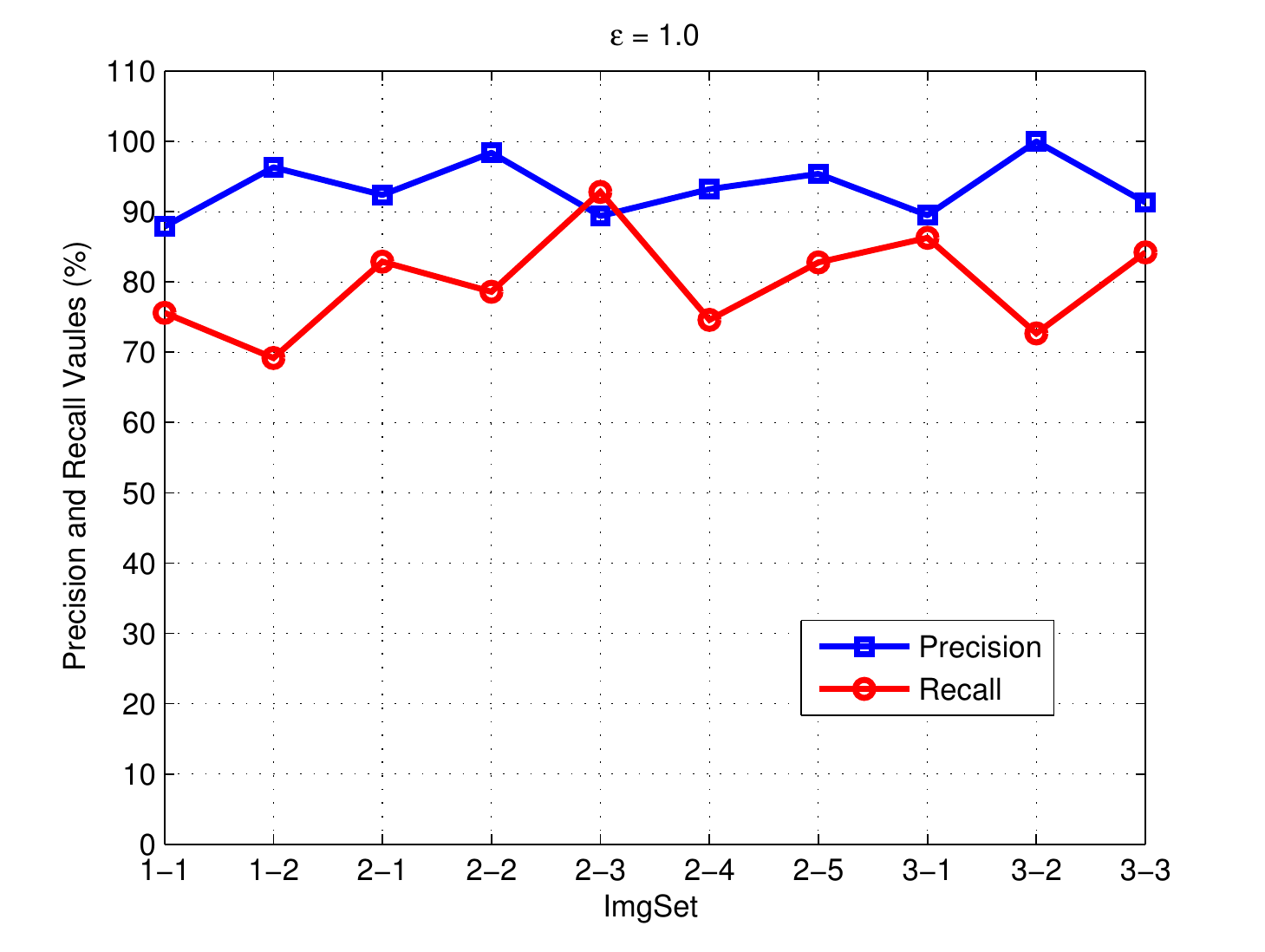}
    \end{minipage}}
  \subfigure[]{
    \label{fig:mini:subfig:a}
    \begin{minipage}[c]{0.3\textwidth}
      \centering
      \includegraphics[width=2in]{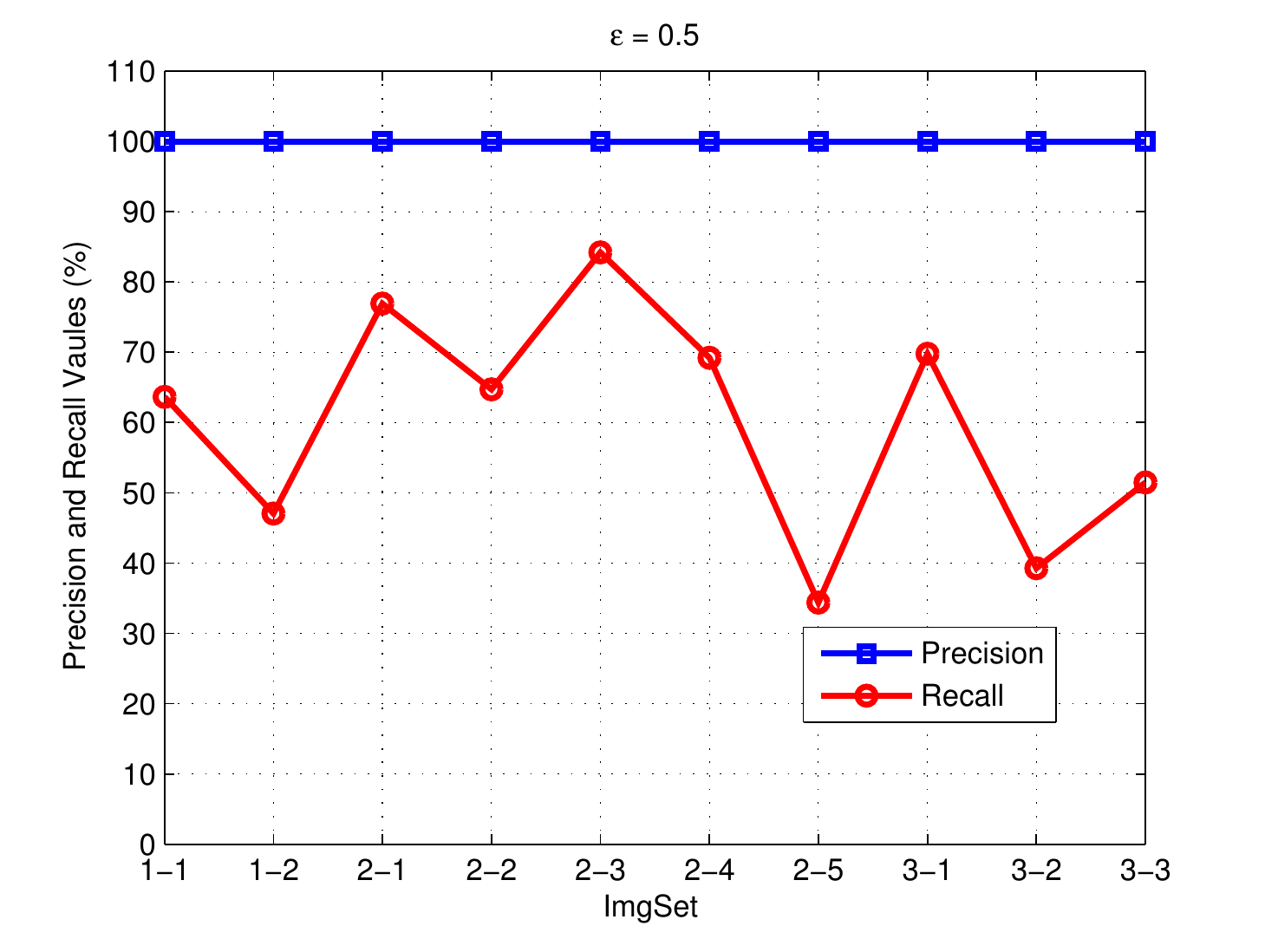}
    \end{minipage}}\\
  \captionstyle{normal}
  \caption{Performance comparisons with different stopping threshold $\epsilon$. (a) $\epsilon=2.0$. (b) $\epsilon=1.0$. (c) $\epsilon=0.5$.}
  \label{fig-e}
\end{figure*}

Fig. \ref{fig-e} shows the values of precision and recall for the three image datasets with different setting of $\epsilon$, i.e., $\epsilon=2.0$, $\epsilon=1.0$, and $\epsilon=0.5$.
It can be observed from Fig. \ref{fig-e} (a)-(c) that the smaller values of $\epsilon$ lead to higher precisions but lower recalls.
In order to achieve a smaller RMSE, more outliers can be identified with more iterations.
Accordingly, more inlier are easily removed in the outlier removal iterations.
Besides that, a smaller stopping threshold with re-extraction and re-matching inevitably increases the time complexity of the proposed algorithm.
Therefore, it is a trade-off to set the values of $\epsilon$ between the precision values, recall values, and the time complexity.

%

\vspace{-0.5cm}

\section{CONCLUSION AND FUTURE WORK}\label{sec:model}
	The paper has reported a novel automatic registration approach for images with affine transformation and inconsistent content. This approach consists of an iterative line-support region segmentation and SIFT matching equipped with geometrical outlier removal. To begin with, line-support regions is proposed to extract linear features.  Next, feature point extraction and matching for the segmented images is implemented by SIFT approach equipped with geometrical outlier removal. The proposed feature matching method GOR rejects outliers and preserves inliers based on comparing the disparity of geometrical relationships. Furthermore, an iterative strategy with multi-resolution is employed to improve the incompleteness of line segments.
In our experiments, we have tested the proposed method on image pairs in the following situations: aerial images with simulated affine deformations, multispectral remote sensing image pairs, and 2-D map image pairs. The experimental results show that the proposed method obviously improves the accuracy of registration and achieves acceptable computational efficiency.
One of our future work includes incorporating proper clustering techniques into outlier removal for a more effective feature point matching.
Regarding the registration of images with the differences of the terrain height, the affine transformation model is not suitable.
Other more reasonable transformation models should be considered to handle the influence of differences in the terrain height.

\vspace{-0.2cm}

\bibliographystyle{IEEEtran}
\bibliography{MyReference}

\begin{IEEEbiography}[{\includegraphics[width=1in,height=1.25in,clip,keepaspectratio]{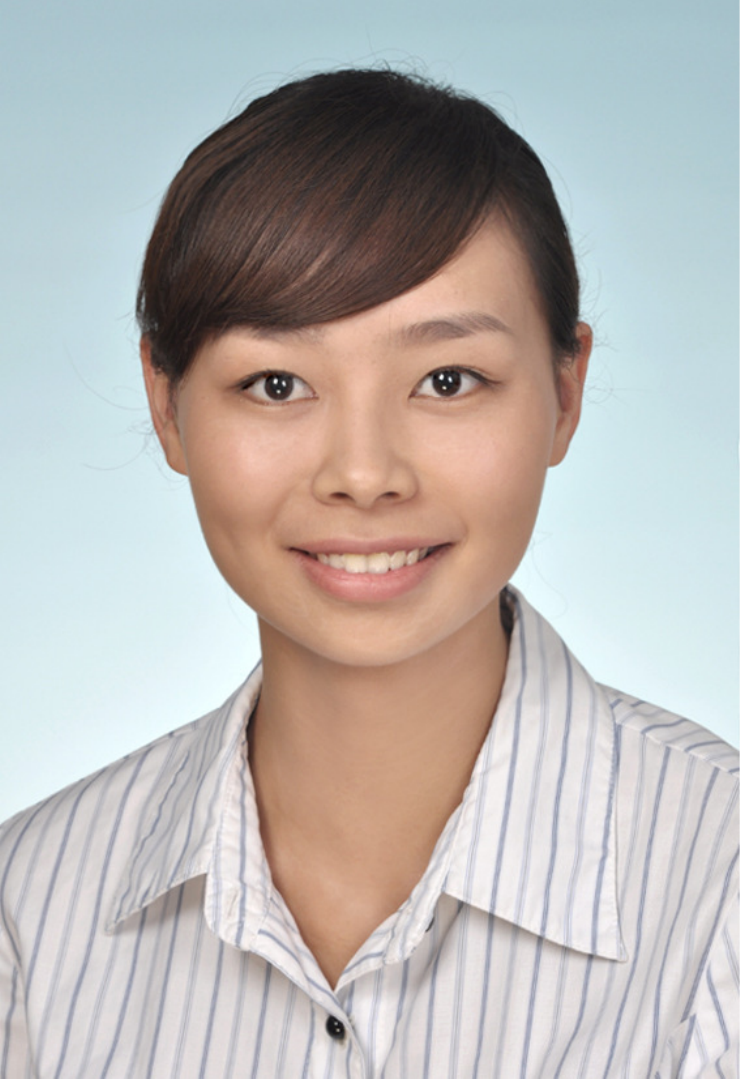}}]
{Ming Zhao}
received the B.S. degree in telecommunication engineering from Wuhan University, Wuhan, China, in July 2007,
the Ph.D. degree in physical electronics with Shanghai Institute of Technical Physics, the institute of Chinese Academy of Sciences (CAS),
Shanghai, China, in June 2012.

She is currently an associate professor with Shanghai Maritime University, China.
From 2015 to 2016, she was a Visiting Scholar with Friedrich-Alexander University Erlangen-N$\ddot{u}$rnberg, Germany.
Her research interests include remote sensing image processing, pattern recognition, and computer vision.
\end{IEEEbiography}

\begin{IEEEbiography}[{\includegraphics[width=1in,height=1.25in,clip,keepaspectratio]{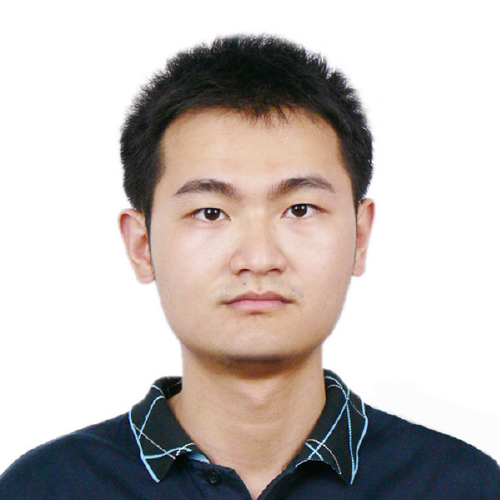}}]
{Yongpeng Wu} (S'08--M'13)
received the B.S. degree in telecommunication engineering from Wuhan University, Wuhan, China, in July 2007, the Ph.D. degree in communication and signal processing with the National Mobile Communications Research Laboratory, Southeast University, Nanjing, China, in November 2013.

Dr. Wu is currently a Tenure-Track Associate Professor with the Department of Electronic Engineering, Shanghai Jiao Tong University, China. Previously, he was a Senior Research Fellow with Institute for Communications Engineering, Technical University of Munich, Germany and the Humboldt Research Fellow and the Senior Research Fellow with Institute for Digital Communications, University Erlangen-N$\ddot{u}$rnberg, Germany. During his doctoral studies, he conducted collaborative research at the Department of Electrical Engineering, Missouri University of Science and Technology, USA. His research interests include massive MIMO/MIMO systems, physical layer security, signal processing for wireless communications, and multivariate statistical theory.

Dr. Wu was awarded the IEEE Student Travel Grant for IEEE International Conference on Communications (ICC) 2010, the Alexander von Humboldt Fellowship in 2014, the Travel Grant for IEEE Communication Theory Workshop 2016, the Excellent Doctoral Thesis Award of China Communications Society 2016, and the Excellent Editor Award of IEEE Communications Letters 2017. He was an Exemplary Reviewer of the IEEE Transactions on Communications in 2015 and 2016. He is the lead guest editor for the special issue ``Physical Layer Security for 5G Wireless Networks'' of the IEEE Journal on Selected Areas in Communications. He is currently an editor of the IEEE Access and IEEE Communications Letters. He has been a TPC member of various conferences, including Globecom, ICC, VTC, and PIMRC, etc. Dr. Wu is an expect on physical layer security for 5G wireless networks and has rich experiences on handing technical papers.
\end{IEEEbiography}

\begin{IEEEbiography}[{\includegraphics[width=1in,height=1.25in,clip,keepaspectratio]{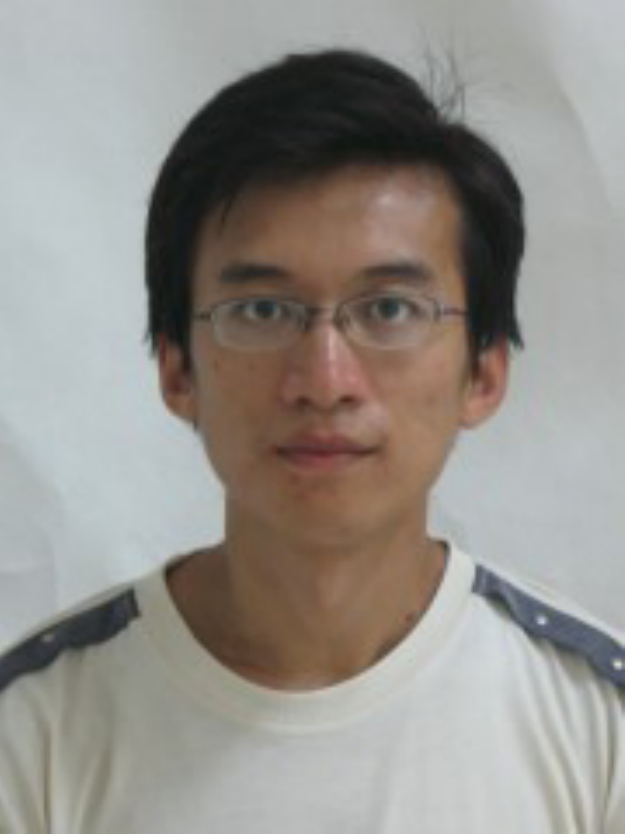}}]
{Shengda Pan}
received his Ph.D. degree in circuits and systems from  University of Chinese Academy of Sciences, Shanghai, China, in 2013. He is currently a lecturer in Shanghai Maritime University , Shanghai, China. His research interests include remote sensing  image processing and pattern recognition.

He is currently a full professor with Shanghai Maritime University, China.
His research interests include photoelectric signal acquisition and remote sensing image processing.
\end{IEEEbiography}

\begin{IEEEbiography}[{\includegraphics[width=1in,height=1.25in,clip,keepaspectratio]{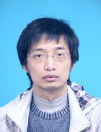}}]
{Fan Zhou}
is a lecturer in Shanghai Maritime University , Shanghai, China. He received his Ph.D. in school of remote sensing and information engineering at Wuhan University in 2014. His research interests lie in computer vision, photogrammetry and Remote sensing

He is currently a full professor with Shanghai Maritime University, China.
His research interests include photoelectric signal acquisition and remote sensing image processing.
\end{IEEEbiography}

\begin{IEEEbiography}[{\includegraphics[width=1in,height=1.25in,clip,keepaspectratio]{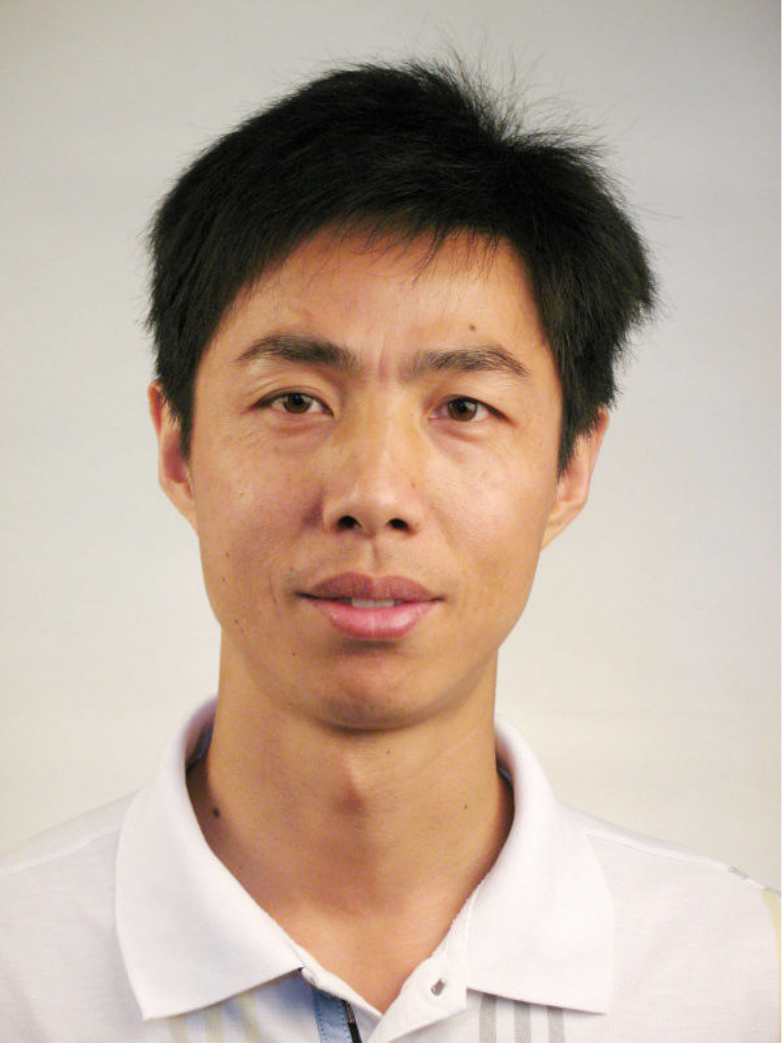}}]
{Bowen An}
received the M.S. degree in communication and information engineering from Wuhan University, Wuhan, China, in April 2004,
the Ph.D. degree in circuits and systems with Shanghai Institute of Technical Physics, the institute of Chinese Academy of Sciences (CAS),
Shanghai, China, in July 2006.

He is currently a full professor with Shanghai Maritime University, China.
His research interests include photoelectric signal acquisition and remote sensing image processing.
\end{IEEEbiography}

\begin{IEEEbiography}[{\includegraphics[width=1in,height=1.25in,clip,keepaspectratio]{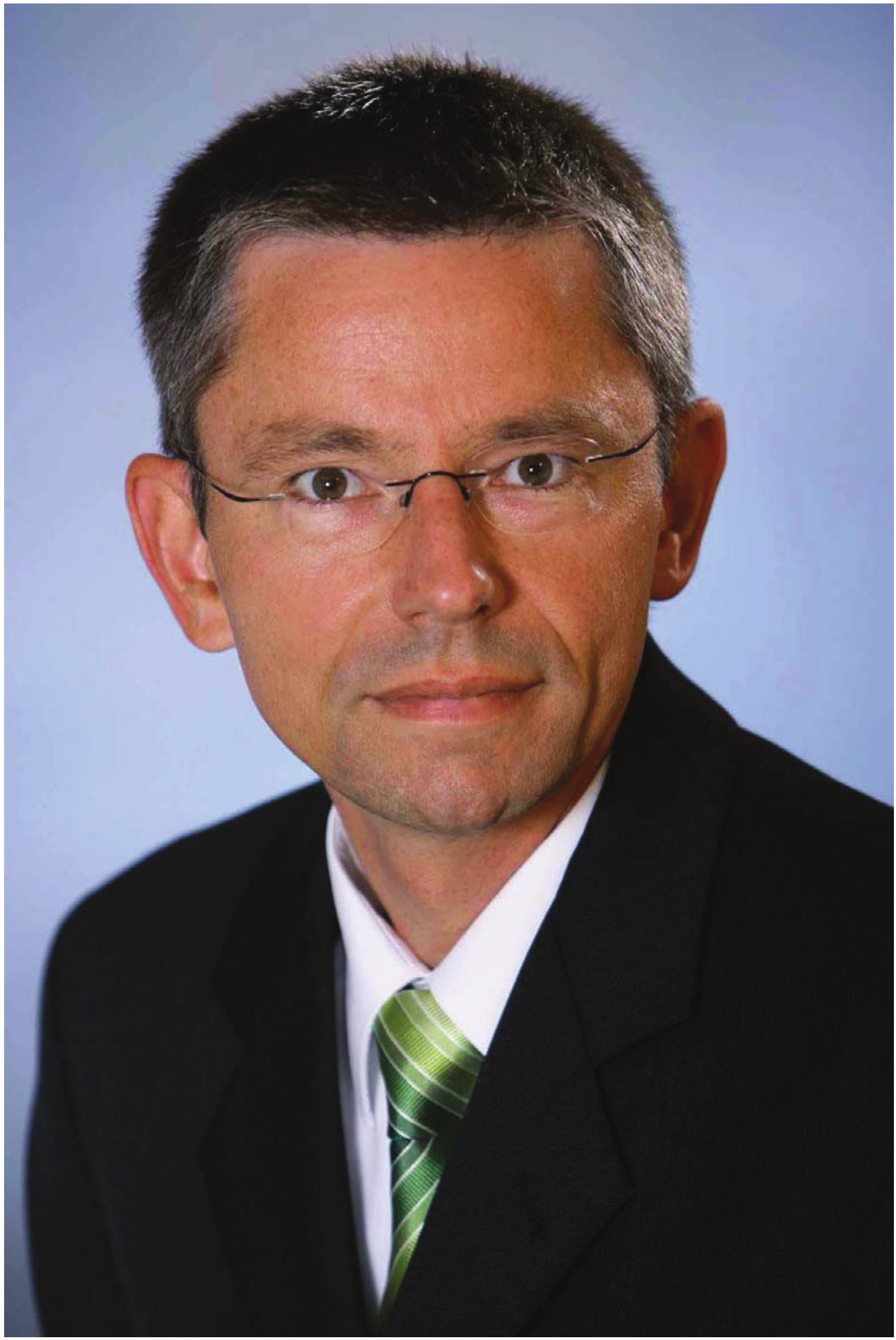}}]
{Andr\'{e} Kaup} (M'96--SM'99--F'13)
received the Dipl.-Ing. and Dr.-Ing. degrees in electrical engineering from Rheinisch-Westf\"{a}lische Technische Hochschule (RWTH) Aachen University, Aachen, Germany, in 1989 and 1995, respectively.

He was with the Institute for Communication Engineering, RWTH Aachen University, from 1989 to 1995. He joined the Networks and Multimedia Communications Department, Siemens Corporate Technology, Munich, Germany, in 1995 and became Head of the Mobile Applications and Services Group in 1999. Since 2001 he has been a Full Professor and the Head of the Chair of Multimedia Communications and Signal Processing, University of Erlangen- Nuremberg, Erlangen, Germany. From 1997 to 2001 he was the Head of the German MPEG delegation. From 2005 to 2007 he was a Vice Speaker of the DFG Collaborative Research Center 603. From 2015 to 2017 he served as Head of the Department of Electrical Engineering and Vice Dean of the Faculty of Engineering. He has authored around 350 journal and conference papers and has over 70 patents granted or pending. His research interests include image and video signal processing and coding, and multimedia communication.

Andr\'{e} Kaup is a member of the IEEE Multimedia Signal Processing Technical Committee, a member of the scientific advisory board of the German VDE/ITG, and a Fellow of the IEEE. He served as an Associate Editor for IEEE TRANSACTIONS ON CIRCUITS AND SYSTEMS FOR VIDEO TECHNOLOGY and was a Guest Editor for IEEE JOURNAL OF SELECTED TOPICS IN SIGNAL PROCESSING. From 1998 to 2001 he served as an Adjunct Professor with the Technical University of Munich, Munich. He was a Siemens Inventor of the Year 1998 and received the 1999 ITG Award. He has received several best paper awards, including the Paul Dan Cristea Special Award from the International Conference on Systems, Signals, and Image Processing in 2013. His group won the Grand Video Compression Challenge at the Picture Coding Symposium 2013 and he received the Teaching Award of the Faculty of Engineering in 2015.

\end{IEEEbiography}

\newpage

\end{document}